\documentclass[aps,preprint,floatfix,nofootinbib,showpacs]{revtex4-1}
\pdfoutput=1
\usepackage{dcolumn}
\usepackage{bm}
\usepackage{graphicx}
\usepackage{amssymb,amsmath}

\usepackage{multirow}
\usepackage{units,changes}
\usepackage{color,url}
\usepackage{tabu}
\usepackage{array}
\usepackage[colorlinks=true,urlcolor=blue,anchorcolor=blue
,citecolor=blue,filecolor=blue,linkcolor=blue,menucolor=blue
,linktocpage=true,pdfproducer=medialab,pdfa=true]{hyperref}
\usepackage{colordvi}
\usepackage{subcaption}
\def\beqa{\begin{eqnarray}}
\def\eeqa{\end{eqnarray}}

\usepackage{soul}
\usepackage{slashed}
\usepackage[labelfont=bf, justification=raggedright, singlelinecheck=false]{caption}

\begin{document}
\preprint{CPTNP-2025-048}

\title{Probing gluons-enriched dark jets from Higgs boson exotic decays at the LHC}
\def\slash#1{#1\!\!\!/}

\author{Wanyun Chen$^1$}
\email{241002026@njnu.edu.cn}
\author{Chih-Ting Lu$^{1,2}$}
\email{ctlu@njnu.edu.cn}
\author{Hanxin Shen$^1$}
\email{241002031@njnu.edu.cn}
\affiliation{$^1$ Department of Physics and Institute of Theoretical Physics, Nanjing Normal University, Nanjing, 210023, P. R. China}
\affiliation{$^2$ Nanjing Key Laboratory of Particle Physics and Astrophysics, Nanjing, 210023, China}

\begin{abstract} 
The dark sector may possess a rich structure yet to be uncovered, and a QCD-like dark sector with GeV-scale dark hadrons can yield novel signatures at the Large Hadron Collider (LHC). In this work, we focus on a light singlet pseudoscalar mediator that connects the QCD-like dark sector to the Standard Model (SM) sector via the Higgs portal. Notably, when the lightest unstable dark meson has a mass of approximately $3$ GeV, it predominantly decays into a pair of gluons and behaves as a long-lived particle, a scenario that has received relatively little attention. We consider various Higgs production channels at the LHC and investigate two processes for generating dark mesons: (1) the cascade decay of the Higgs boson into a pair of light pseudoscalar mediators, which subsequently decay into four dark mesons; and (2) the dark shower and hadronization process whereby the Higgs boson decays into a pair of dark quarks that subsequently evolve into dark mesons. These processes give rise to novel gluon-rich dark jets composed of long-lived dark mesons. Notably, we find that appropriate trigger selection constitutes a crucial factor for detecting these signal signatures in both tracker system and CMS muon system. At the high-luminosity LHC, the exotic Higgs branching ratio to cascade decays (dark showers) can be constrained below $\mathcal{O}(10^{-5}–10^{-1})$ [$\mathcal{O}(10^{-5}–10^{-2})$] for dark meson proper lifetimes $c\tau$ ranging from $1$~mm to $100$~m.
\end{abstract}

\maketitle

\section{Introduction}

Understanding the nature of dark matter (DM) remains one of the greatest mysteries in particle physics, astrophysics, and cosmology~\cite{Bertone:2004pz,Arbey:2021gdg,Cirelli:2024ssz}. Despite years of exploration into DM particle properties, its true nature continues to elude us. Given that the visible sector is characterized by the rich structure of the Standard Model (SM), it is plausible that the dark sector may also exhibit a similarly complex structure, rather than consisting solely of a single DM particle. Among various dark sector models~\cite{Battaglieri:2017aum,Cirelli:2024ssz}, the QCD-like dark sector~\cite{Blinnikov:1982eh,Blinnikov:1983gh,Khlopov:1989fj,Strassler:2006im,Alves:2009nf,SpierMoreiraAlves:2010err,Bai:2013xga,Hochberg:2014kqa}, characterized by a gauge symmetry \(SU(N)_d\) (with \(N \geq 2\)), displays particularly rich phenomenology. In this framework, the DM candidate is a composite particle rather than a fundamental one, and numerous bound states arise at low energies~\cite{Kribs:2016cew,Lonsdale:2017mzg}, while dark quarks produced at high energies can undergo a showering and hadronization processes~\cite{Strassler:2006im,Bai:2013xga}. These unique features, which are difficult to mimic in other dark sector models, have emerged as a distinctive hallmark of the QCD-like dark sector.

Models featuring a QCD-like dark sector have received increasing attention because they can give rise to new DM annihilation mechanisms~\cite{Hochberg:2014kqa,Berlin:2018tvf,Bernreuther:2019pfb}, address the matter–antimatter asymmetry problem~\cite{Bai:2013xga,Lonsdale:2018xwd,Ibe:2018juk,Zhang:2021orr,Bottaro:2021aal}, trigger a first-order phase transition in the early universe~\cite{Schwaller:2015tja,Tsumura:2017knk,Aoki:2017aws,Hall:2019rld,Hall:2021zsk,Reichert:2021cvs}, and offer insights into small-scale astrophysical challenges~\cite{Hochberg:2014kqa,Tsai:2020vpi,Cline:2022leq}. To connect the QCD-like dark sector with the SM sector, one must consider one or more portal interactions, such as scalar~\cite{Strassler:2006ri,Knapen:2021eip}, axion-like~\cite{Hochberg:2018rjs}, dark photon~\cite{Lee:2015gsa,Hochberg:2015vrg}, \(Z'\)~\cite{Strassler:2006im}, or \(t\)-channel~\cite{Bai:2013xga} mediators. For simplicity, in this study we assume that the QCD-like dark sector is not charged under the SM gauge symmetry and that its connection to the SM occurs exclusively through the Higgs portal~\cite{Strassler:2006ri,Knapen:2021eip,Appelquist:2015yfa,Ishida:2016fbp,Lu:2023gjk,Cazzaniga:2025piw}.

To study the QCD-like dark sector in collider experiments, we can classify these scenarios into three types based on the mass spectrum of the mediator and the dark hadrons. First, if the mediator mass is of the same order as the dark hadron masses, one can study either mediator decays into dark hadrons or into SM particles. Investigations of strongly interacting massive particles (SIMPs) in fixed target experiments~\cite{Berlin:2018tvf,Kamada:2021cow,Kuwahara:2023vfc} or electroweak dark hadrons at the LHC~\cite{Kribs:2018ilo,Butterworth:2021jto} fall under this scenario. Second, when the mediator is much heavier than the dark confinement scale \(\Lambda_d\), the produced mediator decays directly into dark quarks, which then undergo dark showers and hadronization to form ``dark jets" including multiple unstable/stable  dark hadrons. Exploring Hidden Valley models at the LHC exemplifies this scenario~\cite{Strassler:2006im}. Depending on the gauge coupling strength and the lifetimes of these dark hadrons, several novel signatures may arise, including emerging jets (EJs)~\cite{Schwaller:2015gea,Renner:2018fhh,CMS:2018bvr,Linthorne:2021oiz,Archer-Smith:2021ntx,Carrasco:2023loy,CMS:2024gxp,ATLAS:2025bsz}, semivisible jets (SVJs)~\cite{Cohen:2015toa,Cohen:2017pzm,CMS:2021dzg,ATLAS:2023swa,ATLAS:2025kuz}, and soft unclustered energy patterns (SUEPs)~\cite{Knapen:2016hky,CMS:2024nca} and others~\cite{Park:2017rfb,Mies:2020mzw,ATLAS:2023kao,Carmona:2024tkg,Carrasco:2025bct}. Finally, scenarios involving mediators at the tens-of-GeV scale and dark hadrons at the GeV scale have received relatively less attention. In this situation, dark jets can be produced via both cascade decays and dark showers~\cite{Lu:2023gjk}. Therefore, we aim to study the last scenario and explore novel signal signatures at the LHC in this work.

Specifically, we consider Higgs exotic decays from four primary Higgs production channels at the LHC: gluon fusion (ggH), vector boson fusion (VBF), associated production with a $W/Z$ (WH/ZH), and associated production with a \(t\bar{t}\) pair ($t\bar{t}H$). We then explore two mechanisms for producing light dark mesons. The first mechanism involves the cascade decay of the Higgs boson into a pair of pseudoscalar mediators, which subsequently decay into four dark mesons. The second mechanism entails the direct decay of the Higgs boson into a pair of energetic dark quarks that undergo dark showers and hadronization to form dark mesons in the final state. Notably, the second process can also produce stable or unstable dark vector mesons, resulting in distinct signatures between the two mechanisms.

In our analysis, we focus on the lightest dark meson ($\sim$3 GeV), which predominantly decays into SM gluon pairs and behaves as a long-lived particle. Both production channels consequently yield gluon-rich jets. Crucially, dark jets from Higgs exotic decays exhibit lower energies than those productions from heavy mediators in previous studies~\cite{Schwaller:2015gea,Cohen:2015toa}. This energy deficit renders existing search strategies inefficient for signal detection and unsuitable for recasting analyses, as established in Refs.~\cite{CMS:2018bvr,Carrasco:2023loy,CMS:2024gxp,ATLAS:2025bsz,Mitridate:2023tbj,CMS:2024bvl,Liu:2025bbc}. We demonstrate that optimized trigger selections significantly enhance detection efficiency for such Higgs boson originating dark jets. The primary objective of this work is therefore to identify these novel signatures and distinguish them from possible backgrounds.

This paper is structured as follows: Section~\ref{sec:model} introduces the dark QCD model with Higgs portal. Section~\ref{sec:constraint} presents relevant constraints and benchmark points. Section~\ref{sec:analysis} explores LHC search strategies for novel gluons-enriched dark jets, including both emerging jet signatures in tracker system and displaced showers in CMS muon system. Our findings are summarized in Section~\ref{sec:conclusion}.

\section{Higgs portal to dark QCD sector}
\label{sec:model}

In the QCD-like dark sector, there are numerous ways to assign the dark quark species and gauge structures. In this study, we focus on a single-flavor Dirac fermionic dark quark field \(q_d\) with gauge symmetry \(SU(3)_d\), which serves as a simplified version of SM QCD in the dark sector without loss of generality~\cite{Knapen:2021eip,Lu:2023gjk,Mitridate:2023tbj,Carrasco:2023loy}. Moreover, we consider a SM singlet pseudoscalar \(\phi_p\) as a mediator to connect the QCD-like dark sector with the SM sector. Consequently, the Lagrangian for this model can be written as follows: 
\begin{equation}
\mathcal{L} = \mathcal{L}_{\text{SM}} + \frac{1}{2}\partial_\mu \phi_p  \partial^\mu\phi_p +\bar{q_d}  (\slashed{D} -m_{q_d}) q_d  -(i y' \phi_p\bar{q_d}\gamma_5 q_d +h.c.) -V(\phi_H, \phi_p), 
\label{eq:Ltot}
\end{equation}
where $\mathcal{L}_{\text{SM}}$ includes all SM interactions expect for the scalar potential. The covariant derivative $\slashed{D}=D_\mu\gamma^\mu$ and $D_{\mu} = \partial_{\mu} -i g_d G^d_{\mu}$ with $g_d$, $G^d_{\mu}$ denote the $SU(3)_d$ gauge coupling strength and gauge field, respectively. $y'$ is the dimensionless dark Yukawa coupling. The scalar potential part is  given by 
\begin{align}
V(\phi_H, \phi_p) &= \mu_H^2\phi_H^{\dagger} \phi_H \, +\frac{\lambda_H}{2} (\phi_H^{\dagger} \phi_H)^2 +\mu_1^3 \phi_p \nonumber +\frac{\mu_p^2}{2}\phi_p^2 +\frac{\mu_3}{3!}\phi_p^3 +\frac{\lambda_p}{4!}\phi_p^4 \nonumber \\
&\quad +\mu_{Hp}\phi_p \phi_H^{\dagger} \phi_H +\frac{\lambda_{Hp}}{2} \phi_p^2 \phi_H^{\dagger} \phi_H, 
\label{eq:potential}
\end{align}
where \(\phi_H\) is the complex SM Higgs doublet, all parameters $\mu_i$'s are with the same dimension as mass and $\lambda_i$'s are dimensionless. Then, \( \phi_H \) and  \( \phi_p \) are expanded with their vacuum expectation values (VEVs) as
\begin{equation} 
\phi_H = \frac{1}{\sqrt{2}} \begin{pmatrix} 0 \\ v + h \end{pmatrix},\quad \phi_p = v_p+h_p, 
\end{equation}
where $v = 246$ GeV. Because the scalar potential \(V(\phi_H, \phi_p)\) is invariant under a shift of the singlet pseudoscalar field VEV by \( v_p \) to \( v'_{p} \), we take \( v_p = 0 \) without any loss of generality~\cite{Chen:2014ask,Baek:2017vzd}.

Two tadpole conditions are imposed to eliminate redundant model parameters. The first condition is
\begin{equation}
\left. \frac{\partial V}{\partial h} \right|_{h=0, h_p=0} = 0\quad\rightarrow\quad {\mu_H}^2 v+\frac{\lambda_H}{2}v^3=0,  
\end{equation}
and the second condition is 
\begin{equation}
\left. \frac{\partial V}{\partial h_p} \right|_{h=0, h_p=0} = 0\quad\rightarrow\quad {\mu_1}^3+\frac{\mu_{Hp}}{2}v^2=0. 
\end{equation} 
Notably, terms containing $\mu_3$ and $\mu_{Hp}$ explicitly break parity symmetry. Sebsequently, the mass terms can be written as 
\begin{equation}
\mathcal{L}_m=-\frac{1}{2}(h,h_p)\left(\begin{array}{cc}{{\lambda_{H}v^{2}}}&{{\mu_{H_p}v}}\\{{\mu_{H_p} v}}&{\mu_p^2+\frac{1}{2}\lambda_{H_p}v^2}\end{array}\right) \left(\begin{array}{c}{h}\\{h_p}\end{array}\right).  
\end{equation}
After electroweak symmetry breaking (EWSB), $h$ and $h_p$ mix via the angle $\theta_p$ to form the intermediate mass eigenstates $h_1$ and $h_2$:  
\begin{equation}
\left(\begin{matrix}h_{1}\\h_{2}\end{matrix}\right)=\begin{pmatrix}\cos\theta_p&\sin\theta_p\\-\sin\theta_p&\cos\theta_p\end{pmatrix}\begin{pmatrix}h\\h_{p}\end{pmatrix}. 
\end{equation}
As a result, both $h_1$ and $h_2$ couple to a pair of dark quarks, and trilinear couplings such as $h_1 h_2 h_2$, $h_2 h_1 h_1$ can be derived after mixing.

Furthermore, in the infrared (IR) region, the pseudoscalar dark meson \(\tilde{\eta}\) is defined as the bound state of a Dirac fermionic dark quark pair. We adopt the mapping
\begin{equation}
\bar{q}_d\, \gamma_5 q_d \rightarrow \tilde{\Lambda}\, f_{\tilde{\eta}}\, \tilde{\eta},
\end{equation}
as described in Ref.~\cite{Knapen:2021eip}, which serves as a simplified hadronization scheme to convert dark quark states into dark meson states. Here, \(\tilde{\Lambda}\) denotes the dark sector confinement scale and \(f_{\tilde{\eta}}\) is the dark meson decay constant. The mass terms in the IR region that involve \(\tilde{\eta}\) can be written as 
\begin{equation}
\mathcal{L}^{\text{IR}}_m = -\frac{1}{2}\,(h,\, h_p,\, \tilde{\eta})
\begin{pmatrix}
\lambda_H v^2 & \mu_{Hp}v & 0 \\[1mm]
\mu_{Hp}v & \mu_p^2 + \frac{1}{2}\lambda_{Hp}v^2 & y'\,\tilde{\Lambda}\, f_{\tilde{\eta}} \\[1mm]
0 & y'\,\tilde{\Lambda}\, f_{\tilde{\eta}} & 0
\end{pmatrix}
\begin{pmatrix}
h \\[1mm]
h_p \\[1mm]
\tilde{\eta}
\end{pmatrix}.
\end{equation}
After sequential diagonalization of this mass matrix, the mass-squared values of the resulting mass eigenstates \(h_1\), \(h'_2\), and \(\eta_d\) are given by 
\begin{equation}
m^2_{h_1, h_2} =  
\frac{1}{4} \left( 2\mu_p^2 + 2\lambda_H v^2 + \lambda_{Hp}v^2 \pm \sqrt{\left(-2\mu_p^2 - 2\lambda_H v^2 -\lambda_{Hp}v^2\right)^2 + 8v^2\left(2\mu_{Hp}^2 - 2\lambda_H\mu_p^2 - \lambda_H\lambda_{Hp}v^2\right)} \right)
\end{equation}
and 
\begin{equation}
m^2_{h'_2, \eta_d} = \frac{1}{2}\left( m_{h_2}^2 \pm \sqrt{ m_{h_2}^4 + 4y'f_{\tilde{\eta}}^2\tilde{\Lambda}^2}\, \right).
\end{equation} 
Additionally, two key mixing angles arising during the sequential diagonalization are defined as 
\begin{equation}
\tan 2\theta_p =  \frac{2\mu_{Hp}v}{\lambda_H v^2 - \left(\mu_p^2+\frac{1}{2}\lambda_{Hp}v^2\right)} 
\end{equation}
and 
\begin{equation}
\tan 2\theta_\Lambda =  \frac{2y'f_{\tilde{\eta}}\,\tilde{\Lambda}}{m_{h_2}^2}.
\end{equation}
In this work, we define $h_1$ as the SM-like Higgs boson ($m_{h_1} = 125$ GeV), $h'_2$ as the light scalar mediator ($m_{h'_2} \sim \mathcal{O}(10)$ GeV), and $\eta_d$ as the lightest pseudoscalar dark meson, with the mass hierarchy $m_{h_1} > m_{h'_2} > m_{\eta_d}$ strictly maintained. 

\begin{table}[h!]
\vspace{1.0mm}
\begin{ruledtabular}
 \begin{tabular}{ l c c c c }
 ~$m_{\eta_d}$ (GeV)~ & $Br(gg)$ & $Br(s\overline{s})$ & $Br(\mu^+\mu^-)$ & $\Gamma_{\eta_d}$ (GeV) \\ \hline
 ~$3$~ & $ 73.23\% $ & $ 17.96\% $ & $ 8.81\% $ & ~$2.50\times 10^{-7}\cdot x^2$~ \\
\end{tabular}
\end{ruledtabular} 
\caption{\small  \label{tab:BR-eta}
The decay branching ratio and width for the most three dominant decay modes of $\eta_d$ for $m_{\eta_d} = 3$ GeV~\cite{Winkler:2018qyg,Kling:2022uzy}. Here $x\equiv\sin\theta_{\Lambda}\sin\theta_p$ is the combined mixing angles factor. }
\end{table}

Based on the mixing among \(h\), \(h_p\), and \(\tilde{\eta}\), the SM Yukawa interactions can be expanded as follows\footnote{In the visible sector, $h_1$, $h_2^{\prime}$, and $\eta_d$ couple scalarly to SM fermions through the Higgs portal. In the dark sector, $h_1$ and $h_2^{\prime}$ exhibit pseudoscalar couplings to dark quark pairs via the SM singlet pseudoscalar portal, and the process $\Tilde{\omega}\to\eta_d\eta_d$ is permitted. }:
\begin{equation}
\begin{aligned}
\frac{m_f}{v}h\,\bar{f}f \quad\rightarrow\quad   &\frac{m_f}{v}\left(h_1\cos\theta_p - h_2\sin\theta_p\right)\bar{f}f \\
&= \frac{m_f}{v}\left[h_1\cos\theta_p - \left(h_2'\cos\theta_\Lambda - \eta_d\sin\theta_\Lambda\right)\sin\theta_p\right]\bar{f}f.
\end{aligned}
\end{equation}
Consequently, \(\eta_d\) is unstable and can decay into a pair of SM fermions or gluons, as discussed in Refs.~\cite{Knapen:2021eip,Lu:2023gjk}. Notably, CP-violating effects can arise from interference between tree-level and one-loop contributions in the $\eta_d$ decay process in this model. However, experimental observation of these effects poses significant challenges. In this work, we focus on the specific decay mode \(\eta_d \rightarrow gg\), which becomes the dominant channel when \(m_{\eta_d} \sim 3\,\mathrm{GeV}\). Accordingly, we consider the benchmark point \(m_{\eta_d} = 3\,\mathrm{GeV}\) and adopt the decay branching ratio and width which have been calculated in Refs.~\cite{Winkler:2018qyg,Kling:2022uzy}. We summarize the results in Table~\ref{tab:BR-eta}, where $x\equiv\sin\theta_{\Lambda}\sin\theta_p$ represents the combined mixing angles factor. These branching ratios remain consistent for $2$ GeV $\lesssim m_{\eta_d} \lesssim 3.5$ GeV. Notably, $\eta_d$ becomes a long-lived particle (LLP) when $x$ is sufficiently small. 







\section{Constraints and benchmark points}
\label{sec:constraint}

In this section, we first examine existing collider constraints on the scalar mass eigenstates $h_1$, $h_2^{\prime}$, and $\eta_d$. Based on these constraints, we then define some representative benchmark points (BPs) for investigating signal signatures at the LHC in subsequent sections.

First of all, precision measurements of the SM-like Higgs boson $h_1$ at the LHC have already constrained its properties~\cite{CMS:2018uag,ATLAS:2019nkf}. The most relevant limits involve the mixing angle between $h_1$ and other scalars, as well as exotic Higgs decay channels. The mixing angle is constrained to $\sin^2\theta_p < 0.12$ at $95\%$ confidence level (C.L.)~\cite{ATLAS:2015ciy,ATLAS:2016neq}, while the branching ratio for exotic decays satisfies $BR(h_1\to\text{undetected}) < 19\%$ at $95\%$ C.L.~\cite{ATLAS:2020qdt}. In our model, undetected channels include $h_1\to h_2^{\prime}h_2^{\prime}$ and/or $h_1\to q_d\overline{q_d}$.

Detecting $h_2^{\prime}$ with mass $\mathcal{O}(10)$ GeV presents significant challenges. One approach is through exotic Higgs decays as mentioned above, if $m_{h_2^{\prime}} < m_{h_1}/2$. Additionally, the cleaner environment of lepton colliders enables additional constraints; for examples, LEP experiment has set limits on $h_2^{\prime}\to b\overline{b}$~\cite{LEPWorkingGroupforHiggsbosonsearches:2003ing}. At the LHC, 
detecting low-mass resonances is notoriously difficult~\cite{CMS:2019emo}. Although triggers with an energetic photon can improve sensitivity, the production cross-section is simultaneously suppressed~\cite{CMS:2019xai,ATLAS:2024bms}. Furthermore, $h_2^{\prime}$ can decay not only to SM fermion pairs but also to dark meson pairs. The branching ratios for these channels depend on model parameters, leading to relatively weak constraints on $h_2^{\prime}$.

Finally, the dark meson $\eta_d$ is subject to constraints from LEP experiments and precision measurements of various decay channels of SM hadrons. The L3 Collaboration studied $e^+ e^-\to Z\eta_d$ for $0.01~\text{GeV}\lesssim m_{\eta_d}\lesssim 60~\text{GeV}$, where $\eta_d$ is identified via a pair of charged particles and missing energy~\cite{L3:1996ome}. Note that this constraint can also be applied to $h_2^{\prime}$ with mass $\mathcal{O}(10)$ GeV. For $m_{\eta_d}\sim 3$ GeV as considered here, searches focus on exotic decays of $\Upsilon$ and $B$ mesons, particularly the golden channels $\Upsilon\to\eta_d\gamma$~\cite{BaBar:2011kau,BaBar:2012sau} and $B\to K\eta_d$~\cite{LHCb:2015nkv,LHCb:2016awg,Belle-II:2023ueh}. These constraints are, however, highly sensitive to the $\eta_d$ lifetime. The bounds effectively require the combined mixing angles factor to satisfy $x \lesssim \mathcal{O}(10^{-3})$, which corresponds to a decay width of $\Gamma_{\eta_d} \lesssim \mathcal{O}(10^{-13})$ GeV. This naturally leads to the investigation of long-lived $\eta_d$ in the subsequent analysis.

In this work, we fix \( m_{\eta_d} = 3~\text{GeV} \) to ensure the decay \( \eta_d \to gg \) dominates~\cite{Winkler:2018qyg,Kling:2022uzy}. The number of dark colors is set to \( N^D_c = 3 \), and the number of dark quark flavors to \( n^D_F = 1 \)~\cite{Knapen:2021eip,Lu:2023gjk,Mitridate:2023tbj,Carrasco:2023loy,CMS:2024bvl}. We classify the signal processes into two scenarios:

\textbf{1. Cascade decay process:} \( h_1 \) decays to a pair of \( h_2' \) particles, each of which subsequently decays to \( \eta_d \) pairs, \( h_1\to h_2' h_2'\to (\eta_d \eta_d) (\eta_d \eta_d) \). We define two BPs: 
   \begin{itemize}
     \item cascade-1: \( m_{h_2'} = 5m_{\eta_d} \), 
     \item cascade-2: \( m_{h_2'} = 10m_{\eta_d} \).
   \end{itemize}

\textbf{2. Dark shower process:} \( h_1 \) decays to dark quark (\( q_d \)) pairs, which undergo dark showering and hadronization, producing multiple dark mesons, \( h_1\to q_d \bar{q_d}\to \text{dark showers, hadronization} \). The dark quark mass is set to \( m_{q_d} = 0.4m_{\eta_d} \). We define three BPs:
   \begin{itemize}
     \item DS1: \( \Lambda_d = m_{\tilde{\omega}} = 2.5m_{\eta_d} \), \( \texttt{probVector} = 0.32 \) (vector-meson poor, \( \tilde{\omega} \to \eta_d \eta_d \)), 
     \item DS2: \( \Lambda_d = 2m_{\tilde{\omega}} = 5m_{\eta_d} \), \( \texttt{probVector} = 0.54 \) (vector-meson rich, \( \tilde{\omega} \to \eta_d \eta_d \)), 
     \item DS3: \( \Lambda_d = 2m_{\tilde{\omega}} = 2.5m_{\eta_d} \), \( \texttt{probVector} = 0.69 \) (stable \( \tilde{\omega} \)). 
   \end{itemize} 
Here, \( \Lambda_d \) denotes the dark confinement scale, \( \tilde{\omega} \) the dark vector-meson mass, and \texttt{probVector} the fraction of vector to pseudoscalar dark mesons. The \texttt{probVector} values are calculated following Appendix A of Ref.~\cite{Knapen:2021eip}. In the Hidden Valley module~\cite{Carloni:2010tw,Carloni:2011kk}, the dark shower \( p_T \) cutoff is set to \( \texttt{pTminFSR} = 1.1\Lambda_d \). These five BPs will be used in the subsequent analysis. We expect the angular separation of these light $\eta_d$ particles to be wider in cascade decay processes than in dark shower processes. Additionally, the DS3 scenario can produce a novel semi-visible emerging jet (SVEJ) signature~\cite{Bernreuther:2020xus,Carmona:2024tkg,Carrasco:2025bct}, which distinguishes it from the other four BPs.

\section{Exploring gluons-enriched dark jets at the LHC}
\label{sec:analysis}

In this section, we explore gluons-enriched dark jets composed of multiple long-lived dark mesons from Higgs exotic decays at the LHC. Based on dark meson lifetimes, we focus on two detection regions within the CMS detector. For shorter dark meson lifetimes, we consider emerging jet (EJ) signatures in the tracker system~\cite{CMS:2018bvr,Carrasco:2023loy,CMS:2024gxp,ATLAS:2025bsz}; for longer dark meson lifetimes, we study displaced shower signatures in the muon system~\cite{CMS:2021juv,Cottin:2022nwp,Mitridate:2023tbj,CMS:2024bvl,CMS:2024ake,Liu:2024fey,Lu:2024ade,CMS:2025urb,Liu:2025ldf,Liu:2025bbc,CMS:2025rtd}. In the published CMS papers~\cite{CMS:2018bvr,CMS:2021juv}, the events are selected by using a trigger $H_T > 900$ GeV for EJ signatures and a trigger $p^{\mathrm{miss}}_T > 200$ GeV for displaced shower signatures, where $H_T$ denotes the scalar sum of transverse momenta of all hadronic jets and $p_T^{\mathrm{miss}}$ (missing transverse momentum) is defined as the negative vector sum of the transverse momenta of all reconstructed visible particles. However, these trigger conditions are not suboptiomal for Higgs exotic decay analyses, resulting in significant signal loss in previous studies. Consequently, careful designed trigger conditions are essential for probing such gluons-enriched dark jets at the LHC.

Signal events were generated using \texttt{MadGraph5\_aMC@NLO}~\cite{Alwall:2014hca} (v3.4.2),  accounting for multiple Higgs production channels: gluon fusion (ggH), vector boson fusion (VBF)\footnote{We apply a generator-level cut of $\Delta\eta_{jj} > 2.5$ to the tagging jet pair in the forward and backward direction.}, associated production with a $W$ or $Z$ boson (VH, subdivided into hadronic and leptonic decays, denoted as VH-had and VH-lep), and top-associated production ($t\bar{t}H$, similarly subdivided into $t\bar{t}H$-had, $t\bar{t}H$-lep, and $t\bar{t}H$-semi). Parton showering, hadronization, and the underlying event were simulated with \texttt{Pythia8}~\cite{Sjostrand:2014zea}. For the ggH process, matching and merging between the matrix element and parton shower were employed; the matching scale was set to \texttt{xqcut = 25 GeV} at the matrix-element level, with the corresponding parameter in Pythia8 set to \texttt{JetMatching:qCut = 56.25 GeV} to ensure consistency. The Higgs boson mass was fixed at $125$~GeV in the simulation. Both dark shower and cascade decay processes were investigated, with the proper lifetime $c\tau$ of the long-lived dark meson $\eta_d$ ranging from $1$~mm to $1$~m in the tracker system and $50$~mm to $100$~m in the CMS muon system, respectively. 

\begin{table}[h!]
\centering
\begin{tabular}{|>{\centering\arraybackslash}p{3.5cm}|>{\centering\arraybackslash}p{12cm}|}
\hline
\textbf{Trigger} & \textbf{Condition} \\ \hline
Single jet & $p_T^j > 180\,\text{GeV}$, $|\eta_j| < 2.4$ \\ \hline
\multirow{2}{*}{Dijet} & $\lvert \eta_{j_{1,2}} \rvert < 2.4$,  $\Delta\eta = |\eta_{j_1} - \eta_{j_2}| < 1.6$ \\ \cline{2-2}
 & $p_T^{j_1} > 70\,\text{GeV}$ , $p_T^{j_2} > 40\,\text{GeV}$ \\ \hline
\multirow{3}{*}{VBF jet} & $\lvert \eta_{j_{1,2}} \rvert < 5$ , $\eta_{j_1} \times \eta_{j_2} < 0$ , $\Delta\eta > 4.0 $ \\ \cline{2-2}
 & $\Delta\phi = |\phi_{j_1} - \phi_{j_2}| < 2.0 $ \\ \cline{2-2}
 & $M_{j_1 j_2} > 1000\,\text{GeV}$ \\ \hline
Single $e$ & $p_T^e > 36\,\text{GeV}$ , $|\eta_e| < 2.4$ \\ \hline
Double $e$ & $p_T^{e_1} > 25\,\text{GeV}$ , $p_T^{e_2} > 12\,\text{GeV}$ , $\lvert \eta_{e_{1,2}} \rvert < 2.4$ \\ \hline
Single $\mu$ & $p_T^\mu > 22\,\text{GeV}$ , $|\eta_\mu| < 2.4$ \\ \hline
Double $\mu$ & $p_T^{\mu_1} > 15\,\text{GeV}$ , $p_T^{\mu_2} > 7\,\text{GeV}$ , $\lvert \eta_{\mu_{1,2}} \rvert < 2.4$ \\ \hline
MET & $p_T^{\mathrm{miss}} > 200\,\text{GeV}$,
at lease one jet with  $p^j_T > 50\,\text{GeV}$ and $|\eta_j| < 2.4$ passing the “Tight Lepton Veto” jet identification criteria~\cite{CMS:2017wyc}\\ \hline
$H_T$ & $H_T > 900\,\text{GeV}$ \\ \hline
\end{tabular}
\caption{Categorization of trigger conditions in this analysis: hadronic final states (single jet, dijet, VBF jet, MET, and $H_T$) and leptonic final states (single/double electron and muon).}
\label{tab:trigger}
\end{table} 

In Higgs exotic decay analyses, to improve the signal efficiency, we designed and implemented multiple trigger conditions for the $\sqrt{s} = 14$~TeV high-luminosity LHC (HL-LHC). 
Considering the above six major Higgs production channels, we defined nine trigger conditions, categorized into two groups: 
\begin{itemize}
    \item \textbf{Category I (Hadronic Final States):} single jet, dijet, VBF jet, missing transverse energy (MET), and $H_T$.
    \item \textbf{Category II (Leptonic Final States):} single electron, double electron, single muon, and double muon. 
\end{itemize} 
The details of these nice trigger conditions are summarized in Table.~\ref{tab:trigger} based on Refs.~\cite{CMS:2024bvl,Bhattacherjee:2021rml,CMS:2024gxp}. 
At the event level, the ggH, VBF, VH-had, and $t\bar{t}$H-had \& $t\bar{t}$H-semi processes adopt Category I trigger conditions, while the VH-lep and $t\bar{t}$H-lep \& $t\bar{t}$H-semi processes use Category II trigger conditions. This mapping ensures that the most effective set of triggers is applied for each production mode, thereby enhancing the overall signal acceptance and detection efficiency.

\begin{figure}[h!]
\centering
\includegraphics[width=0.8\textwidth]{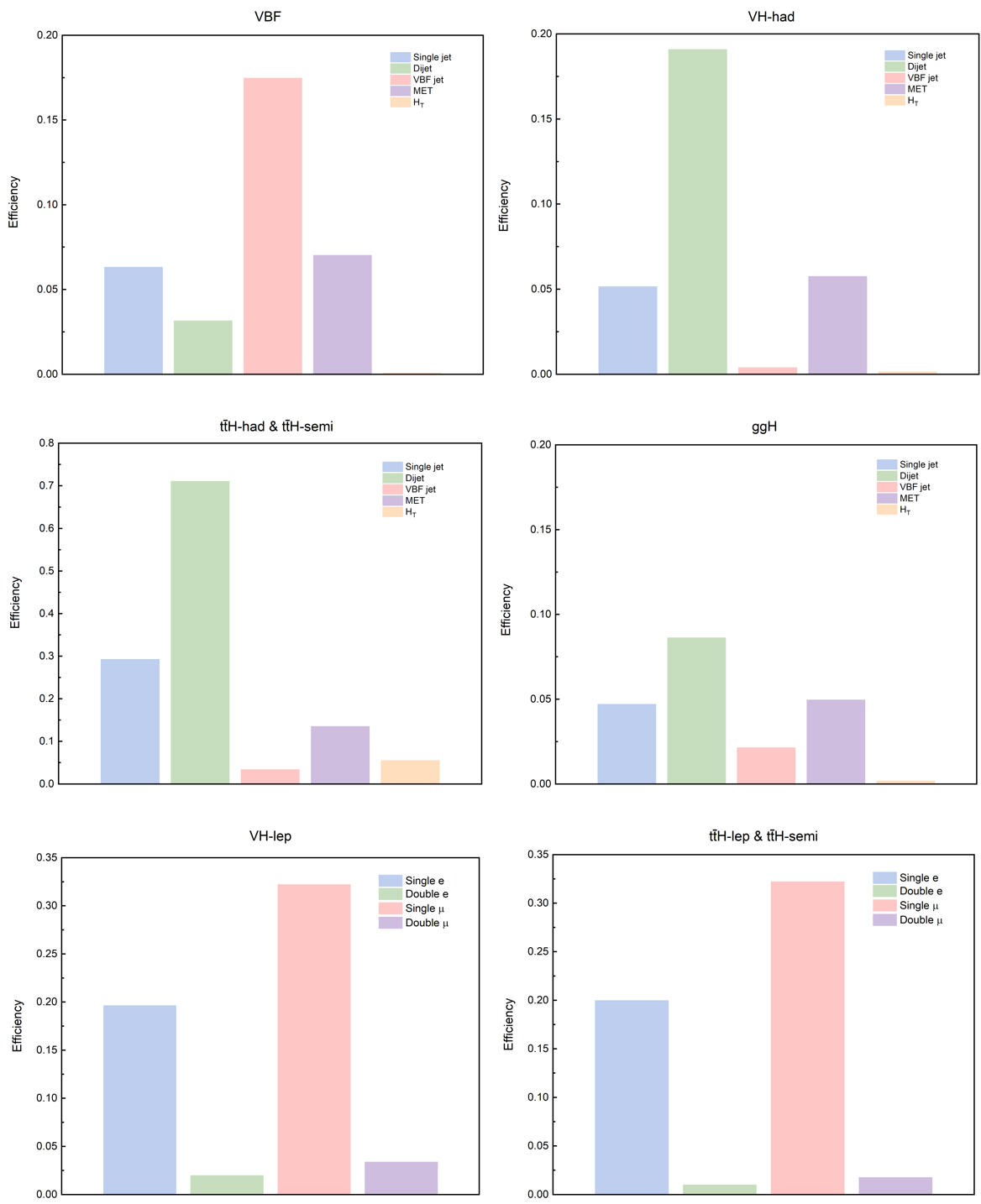}
\caption{Efficiencies for the six major Higgs production processes (ggH, VBF, VH-had, VH-lep, $t\bar{t}H$-had \& $t\bar{t}H$-semi, and $t\bar{t}H$-lep \& $t\bar{t}H$-semi) under individual trigger conditions at the $\sqrt{s} = 14$ TeV HL-LHC. }
\label{fig:trigger_efficiency}
\end{figure}

\begin{figure}[h!]
\centering
\includegraphics[width=0.6\textwidth]{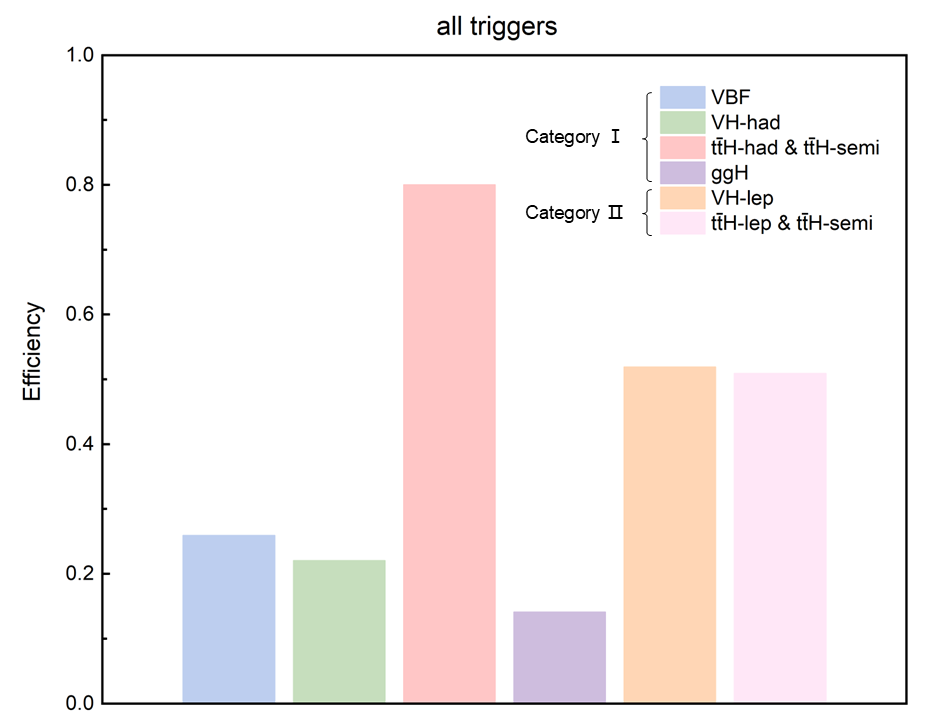}
\caption{Efficiencies for the six major Higgs production processes (ggH, VBF, VH-had, VH-lep, $t\bar{t}H$-had \& $t\bar{t}H$-semi, and $t\bar{t}H$-lep \& $t\bar{t}H$-semi) under combined trigger conditions of Categories I and II at the $\sqrt{s} = 14$ TeV HL-LHC. }
\label{fig:all_triggers}
\end{figure}

Figure~\ref{fig:trigger_efficiency} compares the efficiencies of the six Higgs production channels under individual trigger conditions. As expected, the VBF jet trigger is most efficient for the VBF process, while the dijet trigger dominates for the VH-had and $t\bar{t}$H-had \& $t\bar{t}$H-semi processes. For the VH-lep and $t\bar{t}$H-lep \& $t\bar{t}$H-semi processes, the single muon and single electron triggers provide the leading and sub-leading efficiencies, respectively. For the ggH process, no single trigger (dijet, single jet, or MET) shows a clear dominance. Therefore, to maximize the number of recorded signal events for subsequent analysis, we collect events that satisfy any one of the trigger conditions relevant to each Higgs production channel. Figure~\ref{fig:all_triggers} shows the efficiencies under the combined trigger conditions of Categories I and II. The combination of triggers significantly improves the overall efficiency for each channel, thereby preventing substantial signal loss.

In the following subsections, we investigate the signal signatures for the five BPs from Section~\ref{sec:constraint}, focusing on EJs in the tracker system and displaced showers in the muon system. For comparison, we first recast the existing search strategies from the CMS Collaboration at $\sqrt{s} = 13$ TeV~\cite{CMS:2018bvr,CMS:2021juv}. We then analyze the $\sqrt{s} = 14$ TeV HL-LHC phase with our dedicated trigger strategy enabled. A clear enhancement in the exclusion reach by more than an order of magnitude is observed, demonstrating that the combined trigger strategy significantly improves the overall signal efficiency. 


\subsection{Emerging jet signatures in the tracker system} 
\label{sec:EJ_analysis}

For shorter dark meson $\eta_d$ lifetimes, its decays predominantly occur within the inner tracking detector, giving rise to the characteristic EJ signature, where displaced charged tracks gradually appear inside the jet cone as long-lived $\eta_d$ decays. 
Signal events are generated for all six primary Higgs production channels and the LLP decays are simulated by \texttt{PYTHIA~8.312} for five BPs defined in Sec.~\ref{sec:constraint}. 
The dominant SM background arises from QCD multijet production, in particular heavy-flavor ($pp\to b\bar{b}$) events that can mimic the EJ signature through  the decay of long-lived $B$-mesons. 
We generate background samples at $\sqrt{s} = 14~\mathrm{TeV}$ by explicitly activating the \texttt{HardQCD:hardbbbar} process in \texttt{PYTHIA~8.312} and imposing a minimum $p_T$ threshold of 25 GeV on the hard process, to maximize statistical precision.
The lifetimes of main $B$ mesons are set to their PDG values~\cite{ParticleDataGroup:2024cfk}\footnote{\texttt{511:tau0 = 0.4557}, \texttt{521:tau0 = 0.4914}, \texttt{531:tau0 = 0.4545}, and  \texttt{541:tau0 = 0.1530}.}. 
Parton showering, hadronization, and underlying-event effects are included.

To validate our analysis framework, we first perform a recast of the CMS 13~TeV EJ analysis based on Refs.~\cite{CMS:2018bvr,Carrasco:2023loy} with an integrated luminosity of $\mathcal{L}=16.1~\mathrm{fb}^{-1}$\footnote{Our analysis code is built upon the public LLP Recasting Repository: \url{https://github.com/llprecasting/recastingCodes/tree/main/EmergingJets/CMS-EXO-18-001}.}. Signal events are selected by passing the trigger $H_T > 900$~GeV. The jets are clustered with a distance parameter of $R=0.4$ using the anti-$k_t$ algorithm~\cite{Cacciari:2008gp} applied to all tracks with $p_T > 1$~GeV. 
To define the baseline analysis sample, events are required to have at least four jets with $p_T > 20$~GeV and $|\eta| < 2.0$. These pre-selected events then undergo further selection using delicate kinematical variables to tag EJs.

Accurate reproduction of these EJ tagging variables requires realistic modeling of the detector response to displaced vertices. We implement two parametrizations for the tracking efficiency $\epsilon_{\text{trk}}$ based on Ref.~\cite{Carrasco:2023loy}:
\begin{itemize} 
    \item \textit{$It_5$}: A data-driven efficiency map based on the ``Iteration 5'' tracking step from Ref.~\cite{CMS:2014pgm}.
    \item \textit{R}: A simplified geometric step function where $\epsilon_{\text{trk}} = 1$ for tracks with a transverse production vertex $r_{\text{prod}} < 102$~mm and 0 otherwise. 
\end{itemize}
In both cases, impact parameter smearing is applied to mimic finite detector resolution. For the validation results presented in the main text, we adopt the \textit{$It_5$} strategy as our baseline and provide the results from \textit{R} strategy in Appendix~\ref{appendix:B}. 

\begin{figure}[tbp]
    \centering
    \includegraphics[width=0.95\textwidth]{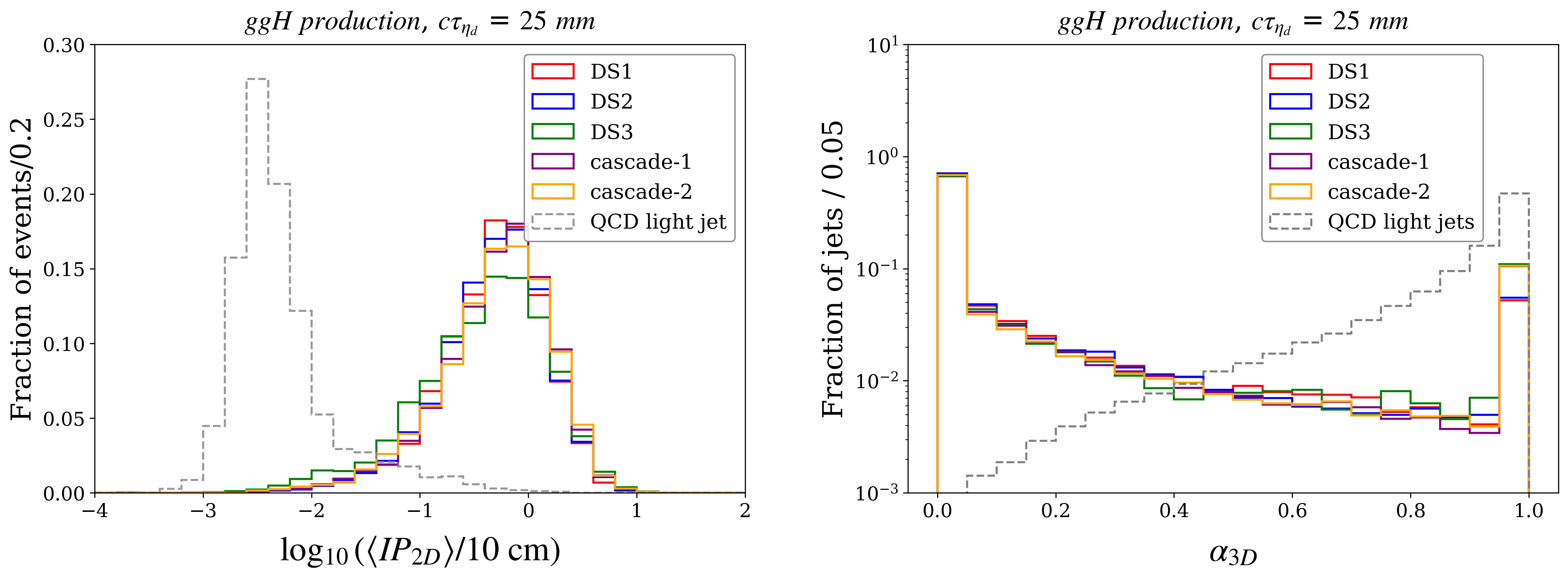}
    \caption{Normalized distributions of the event-level median impact parameter $\langle IP_{2D} \rangle_{\text{event}}$ (left) and the $\alpha_{3D}$ calculated with $D_{\text{cut}}=10$ (right). 
    The five signal BPs in the ggH production mode are distinguished by color: Dark shower scenarios DS1 (red), DS2 (blue), DS3 (green), and cascade decay scenarios cascade-1 (purple), cascade-2 (yellow).
    All BPs correspond to a fixed  $c\tau_{\eta_d} = 25$~mm. The SM QCD light-jet background is shown as the grey dashed line~\cite{CMS:2018bvr}. }
    \label{fig:kinematics_13tev}
\end{figure}

To discriminate EJs from the major QCD background, we construct track-based variables sensitive to the displaced vertices. First, we quantify the displacement significance of each track. The 3D impact parameter significance, $D_N$, is defined as:
\begin{equation}
    D_N = \sqrt{ \left(\frac{z_{\text{trk}} - z_{\text{PV}}}{0.01~\text{cm}}\right)^2 + \left(\frac{d_0}{\sigma_{d_0}}\right)^2 },
\end{equation}
where $d_0$ is the transverse impact parameter (with uncertainty $\sigma_{d_0}$), and $z_{\text{trk}} - z_{\text{PV}}$ represents the longitudinal distance between the track and the primary vertex (corresponding to the variable $PU_{dz}$).
Based on this track-level metric, the variable $\alpha_{3D}$ calculates the scalar $p_T$ fraction of tracks associated with the jet that are compatible with the primary vertex:
\begin{equation}
    \alpha_{3D} = \frac{\sum_{k \in \text{jet}, D_N < D_{\text{cut}}} p_{T,k}}{\sum_{k \in \text{jet}} p_{T,k}}.
\end{equation}
Here, $D_{\text{cut}}$ is a threshold parameter defining prompt-like tracks which takes the values of $D_{\text{cut}} \in \{4, 10, 20\}$ from the CMS analysis~\cite{CMS:2018bvr}. EJs exhibit low values of $\alpha_{3D}$ (typically $< 0.25$), as a significant fraction of the jet energy is carried by displaced tracks ($D_N > D_{\text{cut}}$) originating from $\eta_d$ decays. 
For the kinematic distributions shown in Figure~\ref{fig:kinematics_13tev}, we adopt a fixed threshold of $D_{\text{cut}} = 10$, consistent with the validation plots in Ref.~\cite{Carrasco:2023loy}.

In addition, we utilize the median transverse impact parameter, $\langle IP_{2D} \rangle$. For a given jet, we indicate with $d_k$ the vector of closest approach between the $k$-th constituent track and the primary vertex. The transverse component of this vector is denoted as $(d_k)_r$. The jet-level observable is defined as the median of these absolute transverse values, which provides a measure of the typical displacement of the jet constituents while being less sensitive to outliers than the mean: 
\begin{equation}
    \langle IP_{2D} \rangle = \text{median} \left( \{ |(d_k)_r| \}_{k \in \text{jet}} \right).
\end{equation} 
For signal characterization, we construct an event-level variable, $\langle IP_{2D} \rangle_{\text{event}}$ , defined as the median of the $\langle IP_{2D} \rangle$ values of the two hardest jets.

To verify the discrimination power of these kinematic variables under the 13 TeV validation, we compare their distributions of the signal against the SM background. Figure~\ref{fig:kinematics_13tev} presents the normalized distributions of $\langle IP_{2D} \rangle_{\text{event}}$ (left panel) and $\alpha_{3D}$ (right panel) for events passing the baseline trigger $H_T > 900$~GeV. We compare the five signal BPs (with $c\tau_{\eta_d} = 25$~mm) against the background from QCD light-jet~\cite{CMS:2018bvr}. As shown in Fig.~\ref{fig:kinematics_13tev}, the QCD background (grey dashed line) is peaked near zero for $\langle IP_{2D} \rangle$ and near unity for $\alpha_{3D}$, reflecting the prompt nature of tracks inside the typical SM QCD jets. In contrast, the five signal BPs exhibit significant tails extending to larger impact parameters and cluster at low $\alpha_{3D}$ values. This clear separation confirms that the jet energy in the signal is primarily carried by displaced tracks ($D_N > 10$), validating the effectiveness of these observables for signal and background discrimination.

Based on the discriminating power of these observables, we adopt the six specific EJ tagging groups, labeled EMJ-1 to EMJ-6. These are defined by simultaneous cuts on the four variables $\langle IP_{2D} \rangle$, $\alpha_{3D}$, $D_N$, and $PU_{dz}$, and correspond to the first six rows of Table~1 in the original CMS analysis~\cite{CMS:2018bvr}, as consistently listed in Table~1 of Ref.~\cite{Carrasco:2023loy}. The $PU_{dz}$ variable is employed here to mitigate pile-up effects. Although pile-up was not simulated in this study, its impact can be effectively rejected by using the methods described in Refs.~\cite{Schwaller:2015gea,Linthorne:2021oiz,ATLAS:2025bsz}.
These tagging criteria are paired with seven signal regions, referred to as ``Selection Sets''. The specific kinematic thresholds for these sets (involving $H_T$ and $p^j_T$) are adopted directly from Table~2 of Ref.~\cite{Carrasco:2023loy}.

\begin{figure}[tbp]
    \centering
    \includegraphics[width=0.95\textwidth]{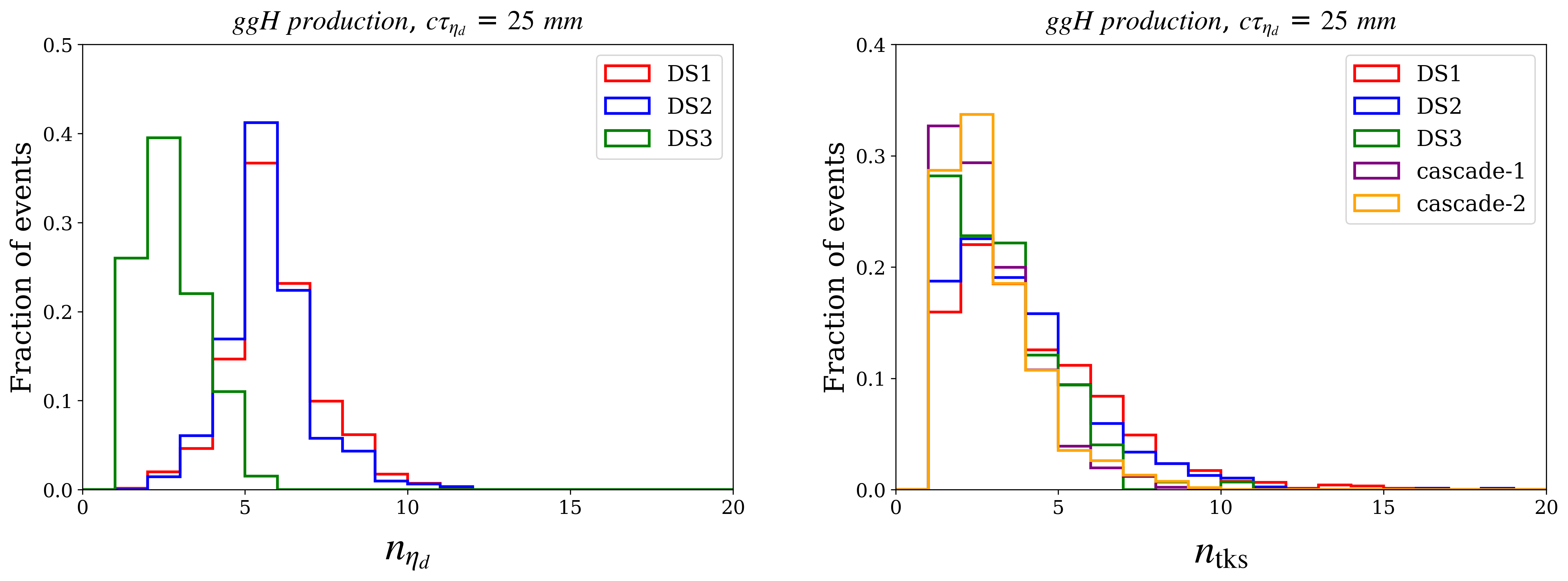}
    \caption{Multiplicity distributions of $\eta_d$'s (left) and reconstructed tracks (right) within the EJ cone ($R=0.4$). 
    The distributions are obtained for EJ candidates in signal events in the ggH production mode and passing the trigger $H_T > 900$~GeV. 
    The five signal BPs are distinguished by color: Dark shower scenarios DS1 (red), DS2 (blue), DS3 (green), and cascade decay scenarios cascade-1 (purple), cascade-2 (yellow).
    All BPs correspond to a fixed $c\tau_{\eta_d} = 25$~mm.}
    \label{fig:multiplicities_13tev}
\end{figure}

The effectiveness of these EJ tagging variables primarily stems from the high multiplicity of dark mesons produced in the dark shower and hadronization processes. To visualize this feature, we present in Fig.~\ref{fig:multiplicities_13tev} the distributions of the number of $\eta_d$'s ($n_{\eta_d}$) and the number of reconstructed tracks ($n_{\text{tracks}}$) within the jet cone ($R=0.4$). 
These distributions are obtained for EJ candidates in events passing the validation baseline trigger ($H_T > 900$~GeV).
The high track multiplicity, therefore, complements the displacement-based tagging criteria. Note that for the cascade decay scenarios (cascade-1/2), the number of dark mesons is strictly fixed by the decay topology ($h \to 4\eta_d$), resulting in discrete peaks rather than a continuous distribution; hence, their $n_{\eta_d}$ distributions are omitted for clarity.

For each of the seven selection sets mentioned above and defined in Table~2 of Ref.~\cite{Carrasco:2023loy}, the expected number of signal events is calculated as:
\begin{equation}
N_{S}^{i}(c\tau_{\eta_d}) 
= \mathcal{L}
\times \mathrm{BR}_{\mathrm{exo}}
\times
\sum_{j}
\left[ \sigma_{\mathrm{proc},j}^{\mathrm{SM}} \times A_{i,j}(c\tau_{\eta_d}) \right],
\label{eq:Ns_sum}
\end{equation}
where $i=1,\dots,7$, $\mathcal{L}$ denotes the integrated luminosity,  $\mathrm{BR}_{\mathrm{exo}}$ the Higgs exotic decay branching ratio (corresponding to either $h_1 \to h_2' h_2'$ or $h_1 \to q_D \bar{q}_D$), and $\sigma^{\mathrm{SM}}_{\mathrm{proc}}$ denotes the corresponding SM Higgs production cross-section for the $j$-th mode. Here, $A_{i,j}$ represents the signal acceptance, defined as the fraction of simulated events from production mode $j$ that satisfy all criteria of the $i$-th selection set.

The $95\%$~C.L. upper limit $S_{95}^{i}$ from the CMS analysis~\cite{CMS:2018bvr} is matched to our simulated yield. 
The minimum excluded branching ratio at a given lifetime is then
\begin{equation}
\mathrm{BR}^{(i)}(c\tau_{\eta_d}) 
= \frac{S_{95}^{i}}
{\mathcal{L}\sum_{j}\sigma_{\mathrm{proc},j}^{\mathrm{SM}}A_{i,j}(c\tau_{\eta_d})},
\end{equation}
and the most sensitive selection (the smallest $\mathrm{BR}^{(i)}$) is taken as the final limit. 
Here, the relevant Higgs production cross-sections are rescaled to state-of-the-art precision: ggH is taken at NNLO QCD with NNLL soft-gluon resummation and includes NLO electroweak corrections, while VBF, VH, and $t\bar{t}H$ are computed at NNLO QCD and NLO EW accuracy~\cite{sigma-13TeV}. 

\begin{figure}[tbp]
    \centering
    \begin{subfigure}[b]{0.49\textwidth}
        \includegraphics[width=\textwidth]{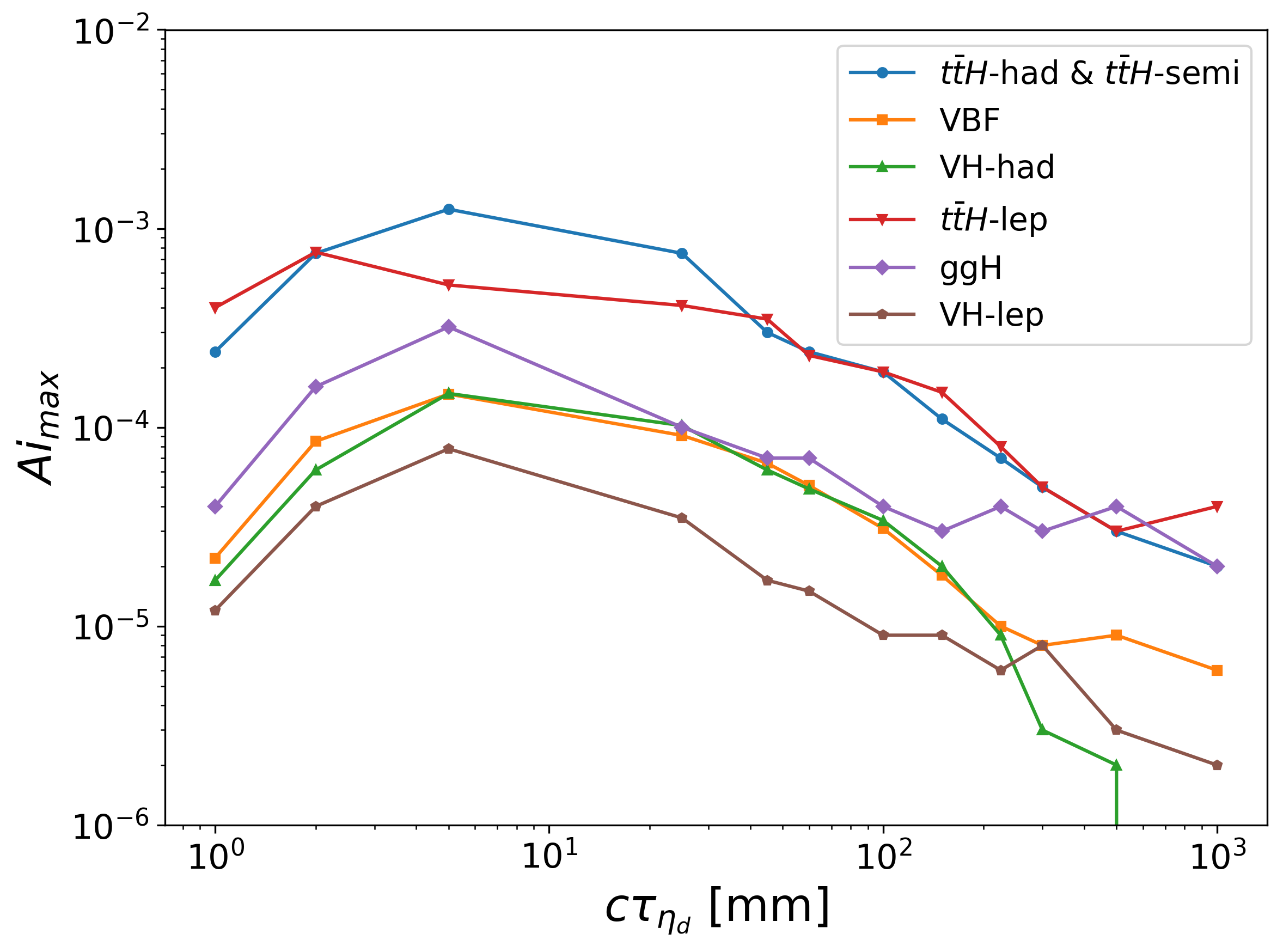} 
        \label{fig:13tev_amax}
    \end{subfigure}
    \hfill
    \begin{subfigure}[b]{0.49\textwidth}
        \includegraphics[width=\textwidth]{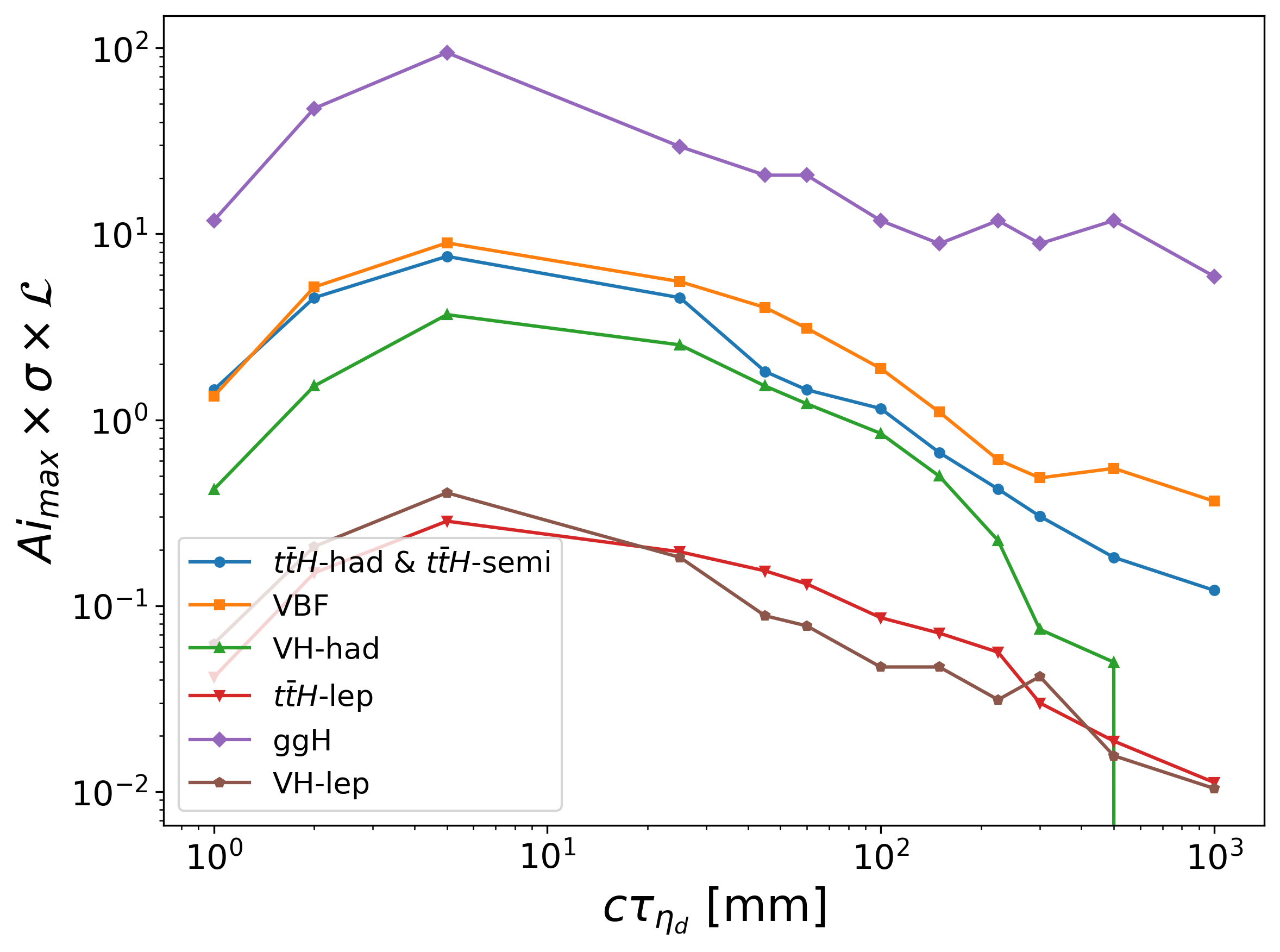}
        \label{fig:13tev_ns}
    \end{subfigure}
    \caption{Maximum signal acceptance $A_{i,\text{max}}$ among the seven selection sets (left) and corresponding expected signal yield $N_S$ for $\mathcal{L}=16.1~\mathrm{fb}^{-1}$ (right) shown as a function of $c\tau_{\eta_d}$ for the DS1. 
    The results are shown for Higgs production through ggH (purple), VBF (orange), VH-had (green), $t\bar{t}H$-had \& $t\bar{t}H$-semi (blue), VH-lep (brown), and $t\bar{t}H$-lep (red). }
    \label{fig:13tev_performance}
\end{figure}

To validate our framework and quantify the limitations of recasting from CMS results~\cite{CMS:2018bvr,Carrasco:2023loy}, we evaluate the signal acceptance and expected yields under $\sqrt{s} = 13$~TeV and $\mathcal{L}=16.1~\mathrm{fb}^{-1}$. 
Figure~\ref{fig:13tev_performance} presents the maximum acceptance ($A_{i,\text{max}}$) among the selection sets (left) and the total expected signal yield $N_S$ (right) as a function of $c\tau_{\eta_d}$ for the DS1 as an example. Specifically, as shown in the left panel, we observe that the maximum signal acceptance is consistently low ($< 1\%$). This suppression is significantly associated with the trigger $H_T > 900$~GeV. As illustrated in Fig.~\ref{fig:trigger_efficiency}, the efficiencies for all Higgs production modes are consistently low under this trigger condition.
This severe signal loss indicates that the current $H_T$-based strategy and the selection sets are too restrictive for probing exotic Higgs decays.

Building upon the validated reinterpretation framework established in the 13~TeV analysis, we extend our study to estimate the sensitivity at the HL-LHC with $\sqrt{s} = 14$~TeV and $\mathcal{L}=3000~\mathrm{fb}^{-1}$. To address the low signal acceptance identified in the validation step and maximize the sensitivity at the HL-LHC, we employ a dedicated parameterized detector simulation that upgrades the particle-level selection to a more realistic CMS detector simulation.  Specifically, we refine the object reconstruction settings, adopting lepton isolation and optimized jet parameters, while maintaining the consistent tracking strategy used in the validation step.

For hadronic jets, in contrast to the validation step ($R=0.4$), we adopt a larger radius of $R=0.5$ and a massive clustering scheme to enhance kinematic precision. Events are selected by passing either the Category I (Hadronic) or Category II (Leptonic) triggers. Furthermore, we require the presence of at least two jets within $|\eta| < 3.0$ and $p_T > 20$~GeV that satisfy the baseline quality criteria (consistent with the validation step).
The $\vec{p}_T^{\,\text{miss}}$, is calculated extending to the forward calorimeter coverage ($|\eta| < 4.9$).

A critical addition to our 14~TeV analysis is the implementation of lepton isolation to suppress the large background of non-prompt leptons from heavy-flavor hadron decays. We implement a track-based relative isolation criterion.
The relative isolation variable, $I_{\text{rel}}$, is defined as the scalar sum of the transverse momenta of all visible particles within a cone of size $\Delta R = 0.5$ around the lepton candidate, excluding the lepton itself, normalized by the lepton's transverse momentum:
\begin{equation}
    I_{\text{rel}} = \frac{\sum_{i \in \Delta R, i \neq \ell} p_T^i}{p_T^{\ell}}, \quad \text{where } \Delta R = \sqrt{(\Delta\eta)^2 + (\Delta\phi)^2} < 0.5.
\end{equation}
Electrons ($|\eta|<2.5$) and muons ($|\eta|<2.4$) with $p_T > 10$~GeV are considered as isolated lepton candidates only if they satisfy $I_{\text{rel}} < 0.12$ and $I_{\text{rel}} < 0.25$, respectively. These thresholds are adopted directly from the standard CMS detector configuration implemented in \texttt{Delphes}.

To discriminate these signals from the SM background at the HL-LHC, we utilize the track-based observables ($\langle IP_{2D} \rangle$, $\alpha_{3D}$, and $D_N$) defined in the 13~TeV validation analysis above. We have verified that the kinematic distributions of these variables at the 14~TeV, specifically for events passing the Category I or II triggers, maintain the discrimination power observed in the 13~TeV validation. The EJ candidates consistently feature large impact parameters and low $\alpha_{3D}$ values, ensuring the effectiveness of these kinematic variables for signal identification. Therefore, we also adopt EMJ‑1 to EMJ‑6 validated in the 13~TeV analysis. To maximize the discovery potential at the HL-LHC, we optimize the selection strategy by categorizing events into two inclusive channels. The selection procedure begins by requiring events to pass the dedicated Category I or II triggers, followed by a pre-selection of at least two reconstructed jets ($N_{\text{jets}} \ge 2$) within the detector acceptance. The analysis is performed by scanning EMJ‑1 to EMJ‑6 tagging working points to determine the optimal configuration for each signal BP.

\begin{table}[htbp]
\centering
\renewcommand{\arraystretch}{1.5} 
\resizebox{\textwidth}{!}{
\begin{tabular}{|c|c|c|c|c|}
\hline
\multirow{2}{*}{\textbf{Cut}} & \multicolumn{2}{c|}{\textbf{Dark Shower}} & \multicolumn{2}{c|}{\textbf{Cascade Decay}} \\
\cline{2-5}
 & \textbf{2-EJ Channel} & \textbf{1-EJ Channel} & \textbf{2-EJ Channel} & \textbf{1-EJ Channel} \\
\hline
\textbf{1} & \multicolumn{4}{c|}{$N_{\text{jets}} \ge 2$} \\
\hline
\textbf{2} & \multicolumn{4}{c|}{Pass \textbf{Category I} (Hadronic) or \textbf{Category II} (Leptonic) Trigger} \\
\hline
\textbf{3} & $N_{\text{EJ}} \ge 2$ & $N_{\text{EJ}} \ge 1$ & $N_{\text{EJ}} \ge 2$ & $N_{\text{EJ}} \ge 1$ \\
\hline
\textbf{4} & 
\begin{tabular}{@{}c@{}} $p_T(J_{1}), ~p_T(J_{2}) > 30$ GeV, \\ $|\eta| < 2.4$ \end{tabular} & 
\begin{tabular}{@{}c@{}} $p_T(J_1) > 40$ GeV, \\ $|\eta| < 2.4$, \\ $p_T^{\text{miss}} > 30$ GeV \end{tabular} & 
\begin{tabular}{@{}c@{}} $p_T(J_{1}), ~p_T(J_{2}) > 30$ GeV, \\ $|\eta| < 2.4$ \end{tabular} & 
\begin{tabular}{@{}c@{}} $p_T(J_1) > 40$ GeV, \\ $|\eta| < 2.4$, \\ $p_T^{\text{miss}} > 30$ GeV \end{tabular} \\
\hline
\textbf{5} & $110 < m_{J_{1}J_{2}} < 140$ GeV & $\Delta\phi(J_1, \vec{p}_T^{\,\text{miss}}) < 3$& $m_{J_1},~m_{J_2} < 8$ GeV & $m_{J_1} < 8$ GeV \\
\hline
\end{tabular}
}
\caption{Summary of the event selection criteria at $\sqrt{s} = 14~\mathrm{TeV}$ HL-LHC. The analysis proceeds sequentially from Cuts 1-5. $J_1$ and $J_2$ denote the leading and sub-leading tagged EJs, respectively.}
\label{tab:selection_logic}
\end{table}

For clarity in the kinematic definitions, jets that satisfy any of the EMJ‑1 to EMJ‑6 criteria are denoted as EJ candidates, ordered by decreasing $p_T$, as $J_1$, $J_2$, $\dots$, etc. Based on the multiplicity of these tagged EJs ($N_{\text{EJ}}$), we define the two analysis channels and their corresponding thresholds: 
\begin{itemize}
    \item \textbf{1-EJ Channel:} To mitigate the larger QCD background associated with single-tag events, we impose tighter kinematic cuts. The leading EJ must satisfy $p_T > 40$~GeV and $|\eta| < 2.4$. Furthermore,  $p_T^{\text{miss}} > 30$~GeV is required to suppress events where $p_T^{\text{miss}}$ arises solely from jet energy mismeasurement.
    \item \textbf{2-EJ Channel:} Exploiting the higher signal purity of double-tag events, we relax the kinematic thresholds to maximize signal acceptance. Both $J_1$ and $J_2$ are required to satisfy $p_T > 30$~GeV and $|\eta| < 2.4$, but no additional $p_T^{\text{miss}}$ cut is applied.
\end{itemize}
The complete event selection sequence, integrating the baseline pre-selection, tagging requirements, and topology-dependent discriminators, is summarized in Table~\ref{tab:selection_logic}.

To further suppress the QCD background, we introduce event-level variables sensitive to the specific BPs. For illustration, we use the ggH production mode as a representative example to show these key kinematic distributions. Concretely, events are first selected with the pre-selection criterion $N_{\text{jets}} \ge 2$ and are then required to satisfy a Category I trigger.

\begin{figure}[tbp]
    \centering
    \includegraphics[width=0.95\textwidth]{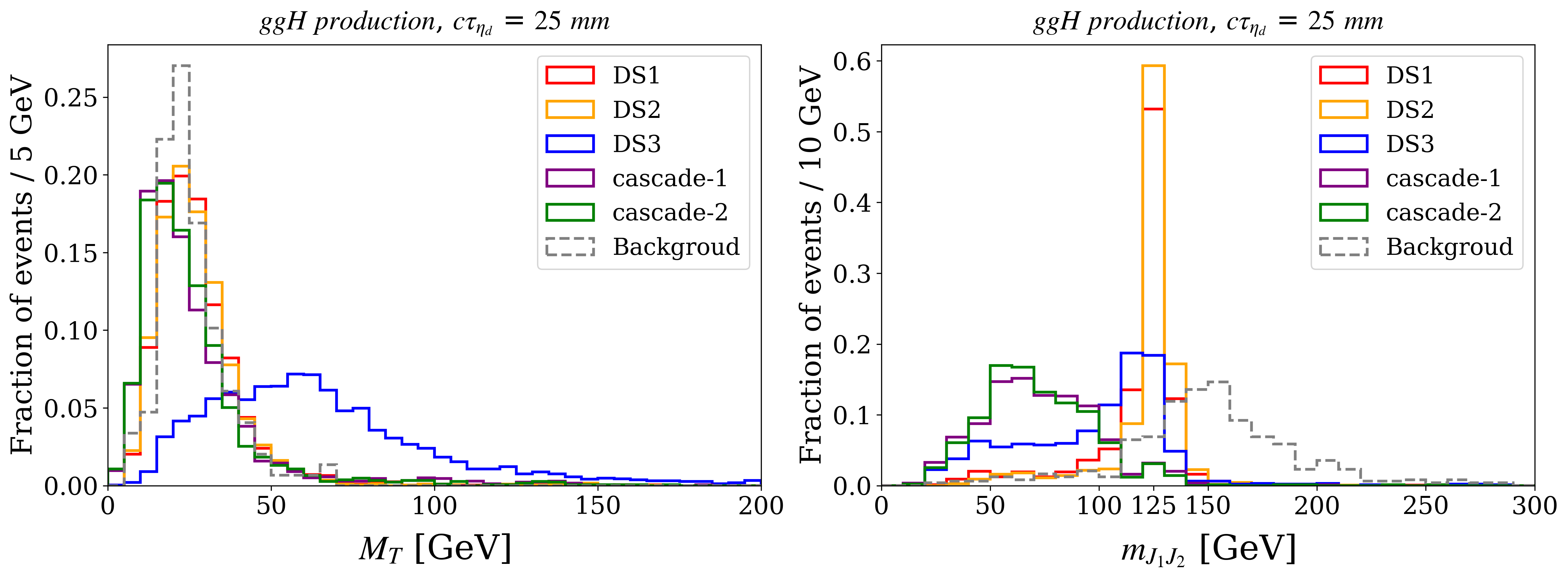}
    \caption{Normalized distributions of the transverse mass $M_T$ (left) and the dijet invariant mass $m_{jj}$ (right) for the ggH production mode at $\sqrt{s}=14$~TeV. 
    The five signal BPs ($c\tau_{\eta_d} = 25$~mm) are distinguished by color: Dark shower scenarios DS1 (red), DS2 (yellow), DS3 (blue), and cascade decay scenarios cascade-1 (purple), cascade-2 (green).
    The dominant SM QCD heavy-flavor ($pp\to b\bar{b}$) background is shown as the grey dashed line. }
    \label{fig:MT_mjj_dist}
\end{figure}

We first consider the transverse mass ($M_T$) and dijet invariant mass ($m_{jj}$). 
For the 1-EJ channel, $M_T$ is constructed from $J_1$ and $p_T^{\text{miss}}$: 
\begin{equation}
    M_T(J_1, p_T^{\text{miss}}) = \sqrt{ m_{J_1}^2 + 2 \left( \sqrt{m_{J_1}^2 + p_{T,J_1}^2} \, p_T^{\text{miss}} - \vec{p}_{T,J_1} \cdot \vec{p}_T^{\,\text{miss}} \right) }.
\end{equation}
For the 2-EJ channel, we reconstruct the dijet invariant mass $m_{jj}$ using the two leading EJ candidates. Figure~\ref{fig:MT_mjj_dist} displays the $M_T$ and $m_{jj}$ distributions, revealing distinct signal behaviors. 
For the fully visible dark shower scenarios (DS1, DS2), 
a distinct resonance peak around $m_{h_1} = 125$~GeV in the $m_{jj}$ distribution can be reconstructed. 
In contrast, the semi-visible dark shower scenario (DS3) is characterized by a broad excess in the $M_T$ distribution due to stable $\tilde{\omega}$'s. A simple cut on $M_T > 30$~GeV, however, provides only limited discrimination against the heavy-flavor QCD background in our analysis region. Consequently, we must turn to other kinematic variables.

\begin{figure}[tbp]
    \centering
    \includegraphics[width=0.9\textwidth]{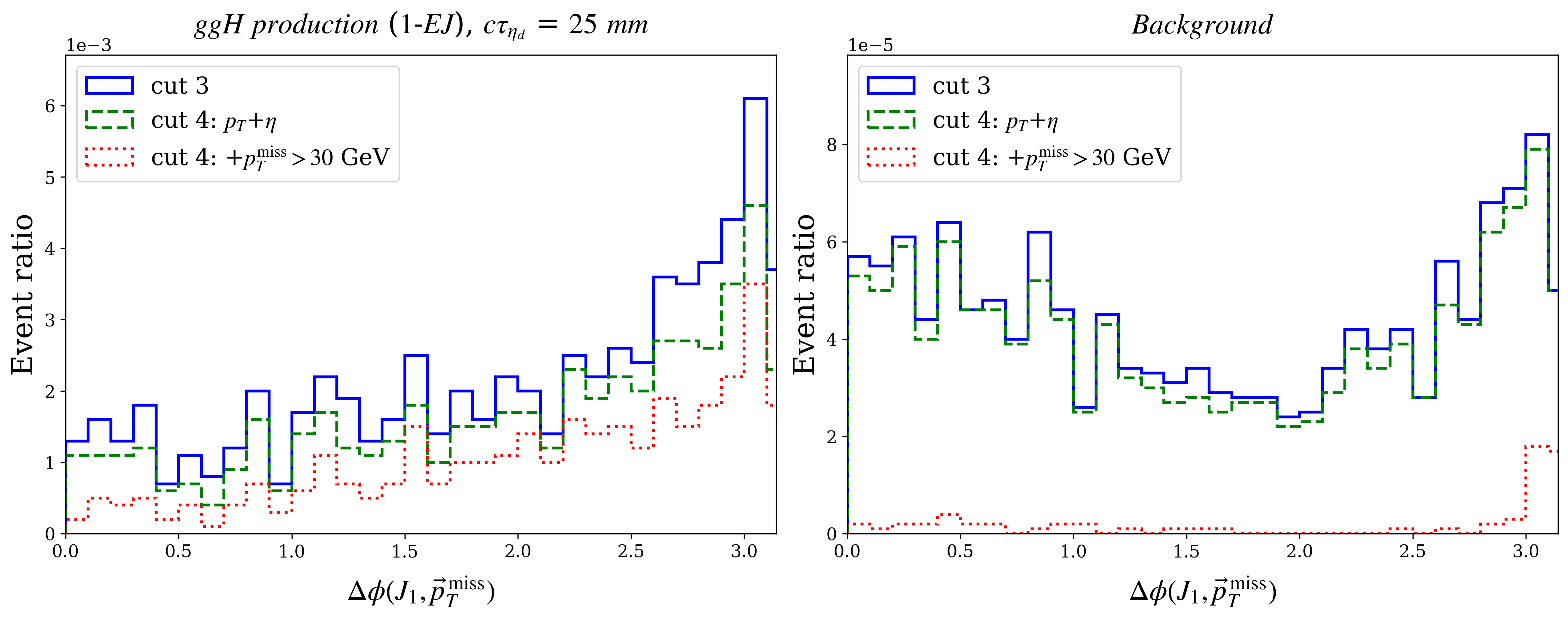} 
    \caption{Distributions of the azimuthal angle difference $\Delta\phi(J_1, \vec{p}_T^{\,\text{miss}})$ for the DS3 and QCD background. The plots show the event ratio to illustrate the evolution across selection stages: after EJ tagging (red solid line), after $p_T/\eta$ cuts (green dashed line), and after the additional $p_T^{\text{miss}}$ cut (red dashed line). The signal corresponds to ggH production with $c\tau_{\eta_d} = 25$~mm.}
    \label{fig:deltaphi_dist}
\end{figure}

To improve the sensitivity for the DS3, we investigate the azimuthal angular separation between $J_1$ and $p_T^{\text{miss}}$, $\Delta\phi(J_1, \vec{p}_T^{\,\text{miss}})$. 
Figure~\ref{fig:deltaphi_dist} shows the distributions of $\Delta\phi(J_1, \vec{p}_T^{\,\text{miss}})$ for the DS3 and the QCD background. We display the event ratio, defined as the number of events surviving a given selection divided by the total number of events, to explicitly demonstrate the background rejection power at three sequential selection stages: after EJ tagging (Cut-3), after partial kinematic requirements (Cut-4 without $p_T^{\text{miss}}$), and after the full requirement of Cut-4. 
We observe that after the full selection, the residual QCD background is predominantly concentrated in the region $\Delta\phi(J_1, \vec{p}_T^{\,\text{miss}}) \gtrsim 3$. In contrast, the signal exhibits a more distributed profile across the angular range.
Based on this feature, we establish the final discriminator for the 1-EJ channel as $\Delta\phi(J_1, \vec{p}_T^{\,\text{miss}}) < 3$ to further reject the background events.

Based on these features, we establish the final cut (Cut-5) for dark shower scenarios as:
\begin{itemize}
    \item \textbf{1-EJ Channel:} $\Delta\phi(J_1, \vec{p}_T^{\,\text{miss}}) < 3$; 
    \item \textbf{2-EJ Channel:} $110 < m_{J_1 J_2} < 140$~GeV.
\end{itemize} 

\begin{figure}[tbp]
    \centering
    \includegraphics[width=0.95\textwidth]{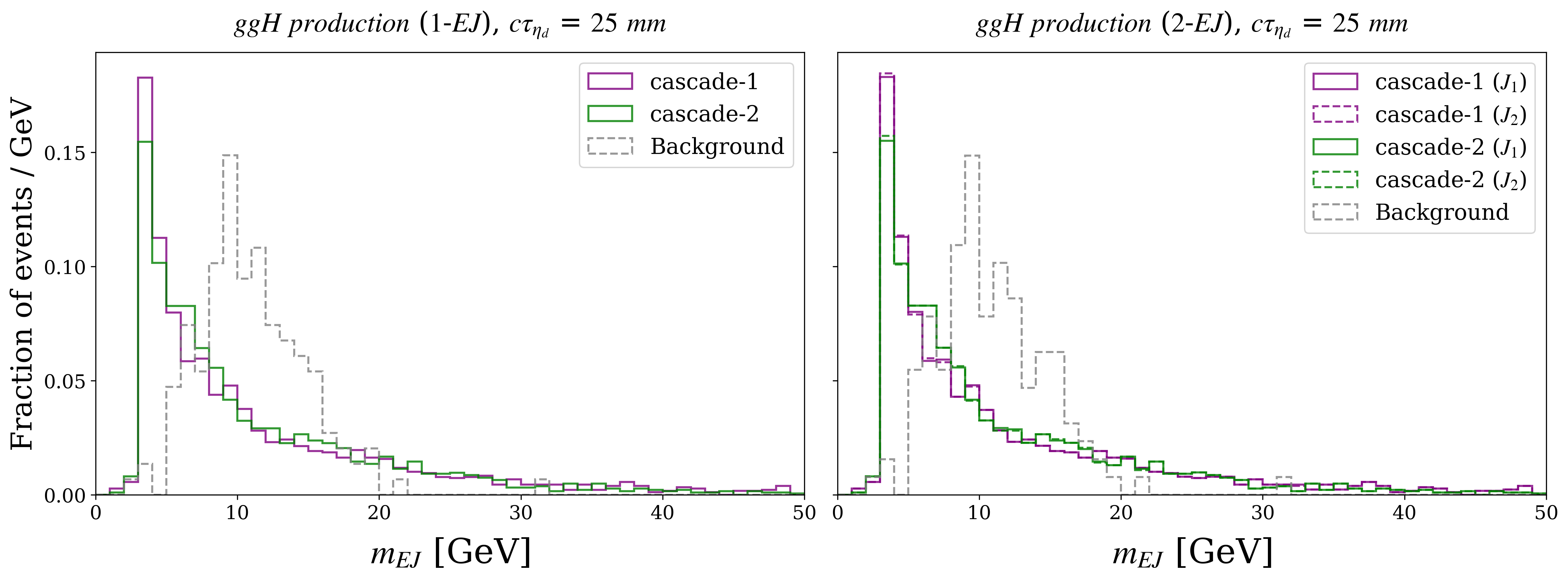}
    \caption{Normalized distributions of the EJ mass ($m_{\text{EJ}}$) for the cascade decay scenarios with $c\tau_{\eta_d} = 25$~mm in the ggH production mode at $\sqrt{s}=14$~TeV. \textbf{Left:} The jet mass distribution of the leading EJ ($J_1$) in the 1-EJ channel. \textbf{Right:} The jet mass distributions of the leading EJ ($J_1$, solid lines) and sub-leading EJ ($J_2$, dashed lines) in the 2-EJ channel.
    The signal benchmarks are distinguished by color: cascade-1 (purple) and cascade-2 (green).
    The dominant SM QCD heavy-flavor ($pp\to b\bar{b}$) background is shown as the grey dashed line. }
    \label{fig:mjet_dist}
\end{figure}

Finally, regarding the cascade decay scenarios (cascade-1, cascade-2), we find that the event-level topological discriminators established above, $m_{jj}$, $M_T$, and $\Delta\phi$, offer insufficient discrimination power. 
This limitation arises from the distinct multi-jet topology: as detailed in Appendix~\ref{appendix:A}, the large $\Delta R$ between the final state $\eta_d$'s prevents them from being clustered into just two jets, thereby rendering these variables ineffective against the SM background.
To address this, we utilize the jet mass of EJ, $m_{\text{EJ}}$. Since the EJ candidates in the cascade decay scenarios originate mainly from light $\eta_d$'s ($m_{\eta_d} = 3$~GeV), they tend to exhibit a narrower mass distribution peaking at lower values compared to the broader mass spectrum of the dominant QCD background. Figure~\ref{fig:mjet_dist} presents the $m_{\text{EJ}}$ distributions for the cascade decay scenarios against the background. The signal distributions distinctly peak at low mass regions ($\sim 3-5$~GeV), while the background extends to significantly larger ones. Exploiting this characteristic feature, we set the final cut (Cut-5) for the cascade decay scenarios as:
\begin{itemize}
    \item \textbf{1-EJ and 2-EJ Channels:} $m_{\text{EJ}} < 8$~GeV for the relevant EJ candidate(s).
\end{itemize}

To evaluate the discovery potential at the HL-LHC, we translate the signal efficiencies into projected upper limits on  $\mathrm{BR}_{\text{exo}}$.  
The statistical analysis requires the expected number of events for both signal ($N_S$) and background ($N_b$) under identical selection criteria. These are evaluated separately for the Category I (Hadronic) and Category II (Leptonic) trigger strategies. In our 14~TeV analysis, we optimize the search by defining 12 distinct \texttt{setN} cases (indexed by $i=1,\dots,12$), combining the analysis channels (1-EJ, 2-EJ) with the EMJ-1 to EMJ-6. The total signal yield for the $i$-th \texttt{setN} is computed using Eq.~\eqref{eq:Ns_sum}. We adopt an integrated luminosity of $\mathcal{L} = 3000~\mathrm{fb}^{-1}$ and update the SM Higgs production cross sections $\sigma_{p}^{\mathrm{SM}}$ to $\sqrt{s}=14$~TeV values~\cite{sigma-14TeV}. The corresponding signal efficiency $A_{i,p}$ follows the optimized selection sequence defined in Table~\ref{tab:selection_logic}. The expected background yield for the same $i$-th \texttt{setN} is estimated as:
\begin{equation}
    N_b^{i} = \mathcal{L} \times \sigma_{\text{bkg}}^{\mathrm{LO}} \times A_{i,\text{bkg}},
\end{equation}
where the background is dominated by QCD heavy-flavor events and its cross section $\sigma_{\text{bkg}}^{\mathrm{LO}}$ is obtained directly from the leading-order estimate in \texttt{PYTHIA~8}.



Unlike the 13~TeV recast which used published $S_{95}$ limits from Refs.~\cite{CMS:2018bvr,Carrasco:2023loy}, the 14~TeV projection requires a statistical estimate of the significance. For the $i$-th \texttt{setN}, we calculate the significance $Z_{i}$ as~\cite{Read:2002hq}
\begin{equation}
    Z_{i}(c\tau_{\eta_d}) = \sqrt{2 \left[ (N_s^{i} + N_b^{i}) \ln\left(1 + \frac{N_s^{i}}{N_b^{i}}\right) - N_s^{i} \right]}.
\end{equation}
The final projected $95\%$ C.L. limit is defined as the minimum $\mathrm{BR}_{\text{exo}}$ value satisfying the condition $Z_{i} = 1.96$ across all \texttt{setN}s at each $c\tau_{\eta_d}$ value.

\subsection{Displaced shower signatures in the CMS muon system} 

In this subsection, we focus on LLP decays within the CMS muon system, particularly considering the gluons-enriched dark jet scenario. The CMS endcap muon detectors consist of alternating layers of detection planes and steel yokes that serve as a magnetic return flux. This unique design allows the decay products of LLPs to initiate both hadronic and electromagnetic showers, producing a high density of hits in localized detector regions~\cite{CMS:2021juv,Mitridate:2023tbj,CMS:2024bvl}. Such showers are classified as muon detector showers (MDS).

To simulate the displaced shower signature in the CMS muon system, we apply \texttt{Delphes3}~\cite{deFavereau:2013fsa} with a CMS CSC cluster template~\cite{Mitridate:2023tbj} 
for the fast detector simulation of MDS. We systematically tested the impact of different analysis steps on the overall signal efficiency and found that the dominant effects arise from the geometrical acceptance and the CSC cluster efficiency settings, specifically the \texttt{CutBasedIDEfficiency} and \texttt{CSCClusterEfficiency} settings in \texttt{Delphes3}. Within $\lvert \eta \rvert < 2.4$, the CSC acceptance for signal clusters shows a clear $\eta$-dependent. Based on the CMS cluster reconstruction efficiency as a function of the simulated $r$ and $\lvert z \rvert$ decay positions from Fig.~1 of Ref.~\cite{CMS:2021juv}, where $r$ corresponds to the transverse (radial) displacement from the beamline and $\lvert z \rvert$ denotes the longitudinal displacement along the beam axis, we extracted the relevant cluster efficiency to define the appropriate range of detector-level CSC cluster efficiency parameters. The pseudorapidity acceptance (\texttt{etacut} efficiency) is modified to incorporate both the ``Single CSC'' requirements and the observed detector-level efficiency variation across $\eta$. By combining these modifications, we obtained a more complete configuration in the Delphes card for the subsequent analysis in which the geometric parameters follow the intrinsic instrumentation geometry of the CMS CSC system, corresponding to the description in Ref.~\cite{CMS:2021juv}, namely \( 0 < R < 6955~\mathrm{mm} \), \( 5000 < \lvert z \rvert < 11000~\mathrm{mm} \), and \( 0.9 < \lvert \eta \rvert < 2.4 \).  In addition, the selection on the number of hits ($N_{\mathrm{hits}}$) is also incorporated into the part of CSC cluster efficiency within the Delphes card.


\begin{figure}[htbp]
    \centering
    \includegraphics[width=0.8\textwidth]{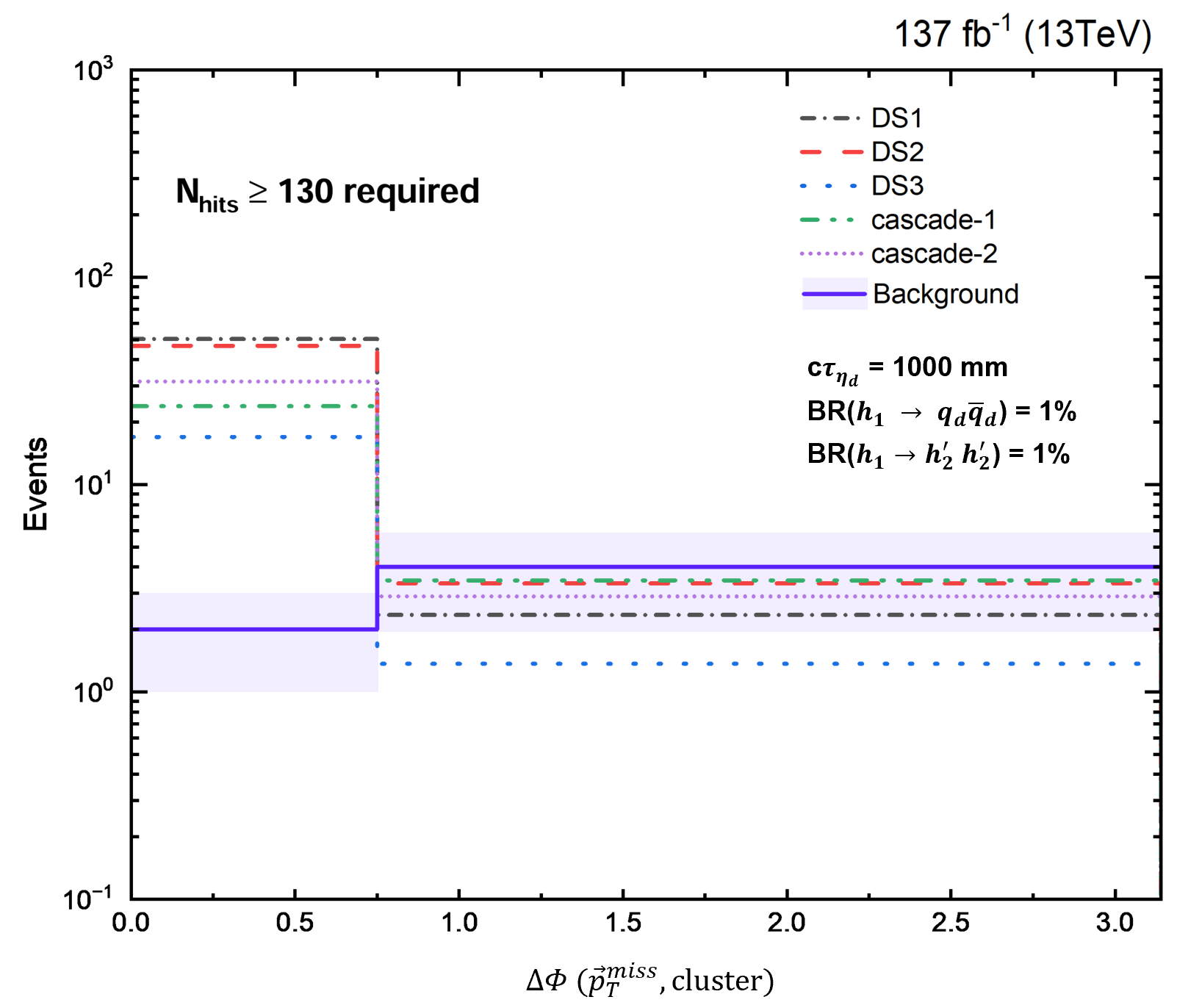}
    \caption{Distributions of $\Delta\phi(\vec{p}_T^{\text{miss}}, \text{cluster})$ for the five signal benchmark points, assuming a long-lived dark meson $\eta_d$ with a proper decay length of $c\tau_{\eta_d} = 1000$~mm. The signal yield normalized to BR$(h_1 \to q_d \bar{q}_d) = 1\%$ for the dark shower scenario: DS1 (black), DS2 (red), DS3 (blue) and BR$(h_1 \to h_2 h'_2) = 1\%$ for the cascade decay scenarios: cascade-1 (green), cascade-2 (purple), respectively. The background is shown as the purple solid line, with the shaded band indicating the uncertainty from Ref.~\cite{CMS:2021juv}. }
    \label{fig:delta phi all-0.75}
\end{figure}

Figure~\ref{fig:delta phi all-0.75} shows the $\Delta\phi(\vec{p}_T^{\text{miss}}, \text{cluster})$ distributions for the five selected BPs, assuming a proper decay length of $\eta_d$ as $c\tau_{\eta_d} = 1000$~mm. These distributions correspond to the final signal predictions obtained by summing over the six dominant Higgs production modes, each normalized to its respective production cross section. Individual contributions from each production mode are detailed in Appendix~\ref{appendix:C}. For the signals, reconstructed clusters generally originate from visible decay products of the LLPs, whereas $\vec{p}_T^{\,\mathrm{miss}}$ arises from invisible particles produced in the same decay or from correlated kinematic effects. Consequently, there is a pronounced azimuthal correlation between the cluster and $\vec{p}_T^{\,\mathrm{miss}}$, favoring small values of $\Delta\phi(\vec{p}_T^{\,\mathrm{miss}}, \text{cluster})$. In contrast, SM backgrounds exhibit little correlation, leading to a nearly uniform $\Delta\phi$ distribution. In cascade decay scenarios, the presence of intermediate states reduces this correlation somewhat, producing a broader $\Delta\phi$ distribution than in dark shower scenarios. Nevertheless, the signals remain strongly concentrated at low $\Delta\phi$, and the selection requirement $\Delta\phi(\vec{p}_T^{\,\mathrm{miss}}, \text{cluster}) < 0.75$ remains effective for signal-background discrimination. Notably, these distributions exhibit the same qualitative behavior as reported in previous CMS measurements, indicating that this angular separation provides a stable and robust shape-based handle for LLP searches.

\begin{figure}[htbp]
    \centering
    \includegraphics[width=1\textwidth]{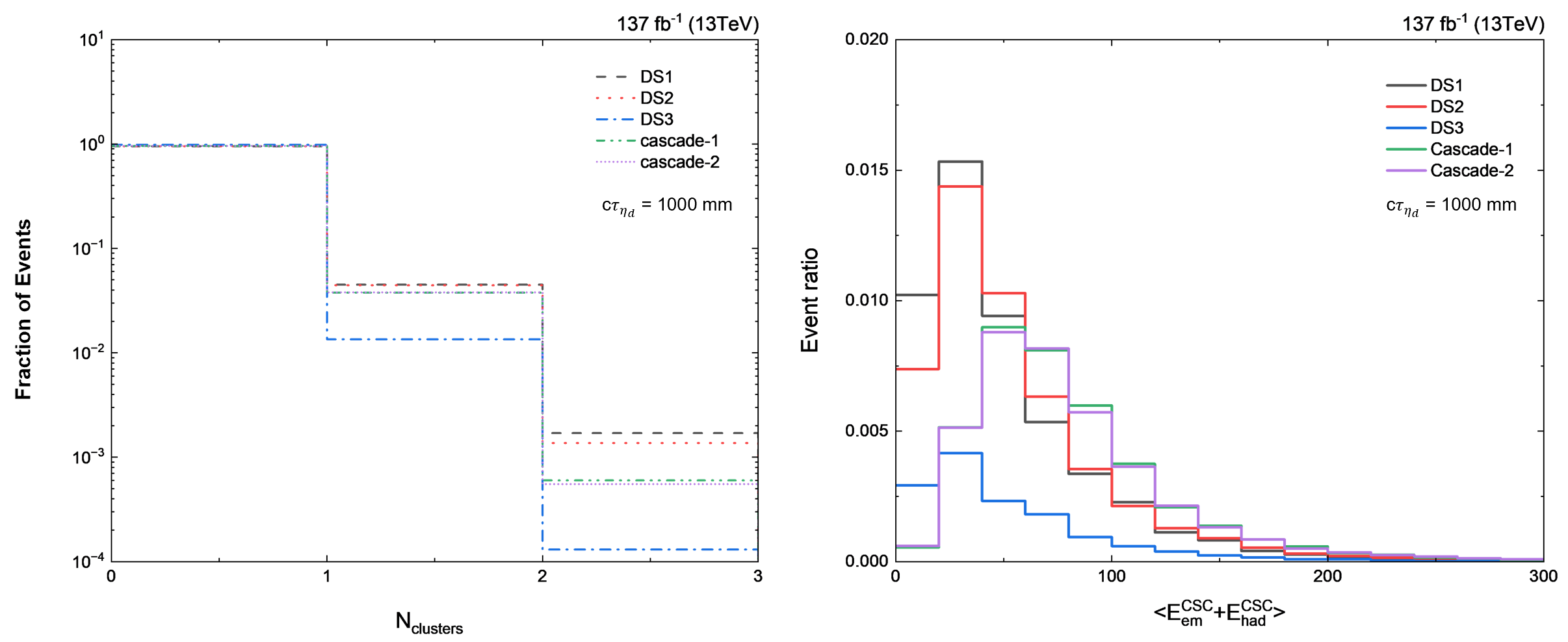}
    \caption{Signal distributions of the number of clusters ($N_{\text{clusters}}$) (left) and the distribution of the average cluster energy, $\langle E_{\text{CSC}}^{\text{em}} + E_{\text{CSC}}^{\text{had}} \rangle$ (right) for all benchmark points with $c\tau_{\eta_d} = 1000$~mm which are color-coded as follows: dark shower scenarios DS1 (black), DS2 (red), and DS3 (blue), alongside cascade decay scenarios cascade-1 (green) and cascade-2 (purple). } 
    \label{fig:Ncluster&Esum}
\end{figure}

In addition, $N_{\mathrm{clusters}}$ and $\langle E^{\text{CSC}}_{\text{em}} + E^{\text{CSC}}_{\text{had}} \rangle$ distributions are presented in Fig.~\ref{fig:Ncluster&Esum}. The requirement $N_{\text{hits}} > 130$ is already implemented in the Delphes card through the cluster efficiency setting. As a result, a considerable fraction of events do not contain any reconstructed clusters, corresponding to $N_{\text{clusters}} = 0$. A notable feature is that DS3 yields significantly fewer reconstructed clusters than DS1 and DS2. This occurs because $m_{\tilde{\omega}} < 2m_{\eta_d}$ along with the high \texttt{probVector} value in DS3 results in a large fraction of stable $\tilde{\omega}$ particles. In contrast, the kinematically allowed $\tilde{\omega}\to\eta_d \eta_d$ with lower \texttt{probVector} values in DS1 and DS2, lead to a greater number of clusters. For the cascade decay scenarios, the LLP multiplicity is fixed at four, resulting in a correspondingly lower total number of reconstructed clusters. Compared to the $\eta_d$-rich in DS1 and DS2, this relatively low multiplicity leads to a significantly smaller number of events satisfying $N_{\rm clusters} > 2$. This trend is evident in the $N_{\text{clusters}}$ distributions shown in the left panel of Fig.~\ref{fig:Ncluster&Esum}.

The right panel of Fig.~\ref{fig:Ncluster&Esum} presents the distribution of $\langle E^{\text{CSC}}_{\text{em}} + E^{\text{CSC}}_{\text{had}} \rangle$ for events with at least one reconstructed CSC cluster. The vertical axis corresponds to the event ratio, defined as the number of events containing at least one CSC cluster divided by the total number of events. The distributions for DS1 and DS2 are nearly identical, with similar peak positions and high event ratio, indicating that the reconstructed cluster energies are comparable in these two dark shower scenarios. DS3, on the other hand, exhibits a significantly lower distribution and smaller event ratio, consistent with the reduced number of reconstructed CSC clusters, reflecting the high fraction of stable $\tilde{\omega}$ particles. In cascade decay scenarios, the distributions lie between those of DS1/DS2 and DS3, with slightly higher peak values, reflecting the energy dispersion across multiple visible decay products in $h'_2$ decays, which leads to moderately higher average cluster energy while decreasing the total number of reconstructed CSC clusters. Moreover, the LLP mass peak can be reconstructed by combining the energy and momentum of the CSC cluster candidates.

In our study, we consider the five selected BPs and begin by recasting the existing CMS search strategies at  $\sqrt{s} = 13~\text{TeV}$ with  $137~\text{fb}^{-1}$~\cite{CMS:2021juv} and $3000~\text{fb}^{-1}$. 
To remain consistent analysis with Ref.~\cite{CMS:2021juv}, we adopt the same event selection procedure, in which all requirements are grouped into two categories: event-level selections and cluster-level selections. The selection criteria are summarized below. 

\textbf{Event-level selections:}
\begin{enumerate}
    \item We require an online trigger requirement of $p_{T}^{\mathrm{miss}} > 120~\mathrm{GeV}$, followed by an offline selection of $p_{T}^{\mathrm{miss}} > 200~\mathrm{GeV}$. When a LLP is produced in association with other prompt and visible objects, or recoils against an initial-state radiation (ISR) jet, and subsequently decays only after traversing the calorimeters, its momentum cannot be properly measured by the detector, leading to a potentially large $p_{T}^{\mathrm{miss}}$. 

    \item We require at least one jet from ISR with $p_T > 50~\mathrm{GeV}$ and $\lvert \eta \rvert < 2.4$ using the anti-$k_t$ algorithm for the reason that signal events that satisfy the requirement of $p_{T}^{\mathrm{miss}}$ are always produced together with an ISR jet. 
    
    \item To suppress SM background events originating from $W$ boson and top-quark decays, we impose a tight lepton veto~\cite{CMS:2017wyc}, which removes any event containing an isolated electron (muon) with $p_T > 35~(25)\,\text{GeV}$ and $|\eta| < 2.5~(2.4)$.
\end{enumerate} 

\begin{figure}[tbp]
    \centering
    \includegraphics[width=1\textwidth]{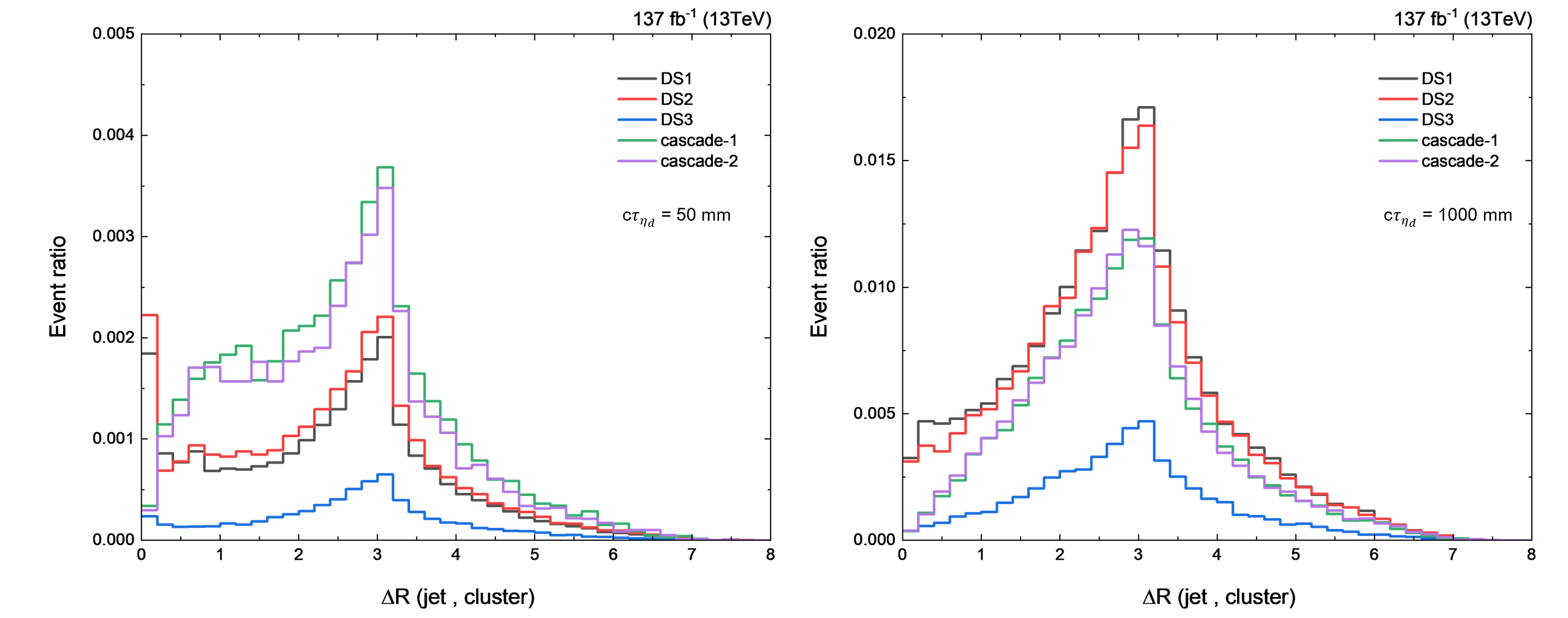} 
    \caption{Distributions of the angular separation $\Delta R(\text{jet, cluster})$ for the five benchmark points, evaluated at $c\tau_{\eta_d} = 50$~mm (left) and $1000$~mm (right). The results are presented for dark shower scenarios DS1 (black), DS2 (red), and DS3 (blue), alongside cascade decay scenarios cascade-1 (green) and cascade-2 (purple). }
    \label{fig:deltaR}
\end{figure}

\textbf{Cluster-level selections:}
\begin{enumerate}
    \item In order to remove backgrounds from punch-through jets and muon bremsstrahlung, we reject all clusters that are geometrically matched to a jet (muon) within a small angular separation, defined as lying within a cone of radius $\Delta R < 0.4$, where the matched jet (muon) is required to have $p_T > 10~(20)~\text{GeV}$. All jet candidates are reconstructed using the particle-flow algorithm and clustered with the anti-\(k_t\) algorithm with a distance parameter $R = 0.4$, combining information from both tracker and calorimeter systems to identify charged and neutral hadrons. In this study, we focus only on signal events that can resemble punch-through jets and explicitly exclude trackless jets, those reconstructed solely from calorimeter information, as discussed in Refs.~\cite{CMS:2021rwb,CMS:2022wjc}. 
    The corresponding $\Delta R$(jet,cluster) distributions are presented in Fig.~\ref{fig:deltaR}.
    
    \item A veto is applied to all clusters with $|\eta| > 2.0$ to further suppress the background from muon bremsstrahlung, as the degraded muon reconstruction and identification performance in the forward region allows some muons to evade the muon veto from previous steps.
    
    \item To mitigate clusters arising from in-time and out-of-time pileup, we impose a requirement on the CSC cluster timing, namely that the average hit time satisfies $-5 <  \Delta t_{\text{cluster}} < 12.5~\text{ns}$, which effectively reduces the associated background.\footnote{Pile-up is not simulated in this study, as its impact is found to be modest according to Ref.~\cite{Mitridate:2023tbj}. Moreover, pile-up effects can be effectively mitigated through the new MIP timing detector and dedicated pile-up removal algorithms.}
    
    \item We impose the requirement $\Delta\phi(\vec{p}_{T}^{\,\mathrm{miss}},\,\mathrm{cluster}) < 0.75$ because clusters originating from LLP decays are typically well aligned with the direction of $\vec{p}_{T}^{\,\mathrm{miss}}$, whereas background clusters exhibit no such correlation and feature an approximately uniform $\Delta\phi$ distribution. As we show in Fig.~\ref{fig:delta phi all-0.75}, by applying this selection, we significantly enhance the probability that the selected cluster arises from signals.

    \item Following Ref.~\cite{CMS:2021juv}, clusters with any hits in the two innermost rings of the ME1 station (ME1/1 and ME1/2) are vetoed, as these rings have the least amount of absorber material in front of them.
\end{enumerate}

To illustrate the effect of jet veto in the first step of cluster-level selections, the distributions of $\Delta R$(jet,cluster) between jets and clusters for the five BPs are shown in Fig.~\ref{fig:deltaR}, where all Higgs production modes are combined according to their relative production rates and the individual contribution for each Higgs production mode are detailed in Appendix~\ref{appendix:C}. The vertical axis is defined identically to that in the right panel of Fig.~\ref{fig:Ncluster&Esum}, representing the event ratio. The distributions correspond to $c\tau_{\eta_d} = 50$~mm (left), with those for $c\tau_{\eta_d} = 1000$~mm (right) included for comparison. For DS1 and DS2 with relatively small $c\tau$, a secondary peak is observed at small $\Delta R$(jet,cluster). This behavior is attributed to the high LLP multiplicity in these channels; specifically, the kinematically allowed decay $\tilde{\omega} \to \eta_d \eta_d$ leads to a large number of clusters. At high boosts, these daughter particles become highly collimated, increasing the probability of multiple clusters being reconstructed within the same jet cone or in close proximity to one another. In contrast, DS3 exhibits only a single peak. This is primarily due to the mass hierarchy $m_{\tilde{\omega}} < 2m_{\eta_d}$ and the high \texttt{probVector} value, which causes a significant fraction of stable $\tilde{\omega}$ particles. For the cascade decay process, the underlying physics reasons responsible for the features observed in the $\Delta R$(jet,cluster) distributions are the same as those governing the $N_{\rm clusters}$ distributions. 

In order to validate our detector simulation settings and analysis strategy, we first test the process of the long-lived scalar with masses of $15$~and $40$~GeV in the decay channel $S\to d\bar{d}$ from the Higgs exotic decay, $h \to SS$, to which the full set of selection criteria introduced above is applied. The derived exclusion limits on the branching ratio $\mathrm{BR}(h \to SS)$ as a function of the proper decay length $c\tau$ in our recast are in good agreement with those reported by the CMS Collaboration~\cite{CMS:2021juv}, thereby validating the reliability of our simulation and analysis procedure.

For the HL-LHC operating at $\sqrt{s} = 14~\mathrm{TeV}$, the event-level selection is superseded by dedicated trigger strategies, introduced to improve the signal acceptance and detailed in Table~\ref{tab:trigger}. The cluster-level selection criteria, however, remain unchanged. It is worth emphasizing that for signals with leptonic final states, the "Tight Lepton Veto" is not imposed in this part of the analysis, since this requirement is incompatible with the Category~II triggers designed to select events containing energetic leptons.

\subsection{Numerical results and discussions} 

In this subsection, we present the analysis results corresponding to the five BPs introduced in Section~\ref{sec:constraint} for EJ signatures in tracker system and displaced showers in CMS muon system. We focus on two classes of long-lived dark meson $\eta_d$ production mechanisms: Higgs-portal cascade decay and dark shower processes. Based on the optimized strategies described above, we quantify the relevant signal efficiencies and derive the projected $95\%$ C.L. upper limits on the Higgs exotic decay branching ratios.

We begin by evaluating the sensitivity of the track-based EJ search. For the HL-LHC projections, we adopt the \textit{$It_5$} strategy as our baseline, while the corresponding results using the \textit{R} strategy are provided in Appendix~\ref{appendix:B}. To characterize the performance of the \texttt{setN}s, we evaluate the signal retention from two perspectives: the cumulative efficiency after successive selection stages (cutflow) and the final maximum achievable acceptance. For both analyses, we select the maximum efficiency among the 12 \texttt{setN}s for each Higgs production mode and $c\tau_{\eta_d}$ value.

\begin{figure}[tbp]
    \centering
    \includegraphics[width=0.95\textwidth]{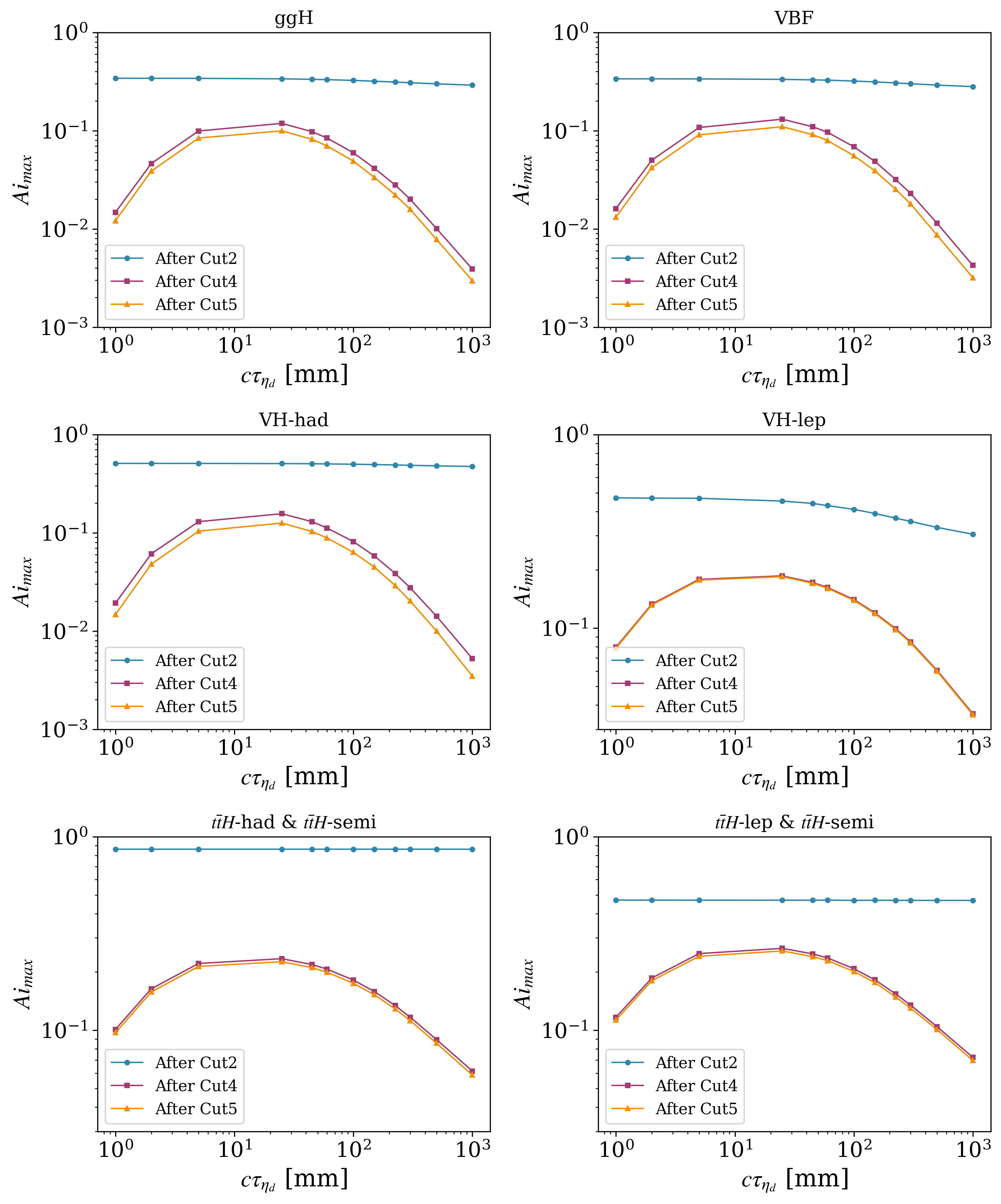} 
    \caption{Signal efficiency at successive selection stages as a function of $\eta_d$ proper lifetime $c\tau_{\eta_d}$ for the DS1. The six panels correspond to the six SM Higgs production modes. 
    In each panel, the curves represent the efficiency remaining after passing the trigger (Cut-2), EJ tagging and kinematics (Cut-4), and final selection (Cut-5).}
    \label{fig:eff_curves}
\end{figure}

Figure~\ref{fig:eff_curves} illustrates the efficiency evolution after the trigger (Cut-2), EJ tagging and kinematics (Cut-4), and the final selection (Cut-5) for the representative benchmark DS1 according to Table.~\ref{tab:selection_logic}. 
The results confirm the robustness of our trigger strategy: by combining Category I and II triggers, we maintain significant acceptance across the diverse Higgs production channels.

\begin{figure}[tbp]
    \centering
    \includegraphics[width=0.6\textwidth]{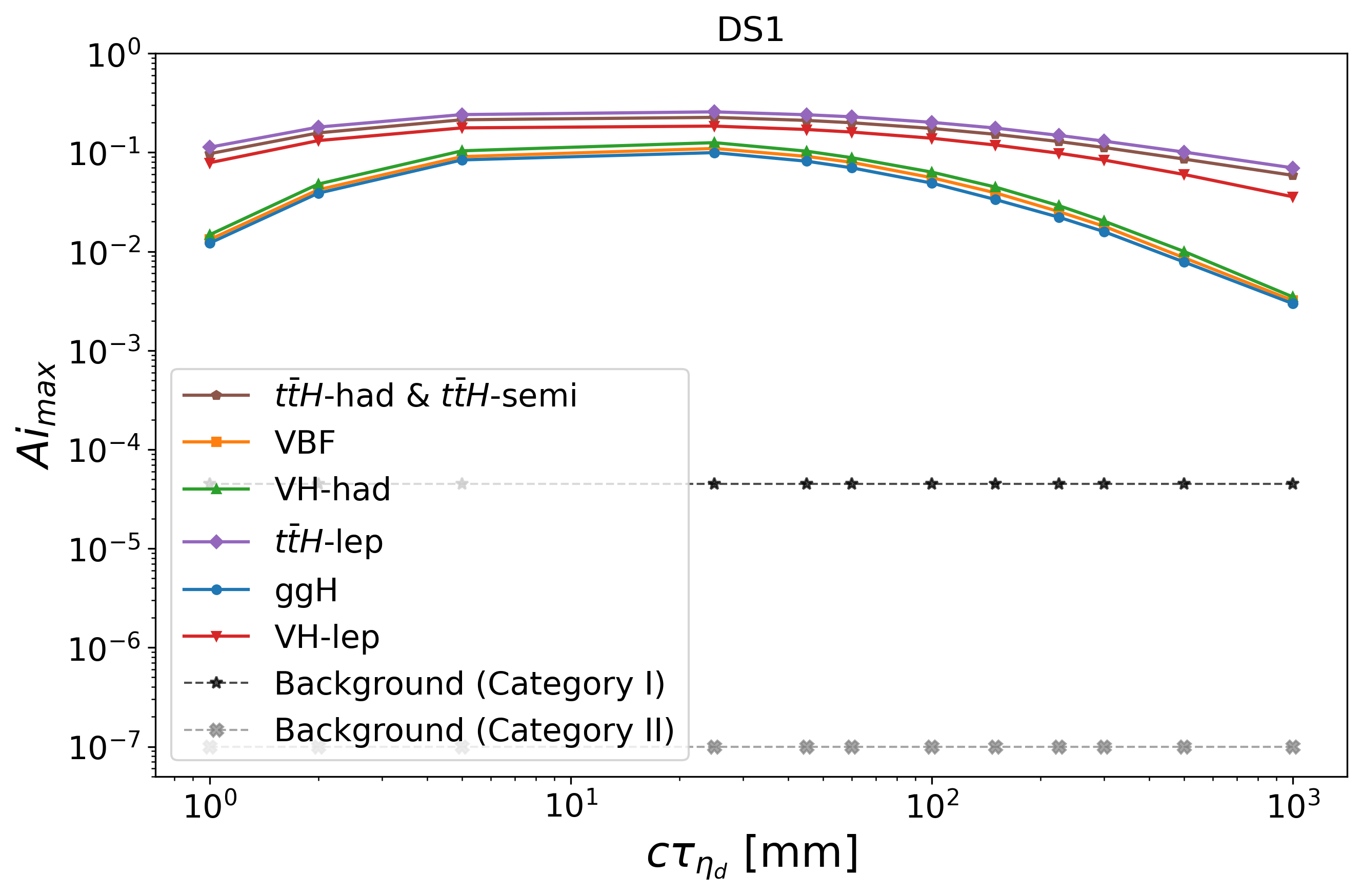} 
    \caption{Maximum signal acceptance ($A_{i,\text{max}}$) as a function of $\eta_d$ proper lifetime $c\tau_{\eta_d}$ for the DS1 at $\sqrt{s}=14$~TeV. 
    The curves represent the optimal efficiency obtained by scanning over all 12 selection sets for each of the six Higgs production modes.
    The corresponding background acceptances for the Category I and II trigger streams are indicated by the horizontal dashed lines (black and grey, respectively). }
    \label{fig:14tev_amax}
\end{figure}

Figure~\ref{fig:14tev_amax} summarizes the final maximum signal acceptance ($A_{i,\text{max}}$) corresponding to the endpoint (Cut-5) of the optimized selection chains shown in Table~\ref{tab:selection_logic}. The acceptance reaches its maximum in $c\tau_{\eta_d} \sim 10\text{--}100$~mm and decreases at both short and long $\eta_d$ lifetimes due to insufficient displacement or decays beyond the tracker volume. 
To demonstrate the discrimination power, the background acceptances for the \texttt{HardQCD} process are superimposed as horizontal lines. The distinct separation of several orders of magnitude between the signal curves ($\mathcal{O}(10^{-1})$) and the background levels ($\sim 10^{-5}$ for Category I and $\sim 10^{-7}$ for Category II) highlights the effectiveness of \texttt{setN}s.

Comparing these Higgs production modes reveals that lepton-associated modes generally maintain higher efficiencies than hadronic modes. This is because leptonic modes benefit from the lower thresholds of Category II triggers and the single-tag requirement of the 1-EJ channel, reaching peak efficiencies of order $\mathcal{O}(10\%)$. In contrast, hadronic modes face stricter Category I trigger requirements and the double-tag penalty of the 2-EJ channel, leading to a steeper efficiency decline at short $\eta_d$ lifetimes 
($c\tau_{\eta_d}<10$~mm).

\begin{figure*}[tbp]
    \centering
    \begin{subfigure}[b]{0.48\textwidth}
        \includegraphics[width=\textwidth]{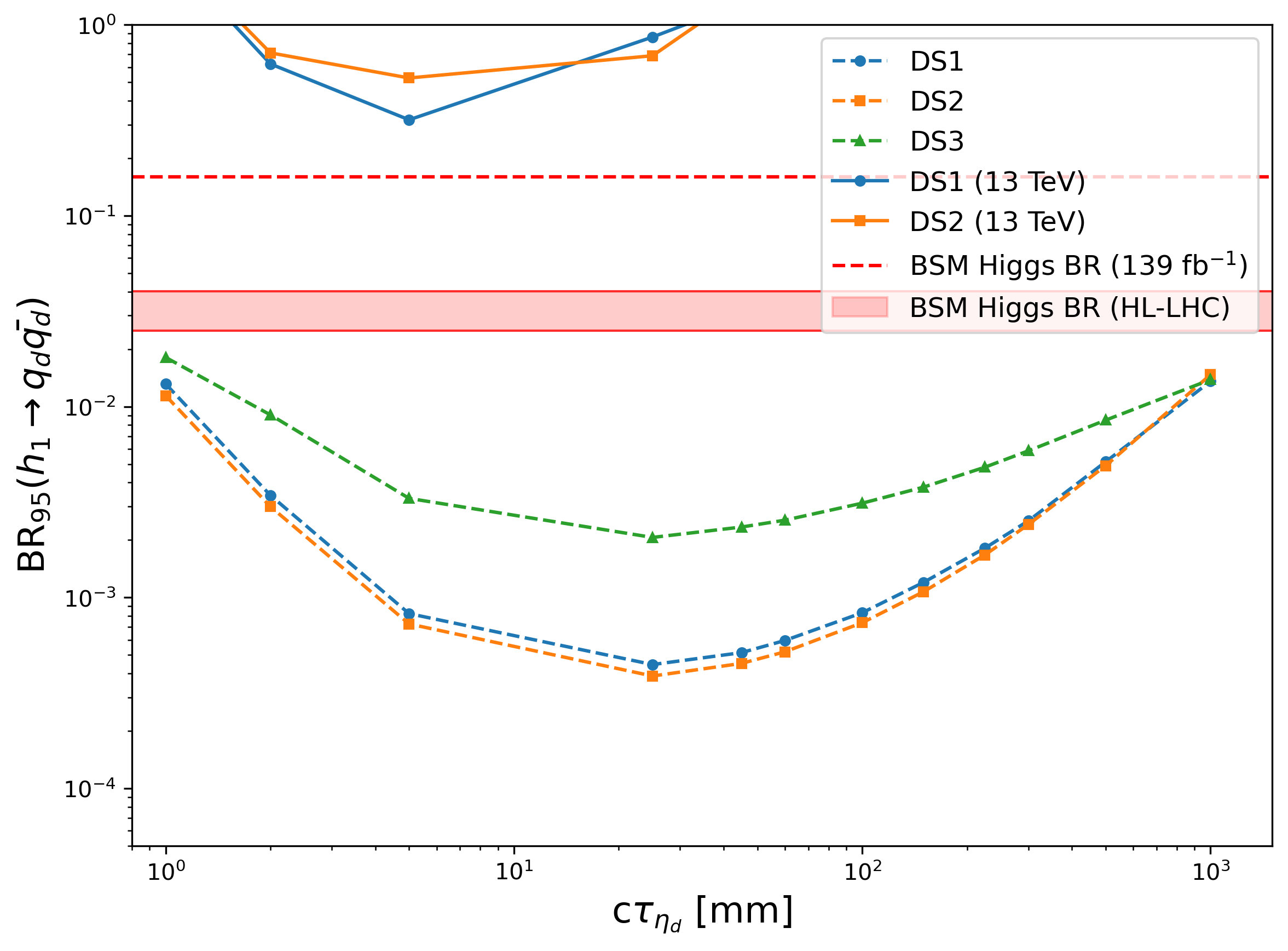} 
        \label{fig:limit_ds_cat1}
    \end{subfigure}
    \hfill
    \begin{subfigure}[b]{0.48\textwidth}
        \includegraphics[width=\textwidth]{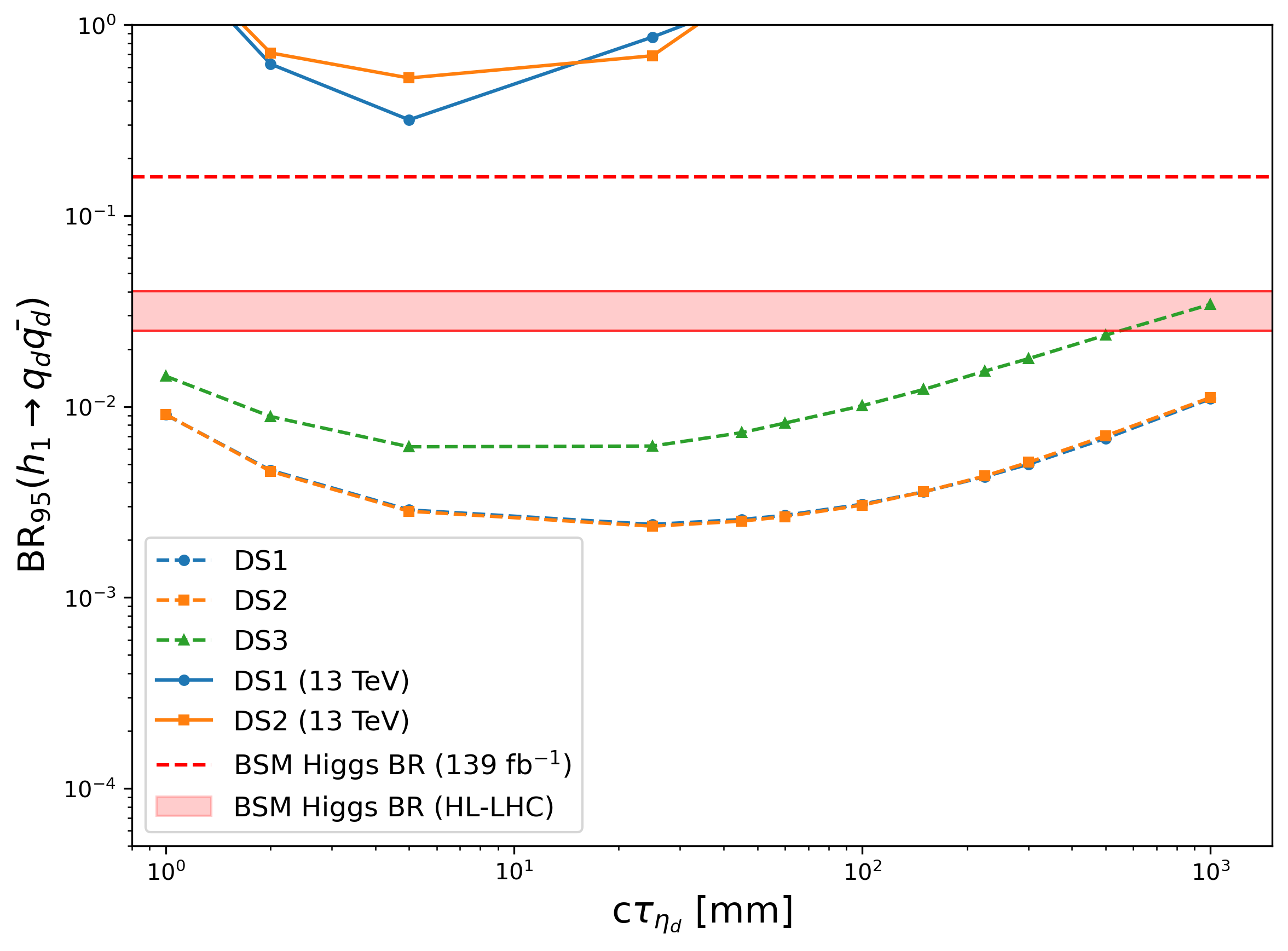} 
        \label{fig:limit_ds_cat2}
    \end{subfigure}
    
    
    \begin{subfigure}[b]{0.48\textwidth}
        \includegraphics[width=\textwidth]{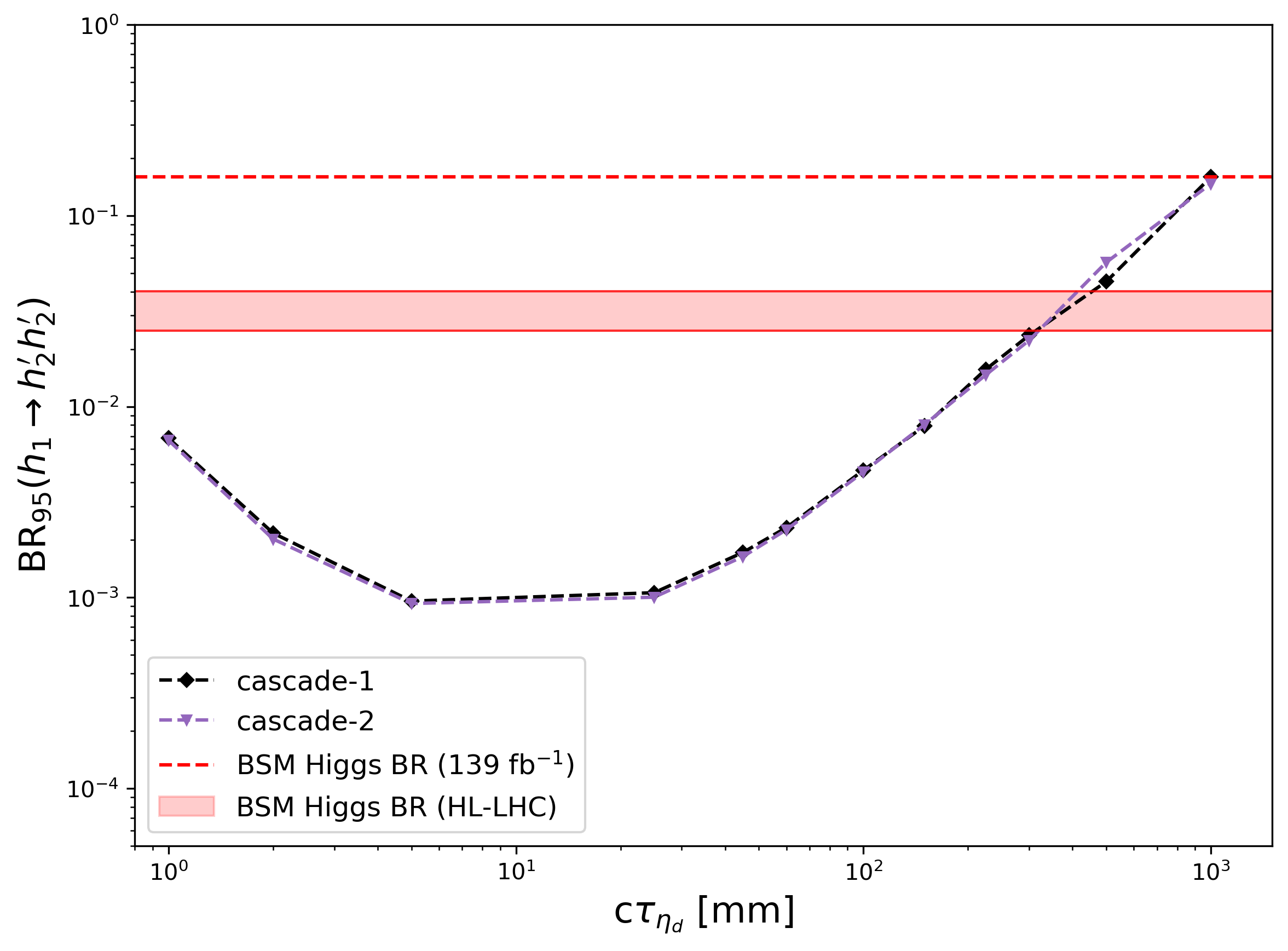} 
        \label{fig:limit_cas_cat1}
    \end{subfigure}
    \hfill
    \begin{subfigure}[b]{0.48\textwidth}
        \includegraphics[width=\textwidth]{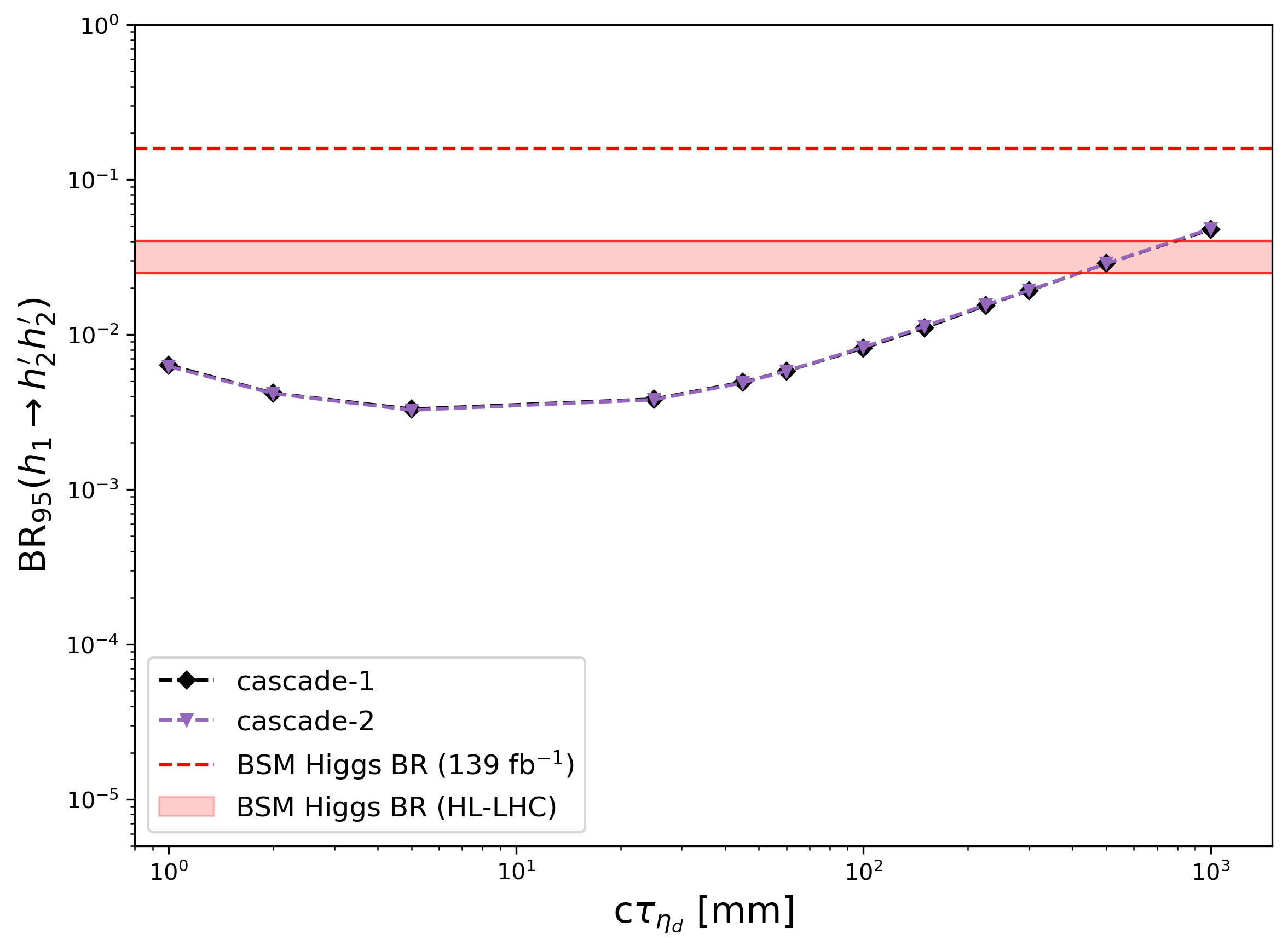} 
        \label{fig:limit_cas_cat2}
    \end{subfigure}
    
\caption{Projected $95\%$ C.L. upper limits on exotic Higgs branching ratios as a function of the $\eta_d$ proper lifetime $c\tau_{\eta_d}$. 
The top (bottom) panels show results for dark shower (cascade decay) scenarios, while the left (right) panels correspond to the Category I (Category II) trigger strategy. 
The solid curves show limits obtained by reinterpreting the CMS EJ search results at $\sqrt{s}=13~\mathrm{TeV}$ with $\mathcal{L}=16.1~\mathrm{fb}^{-1}$~\cite{CMS:2018bvr,Carrasco:2023loy}. The dashed curves correspond to projections for the HL-LHC at $\sqrt{s}=14~\mathrm{TeV}$ with $\mathcal{L}=3000~\mathrm{fb}^{-1}$. 
The red dashed horizontal line indicates the current model-independent exclusion on BSM Higgs decay branching ratios derived from ATLAS and CMS data with $139~\mathrm{fb}^{-1}$~\cite{Cepeda:2019klc}, while the red shaded band shows the projected HL-LHC sensitivity to BSM Higgs decay branching ratios, encapsulating different assumptions on systematic uncertainties~\cite{Carrasco:2023loy}.} 
\label{fig:limits_14tev_grid}
\end{figure*}

Compared to the 13~TeV baseline, the 14~TeV strategy achieves a dramatic improvement, with acceptances reaching $\mathcal{O}(10\%)$ in $c\tau_{\eta_d}\sim 2-50$~mm. This gain is primarily driven by the relaxed jet multiplicity requirement ($N_{\text{jets}} \ge 2$) and the optimized trigger categories. 
The projected $95\%$ C.L. upper limits on the exotic Higgs branching ratio, $\mathrm{BR}(h \to \text{dark sector})$, are shown in Fig.~\ref{fig:limits_14tev_grid}. These limits are derived following the analysis and statistical procedures described in Sec.~\ref{sec:EJ_analysis}. 
The results are organized to illustrate the sensitivity of the event selection sets, including the fixed selection sets at 13~TeV and the optimized \texttt{setN} configurations at 14~TeV. We compare the dark shower (top row) and cascade decay (bottom row) scenarios under the Category~I (left column) and Category~II (right column) trigger strategies. In the figure, the HL-LHC projections (dashed curves) are compared against the recasted 13~TeV limits (solid curves). 
Additionally, we include the current observed upper limit on undetected Higgs decays from ATLAS and CMS ($139~\mathrm{fb}^{-1}$, red dashed line)~\cite{Cepeda:2019klc}  and the projected HL-LHC sensitivity band (red shaded area)~\cite{Carrasco:2023loy}. Limits falling below these reference lines demonstrate the significant added value of our dedicated EJ search.

The comparison between the solid and dashed curves in Fig.~\ref{fig:limits_14tev_grid} highlights the substantial gain in sensitivity achieved by our optimized strategy. The 13~TeV recast limits for DS1 and DS2 lie entirely above the red dashed line, indicating that the previous search strategy with the high $H_T$ trigger was not sensitive enough. In contrast, the 14~TeV projections demonstrate an improvement of more than three orders of magnitude in $c\tau_{\eta_d} \sim 10\text{--}100$~mm, reaching BR$(h_1\to q_d\bar{q_d})$ as low as $\mathcal{O}(10^{-4})$.

Regarding the specific signal BPs, the strongest constraints are observed for the dark shower scenarios under the Category I trigger, where the projected limits remain well below the red shaded band across $c\tau_{\eta_d}=1-1000$~mm. Under the Category II trigger, the limits are similarly strong for DS1 and DS2; however, for DS3 scenario, the sensitivity degrades at $c\tau_{\eta_d}=500-1000$~mm where the limit curve rises into the red shaded band due to the reduced tracking efficiency for displaced decays associated with stable $\tilde{\omega}$'s. 
For the cascade decay scenarios, the Category I trigger yields the strongest constraints at short to intermediate $\eta_d$ lifetimes, with limits well below the red shaded band and reaching BR$(h_1\to h'_2 h'_2)$ of $\mathcal{O}(10^{-3})$ for $c\tau_{\eta_d}<300$~mm. 
At larger $\eta_d$ lifetimes, the sensitivity of this strategy deteriorates, while the Category II trigger becomes more effective, providing stronger constraints in the long-lifetime region and entering the red shaded band only at the highest $c\tau_{\eta_d}$ values.

\begin{figure}[htbp]
    \centering
    \includegraphics[width=0.95\textwidth]{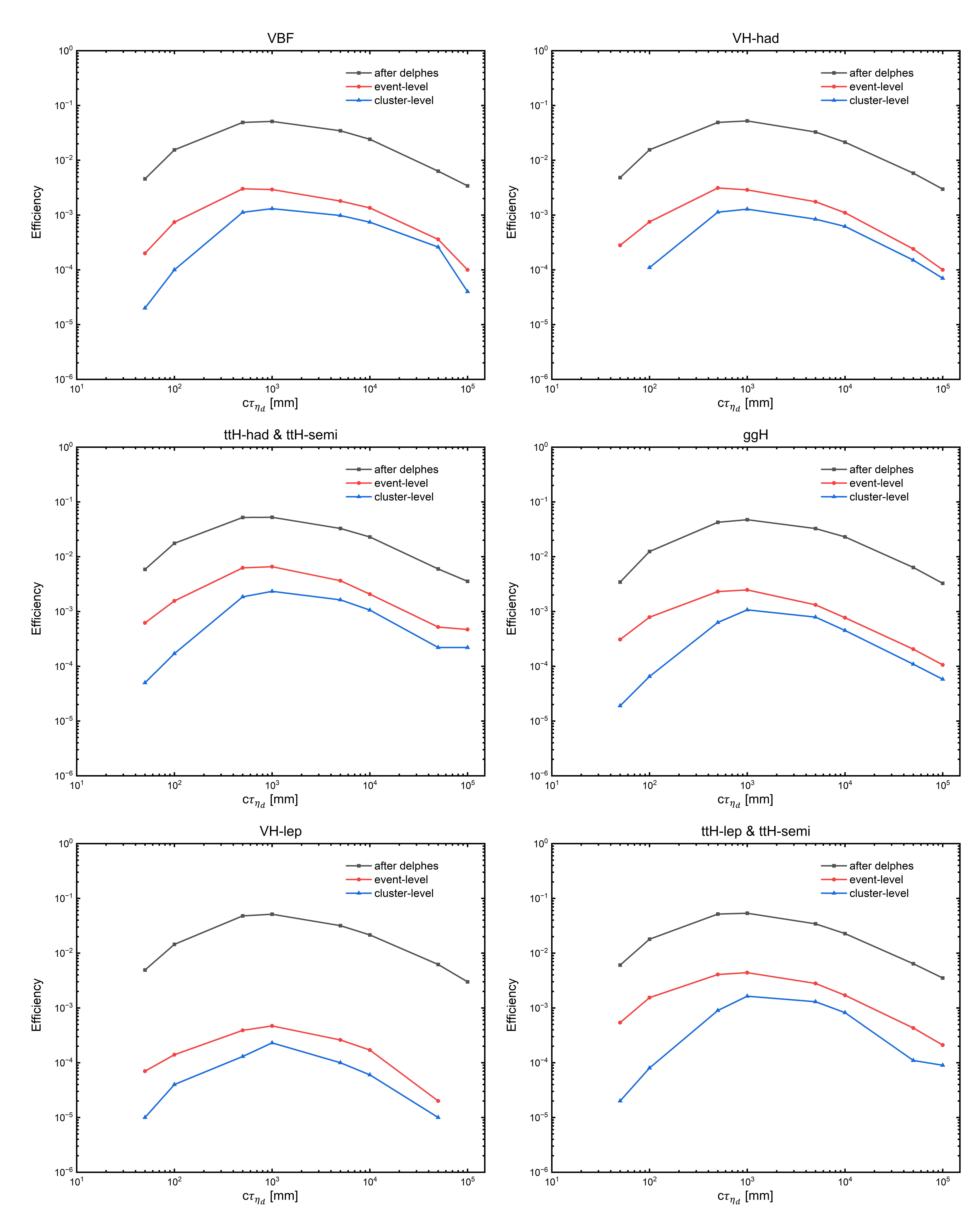}
    \caption{Validation of the geometric acceptance and cluster efficiency using a three-layer procedure (after delphes, event-level, and cluster-level) for the six major Higgs production modes (ggH, VBF, VH-had, VH-lep, $t\bar{t}H$-had \& $t\bar{t}H$-semi, and $t\bar{t}H$-lep \& $t\bar{t}H$-semi) at $\sqrt{s} = 13$~TeV for the DS1: \( \Lambda_d = m_{\tilde{\omega}} = 2.5m_{\eta_d} \), \( \texttt{probVector} = 0.32 \). }
    \label{fig:BP1-CUTS}
\end{figure}

Next, we evaluate the sensitivity of displaced shower in the CMS muon system. To quantify the impact of event-level and cluster-level selections on the overall signal acceptance, we divide the analysis into three steps using benchmark DS1 as an example: (1) acceptance after the Delphes simulation; (2) efficiency after applying the event-level selections; (3) efficiency after imposing the cluster-level requirements. The resulting acceptances for the six dominant Higgs production modes (ggH, VBF, VH-had, VH-lep, $t\bar{t}H$-had \& $t\bar{t}H$-semi, and $t\bar{t}H$-lep \& $t\bar{t}H$-semi) at $\sqrt{s}=13~\mathrm{TeV}$ are presented in Fig.~\ref{fig:BP1-CUTS}.

It should first be noted that the \texttt{Delphes} simulation framework accounts for the overall cluster reconstruction efficiency, which encapsulates several experimental effects and selection criteria. These include the intrinsic cluster reconstruction efficiency, the muon and active rechit (Reconstruction Hits) vetos, time spread requirements, and the $N_{\text{hit}} \ge 130$ threshold. Furthermore, the \texttt{Delphes} implementation incorporates cut-based identification efficiencies. Due to the displacement-dependent nature of these instrumental vetos and the geometric requirements of cluster reconstruction, the resulting signal acceptance following the \texttt{Delphes} simulation exhibits a characteristic dependence on $c\tau_{\eta_d}$. In addition, the event-level selection criteria (most notably the $p_T^{\text{miss}} > 200$~GeV requirement) result in a significant suppression of the signal acceptance, typically exceeding $90\%$. This loss will be significantly mitigated by the implementation of our dedicated trigger strategy. Within this selection chain, the ``Tight Lepton Veto'' exerts a particularly strong impact on channels characterized by isolated leptons in the final state. Specifically, for the VH-lep, $t\bar{t}H$-lep \& $t\bar{t}H$-semi processes, this veto leads to a much more pronounced reduction in acceptance compared to the purely hadronic channels.

\begin{figure}[htbp]
    \centering
    \includegraphics[width=1.0\textwidth]{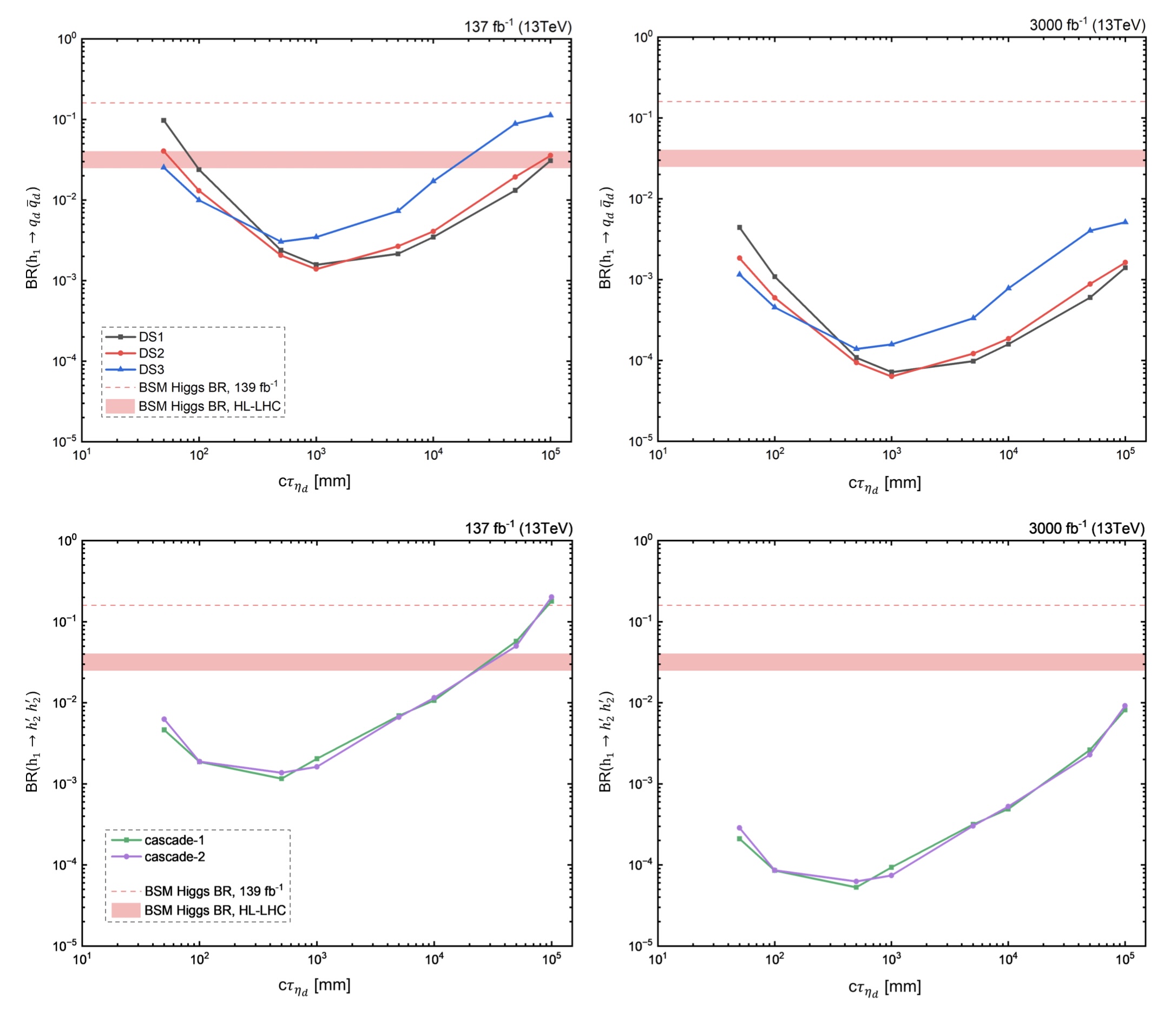}
    \caption{The 95\% CL upper limits on the branching ratios $\text{BR}(h_1 \to q_d \bar{q}_d)$ for dark shower processes (top) and $\text{BR}(h_1 \to h'_2 h'_2)$ for cascade decay processes (bottom), shown as functions of the $\eta_d$ proper lifetime $c\tau_{\eta_d}$ (mm). The exclusion limits are derived by recasting recent CMS results at $\sqrt{s} = 13$~TeV with $137~\text{fb}^{-1}$ (left) and are projected for a future integrated luminosity of $3000~\text{fb}^{-1}$ (right), evaluated across the five benchmark points. The signal benchmarks are color-coded: dark shower scenarios DS1, DS2, and DS3 are shown in black, red, and blue, respectively, while cascade decay scenarios cascade-1 and cascade-2 are depicted in green and purple. For comparison, the red dashed horizontal line indicates the current model-independent exclusion on BSM Higgs decay branching ratios from ATLAS and CMS ($139~\text{fb}^{-1}$)~\cite{Cepeda:2019klc}, while the red shaded band denotes the projected HL-LHC sensitivity to BSM Higgs decays, encompassing various systematic uncertainty assumptions~\cite{Carrasco:2023loy}. }
    \label{fig:13TeV-all}
\end{figure}

Figure~\ref{fig:13TeV-all} presents the current exclusion limits alongside the estimated sensitivities extending to $\mathcal{L}=3000~\mathrm{fb}^{-1}$. The current limits are derived from CMS results with $\sqrt{s} = 13$~TeV and $\mathcal{L}=137~\mathrm{fb}^{-1}$~\cite{CMS:2021juv}. 
For the current dataset, the exclusion reach is estimated by requiring a number of signal events $N_S = 6$, which corresponds to the $95\%$ CL. For the projection with $\sqrt{s} = 13$~TeV and $\mathcal{L}=3000~\mathrm{fb}^{-1}$, we instead require $N_S = 13.6$ to define the 95\% CL exclusion. These choices follow the same statistical prescription as used in the baseline analysis
and allow for a consistent comparison between the current limits and future sensitivities.

In the DS1 and DS2 scenarios, the $\Delta R(\text{jet, cluster})$ distribution at $c\tau_{\eta_d} = 50$~mm (left panel of Fig.~\ref{fig:deltaR}) demonstrates that the jet veto requirement suppresses a substantial fraction of signal events within the $\eta_d$ short-lifetime regime. This attenuation of signal acceptance at small $c\tau_{\eta_d}$ directly results in a weakened exclusion reach for $\text{BR}(h_1 \to q_d \bar{q}_d)$ in Fig.~\ref{fig:13TeV-all}. 
Conversely, the DS3 scenario yields a lower signal efficiency across the majority of the considered $\eta_d$ lifetime range due to the additional presence of stable $\tilde{\omega}$ particles, which leads to generally less stringent projected $\text{BR}(h_1 \to q_d \bar{q}_d)$ exclusion limits compared to those for DS1 and DS2. Regarding cascade decay scenarios, the fixed LLP multiplicity ($N_{\eta_d}=4$) and broader angular distributions among $\eta_d$'s facilitate higher survival rates under the jet veto requirement, particularly at short $\eta_d$ lifetimes. However, as $c\tau_{\eta_d}$ increases, the sequential decay of multiple intermediate states significantly constrains the geometric acceptance; it becomes increasingly improbable for all final-state products to remain within the sensitive volume of CMS endcap muon detectors. As a result, the exclusion sensitivity for cascade decays diminishes more precipitously than in dark shower scenarios at large $c\tau_{\eta_d}$. 

\begin{figure}[htbp]
    \centering
    \includegraphics[width=1.0\textwidth]{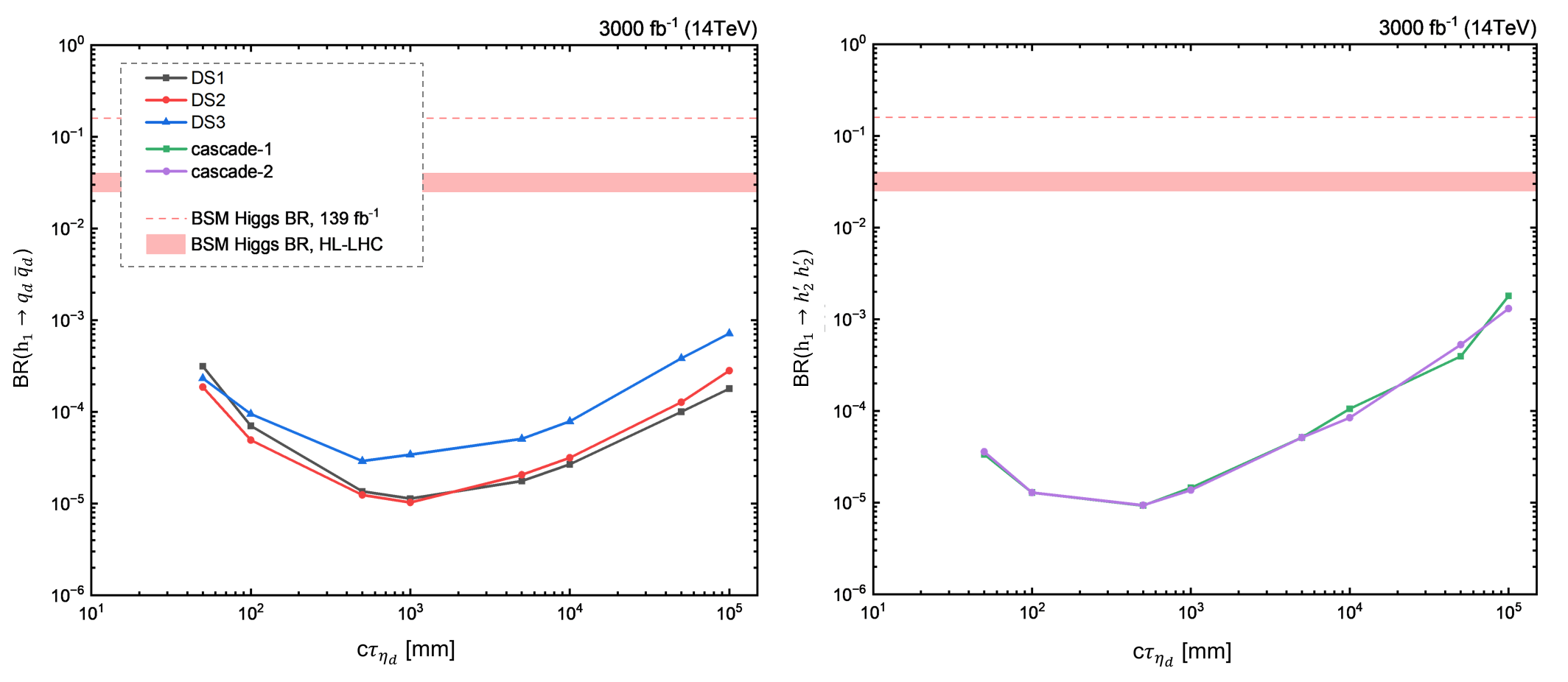}
    \caption{Similar to Fig.~\ref{fig:13TeV-all}, but for the HL-LHC ($14$ TeV, $3000$~fb$^{-1}$) analysis under the condition that events satisfy at least one trigger from either Category I or II (see Table~\ref{tab:trigger}). An optimistic background-free assumption is adopted here. Different colors indicate the signal benchmarks: black (DS1), red (DS2), and blue (DS3) for dark shower processes, and green (cascade-1) and purple (cascade-2) for cascade decays. }
    \label{fig:14TeV-all}
\end{figure}

Moreover, Figure~\ref{fig:14TeV-all} presents the projected exclusion limits at $\sqrt{s} = 14$~TeV. For this HL-LHC sensitivity study, an optimistic, background-free scenario is adopted, wherein the $95\%$ CL upper limit is defined by the requirement of $N_S = 3$ signal events. It is obvious to find that the optimistic HL-LHC exclusion limits demonstrate a consistent enhancement across the entire $c\tau_{\eta_d}$ range for all five BPs relative to the 13~TeV projections. This improvement demonstrate that dedicated trigger strategies, specifically those optimized for exotic Higgs decays and displaced signatures, will play a pivotal role during HL-LHC stage.

To further contextualize these findings, our results are presented in conjunction with the current observed upper limits on undetected Higgs decays from ATLAS and CMS at $\sqrt{s} = 13$~TeV ($139~\mathrm{fb}^{-1}$, red dashed line)~\cite{Cepeda:2019klc} and the projected HL-LHC sensitivity band (red shaded area)~\cite{Carrasco:2023loy}. For the recasting of current results ($\sqrt{s} = 13$~TeV with $137~\text{fb}^{-1}$), several exclusion curves remain above the red shaded band, particularly for $\eta_d$ with the short- and long-lifetime regimes, indicating that contemporary search strategies lack sufficient sensitivity in these regions. In contrast, the projected HL-LHC limits, leveraging our optimized trigger strategies, fall entirely below the red shaded band across the full parameter space examined. Specifically, the branching ratios $\text{BR}(h_1 \to q_d \bar{q}_d)$ (dark showers) and $\text{BR}(h_1 \to h'_2 h'_2)$ (cascade decays) are projected to be constrained to $\mathcal{O}(10^{-5})$ at the HL-LHC. 
This signifies an enhancement in exclusion power by approximately two orders of magnitude relative to the current constraints derived from the $13~\text{TeV}$ dataset with $137~\text{fb}^{-1}$ of integrated luminosity. This systematic enhancement demonstrates the substantial added value of our proposed search and underscores its potential to probe previously inaccessible regions of the Higgs exotic decay branching ratio.

\section{Conclusion} 
\label{sec:conclusion} 

In this work, we have presented a comprehensive study of novel signatures arising from Higgs exotic decays into a QCD-like dark sector via the Higgs portal. Focusing on a scenario where the lightest dark meson \(\eta_d\) has a mass of 3 GeV, dominantly decays to gluon pairs, and is long-lived, we investigated two distinct production mechanisms: the cascade decay \(h_1 \to h'_2 h'_2 \to 4\eta_d\) and the dark shower process \(h_1 \to q_d \bar{q}_d \to \text{multiple $\eta_d$'s}\). The resulting final states consist of gluon-rich ``dark jets" that present unique experimental challenges and opportunities at the LHC.

A central finding of our analysis is that the low transverse momentum characteristic of dark jets from Higgs decays renders conventional high-\(H_T\) or high-\(p_T^{\text{miss}}\) trigger strategies highly inefficient, leading to severe signal loss. To overcome this, we designed and implemented a dedicated, channel-optimized trigger strategy for the HL-LHC, categorizing events into hadronic (Category I) and leptonic (Category II) final states tailored to the major Higgs production modes. This strategy enhances the signal acceptance by more than an order of magnitude, establishing optimized triggering as a crucial prerequisite for probing such exotic Higgs decays.

For shorter-lived \(\eta_d\) (\(c\tau_{\eta_d} \sim 1\ \text{mm} - 1\ \text{m}\)), we analyzed emerging jet (EJ) signatures within the tracker system. Utilizing track-based observables such as the median impact parameter \(\langle IP_{2D} \rangle\) and the displaced track fraction \(\alpha_{3D}\), and defining optimized 1-EJ and 2-EJ event selection channels with topological discriminators (\(m_{jj}\), \(M_T\), \(m_{\text{EJ}}\), \(\Delta\phi\)), we achieved powerful separation from the dominant QCD heavy-flavor background. Our HL-LHC projections (\(\sqrt{s}=14\ \text{TeV}\), \(\mathcal{L}=3000\ \text{fb}^{-1}\)) demonstrate the ability to probe exotic Higgs decay branching ratio sensitivities of \(\mathcal{O}(10^{-4}-10^{-2})\) for dark shower scenarios and \(\mathcal{O}(10^{-3}-10^{-1})\) for cascade decays across a wide range of $\eta_d$ lifetimes, significantly surpassing future model-independent limits.

For longer lifetimes (\(c\tau_{\eta_d} \sim 50\ \text{mm} - 100\ \text{m}\)), we explored displaced showers in the CMS muon system, characterized by clustered hits (CSC clusters) from late-decaying \(\eta_d\). Key discriminators such as the cluster multiplicity \(N_{\text{clusters}}\), cluster energy, and the azimuthal correlation \(\Delta\phi(\vec{p}_T^{\text{miss}},\text{cluster})\) effectively suppress backgrounds from punch-through jets and muon bremsstrahlung. Projections show that with our dedicated trigger strategy, the HL-LHC can reach exotic Higgs decay branching ratios down to \(\mathcal{O}(10^{-5}-10^{-3})\) across a wide range of $\eta_d $ lifetimes.

Our study highlights distinctive features of the considered benchmark scenarios: Dark shower processes (DS1, DS2, DS3) yield high track/cluster multiplicities, aiding reconstruction. The semi-visible scenario DS3, with stable dark vector mesons, produces a missing transverse momentum signature requiring specific kinematic cuts. Cascade decays lead to a multi-jet topology with large angular separation among \(\eta_d\) particles, rendering dijet mass reconstruction ineffective but allowing discrimination via low jet mass \(m_{\text{EJ}}\).

In summary, this work provides a robust, optimized framework for probing gluons-enriched dark jets from Higgs exotic decays at the HL-LHC. The proposed trigger and analysis strategies dramatically improve sensitivity to these challenging signatures, opening a concrete pathway to explore the rich phenomenology of a composite dark sector. Future studies may extend this methodology to heavier long-lived $\eta_d$ particles decaying into charm quarks or tau leptons~\cite{Knapen:2021eip,Lu:2023gjk}, further broadening the scope of LLP searches in QCD-like dark sectors. 

\section*{Acknowledgments}
We thank Juliana Carrasco, Jose Zurita, Christina W. Wang, Dongjoo Kim, Soojin Lee, Jeonghyeon Song for helpful discussions. The authors gratefully acknowledge the valuable discussions and insights provided by the members of the China Collaboration of Precision Testing and New Physics. C.T.L., W.Y.C. and H.X.S. are supported by the National Natural Science Foundation of China (NNSFC) under grants No.~12335005, No.~12575118, and the Special funds for postdoctoral overseas recruitment, Ministry of Education of China. 

\appendix
\section{Kinematic analysis of the cascade decay scenario}
\label{appendix:A} 

In Section~\ref{sec:analysis}, we noted that the cascade decay scenario lacks a reconstruction peak at the Higgs mass in the $m_{jj}$ distribution. In this appendix, we investigate the origin of this behavior by analyzing the jet multiplicity and the angular distance $\Delta R$ of $\eta_d$'s. We utilize the ggH production mode with $c\tau_{\eta_d} = 25$~mm as a representative example, noting that other production modes exhibit similar kinematic trends.

\begin{figure}[htbp]
    \centering
    \includegraphics[width=0.9\textwidth]{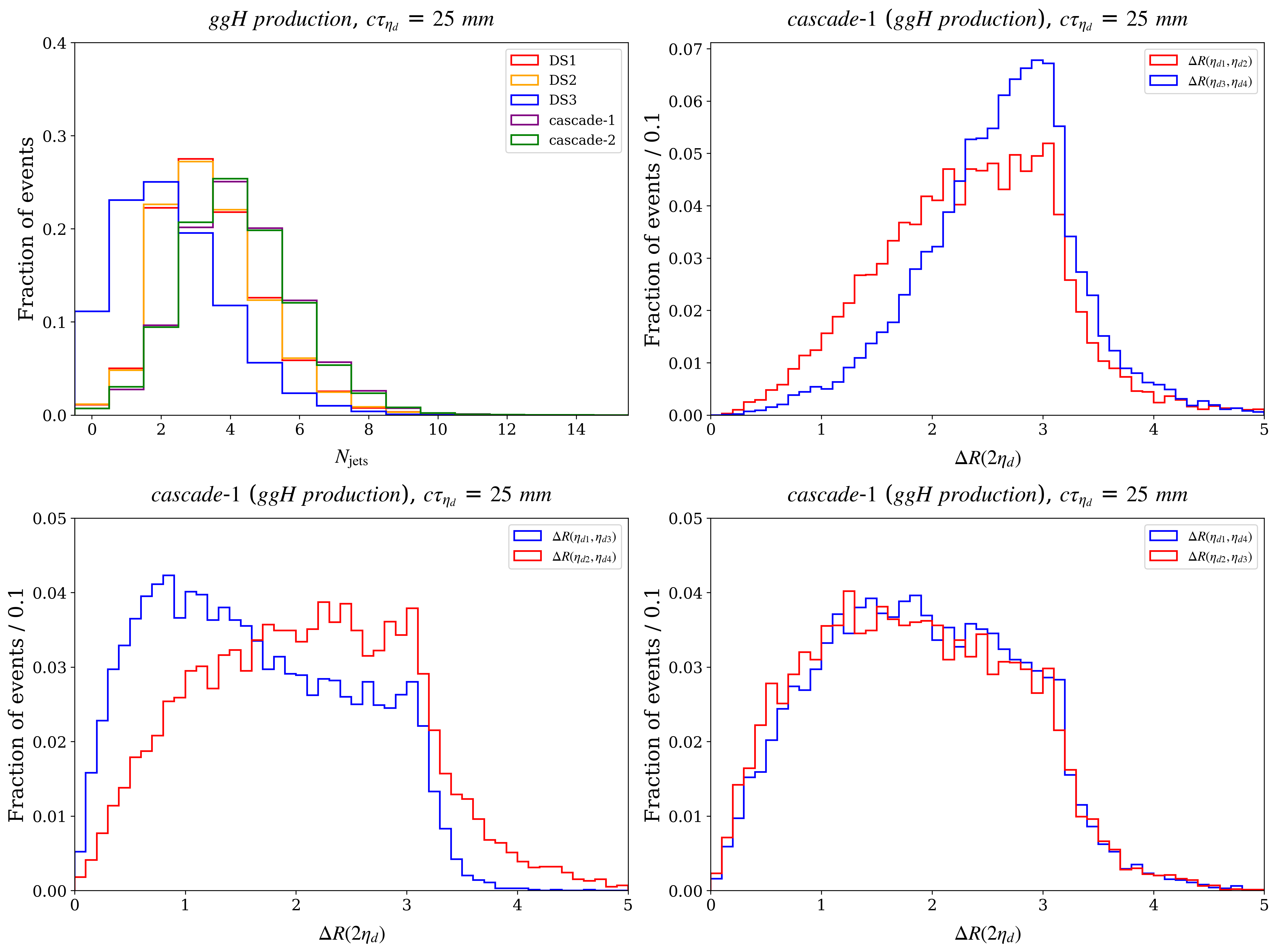}
    \caption{Kinematic distributions for the signal benchmarks in the ggH production mode ($\sqrt{s}=14$~TeV, $c\tau_{\eta_d} = 25$~mm).
    We present the number of reconstructed jets $N_{\text{jets}}$ for all five BPs (upper-left); and for the cascade-1 scenario, the angular distance $\Delta R$ between the same-parent pairs $\eta_{d1}\text{-}\eta_{d2}$ and $\eta_{d3}\text{-}\eta_{d4}$ (upper-right), the cross-parent pairs $\eta_{d1}\text{-}\eta_{d3}$ and $\eta_{d2}\text{-}\eta_{d4}$ (lower-left), and the cross-parent pairs $\eta_{d1}\text{-}\eta_{d4}$ and $\eta_{d2}\text{-}\eta_{d3}$ (lower-right). }
    \label{fig:Njet+deltaR}
\end{figure}

In the upper-left panel of Fig.~\ref{fig:Njet+deltaR}, we compare the exclusive jet multiplicity ($N_{\text{jets}}$) for the five signal BPs. While events from dark showers are predominantly characterized by 2-jet topologies, events from cascade decays ($h \to 2h_2' \to 4\eta_d$) exhibit a clear tendency towards higher multiplicities, frequently yielding three or more reconstructed jets.

To understand this multi-jet structure, we analyze the angular distance $\Delta R$ between $\eta_d$'s in the cascade-1 scenario. We strictly define the four final-state dark mesons ($\eta_{d1}, \eta_{d2}, \eta_{d3}, \eta_{d4}$) based on the energy hierarchy:
\begin{itemize}
    \item The leading (higher energy) intermediate scalar $h_2'$ decays into $\eta_{d1}$ and $\eta_{d2}$.
    \item The sub-leading (lower energy) intermediate scalar $h_2'$ decays into $\eta_{d3}$ and $\eta_{d4}$.
\end{itemize} 
The $\Delta R$ distributions for all pairwise combinations are shown in the upper right, lower left, and lower right panels of Fig.~\ref{fig:Njet+deltaR}. 
Crucially, the upper-right panel shows the separation between $\eta_d$ pairs originating from the same parent ($\eta_{d1}\text{-}\eta_{d2}$ and $\eta_{d3}\text{-}\eta_{d4}$). The distributions peak at $\Delta R \approx 1-3$, which is significantly larger than the jet clustering radius ($R=0.5$). 
This large angular distance implies that even the $\eta_d$'s from the same parent cannot be captured within a single jet cone. Consequently, the two leading jets ($J_1, J_2$) used in the main analysis capture only a fraction of the Higgs decay products, resulting in the failure of $m_{h_1}$ reconstruction of the $m_{jj}$ distribution.

\section{Supplementary results with \textit{R} strategy in the EJ analysis}
\label{appendix:B} 

In the main text, we utilized the data-driven \textit{It5} tracking efficiency map to estimate the HL-LHC sensitivity. For completeness and to assess the impact of tracking modeling assumptions, we present in this appendix the corresponding results using the simplified geometric \textit{R} strategy.

\begin{figure}[htbp]
    \centering
    \includegraphics[width=0.6\textwidth]{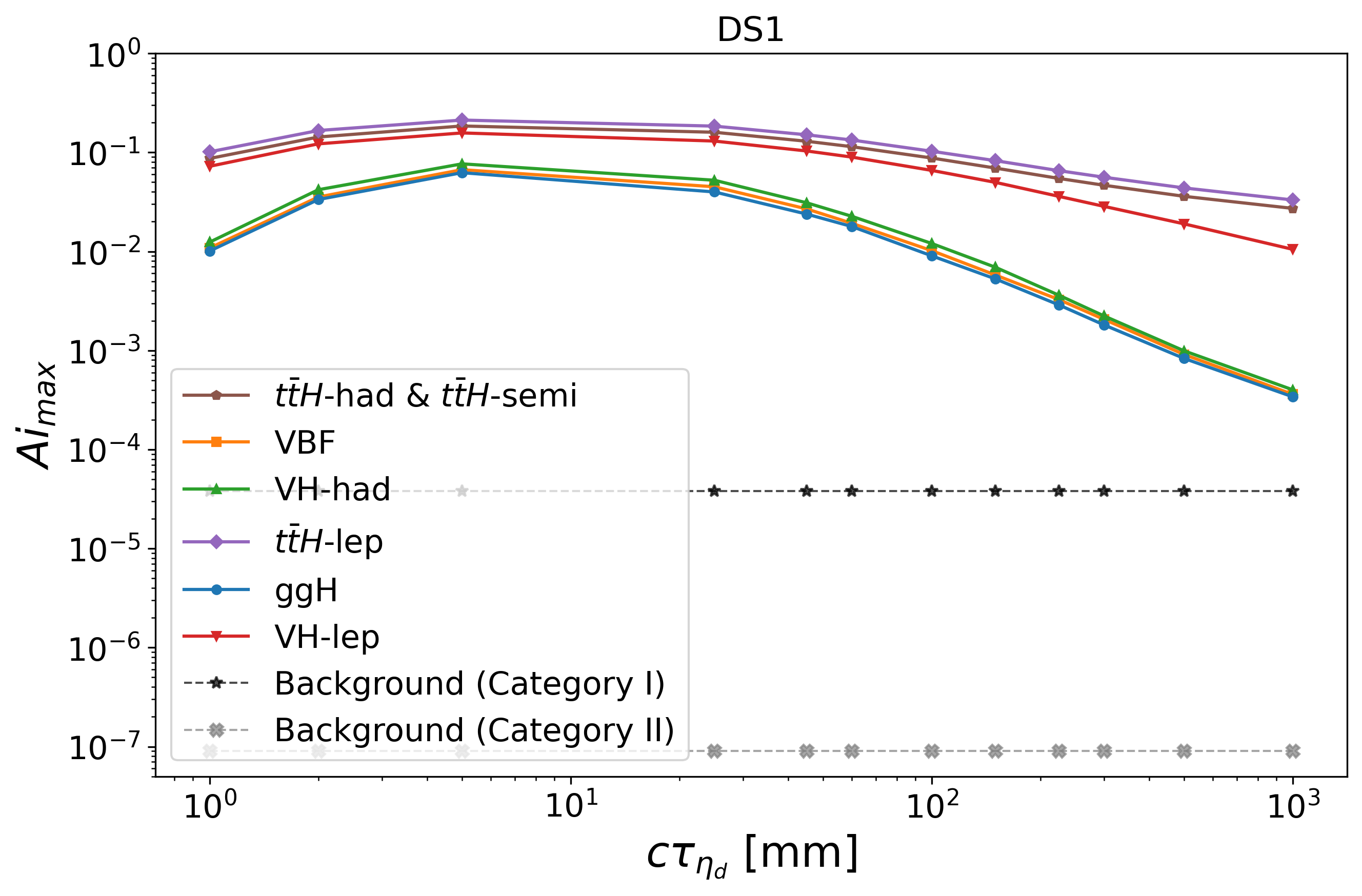} 
    \caption{Maximum signal acceptance ($A_{i,\text{max}}$) as a function of proper lifetime $c\tau_{\eta_d}$ for the DS1 at $\sqrt{s}=14$~TeV, obtained using the \textit{R} strategy.}
    \label{fig:14tev_amax_R}
\end{figure}

\begin{figure*}[htbp]
    \centering
    \begin{subfigure}[b]{0.48\textwidth}
        \includegraphics[width=\textwidth]{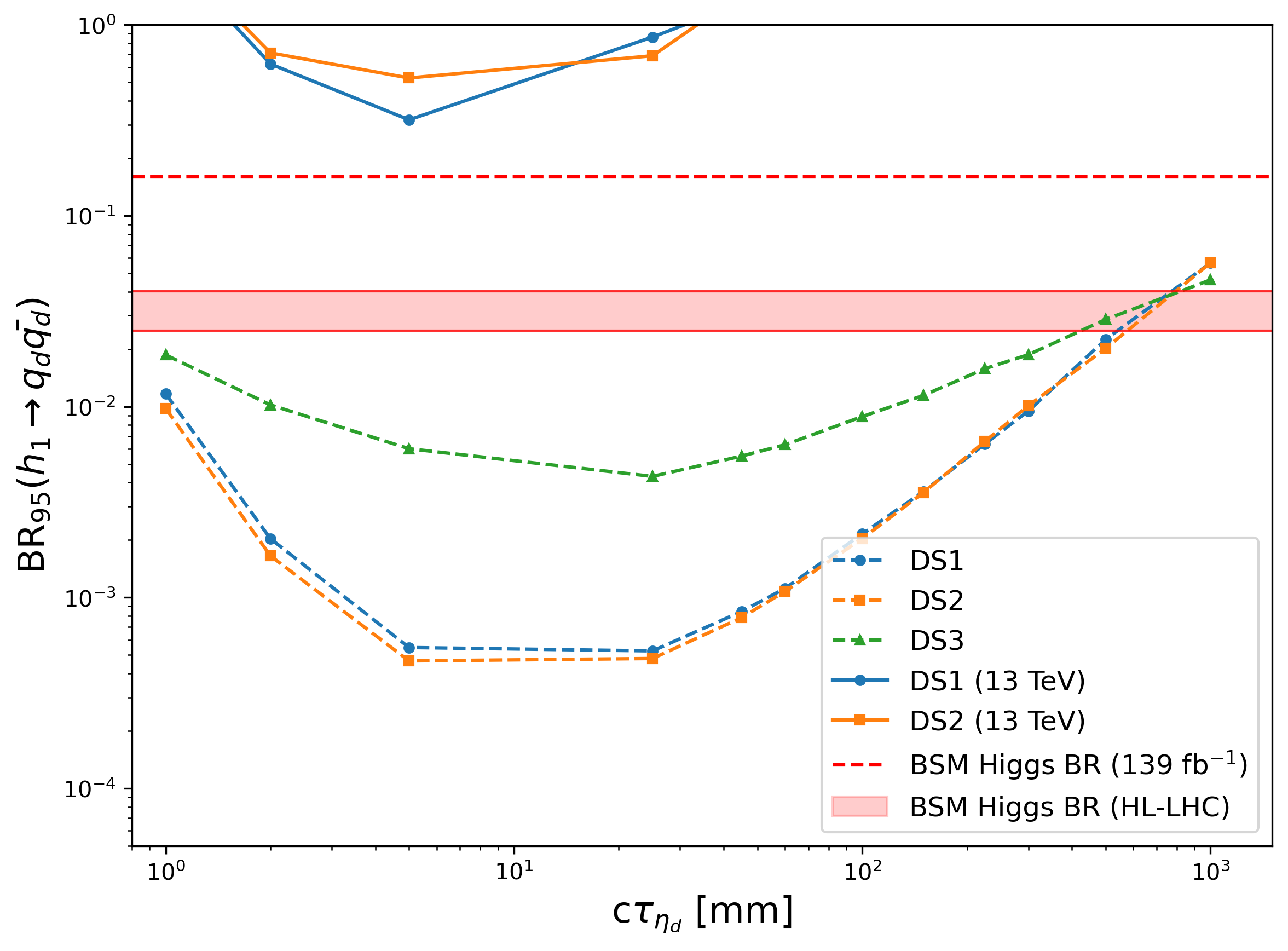} 
        \label{fig:limit_ds_cat1_R}
    \end{subfigure}
    \hfill
    \begin{subfigure}[b]{0.48\textwidth}
        \includegraphics[width=\textwidth]{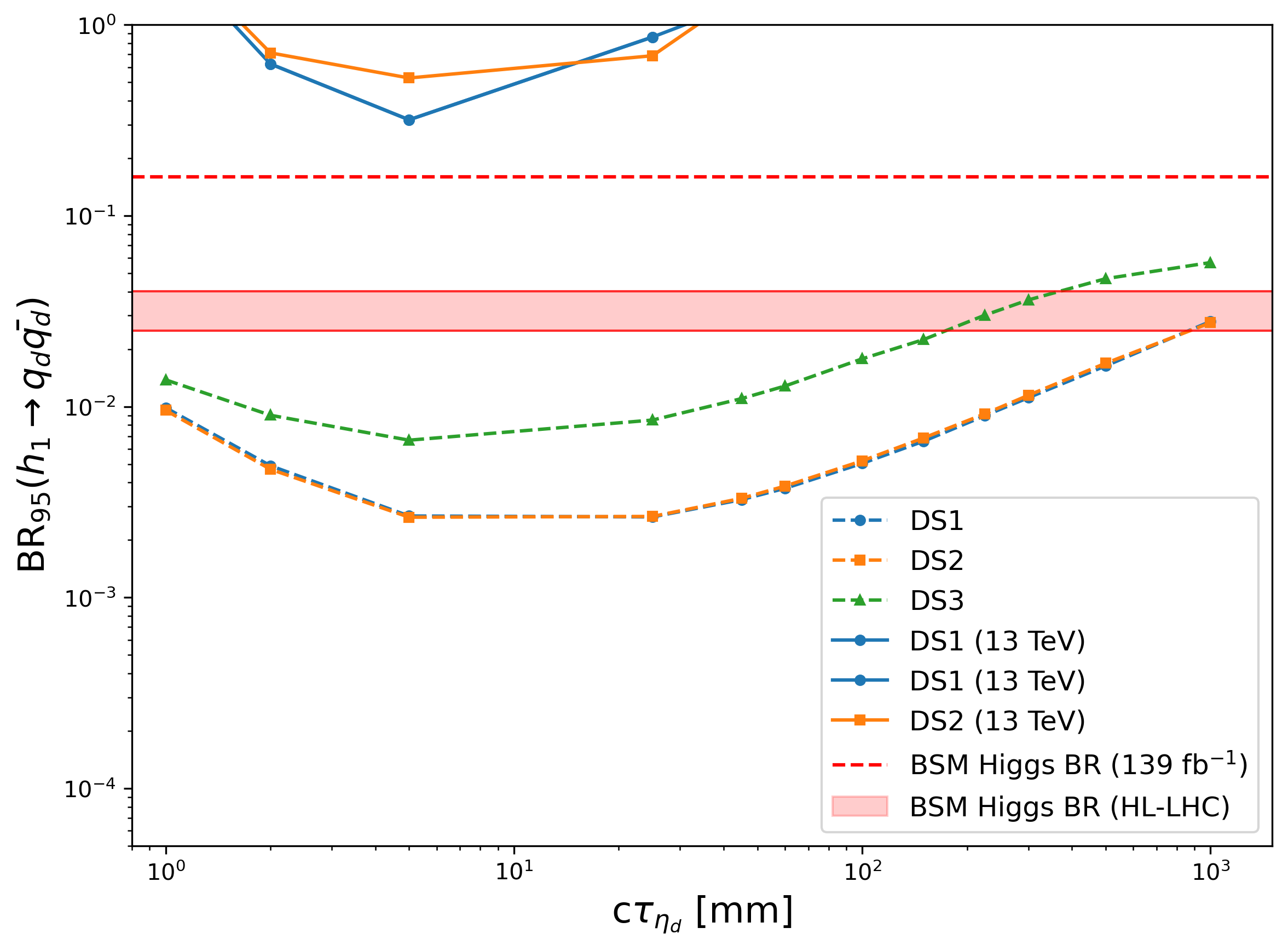} 
        \label{fig:limit_ds_cat2_R}
    \end{subfigure}
    
    \vspace{0.2cm}
    
    \begin{subfigure}[b]{0.48\textwidth}
        \includegraphics[width=\textwidth]{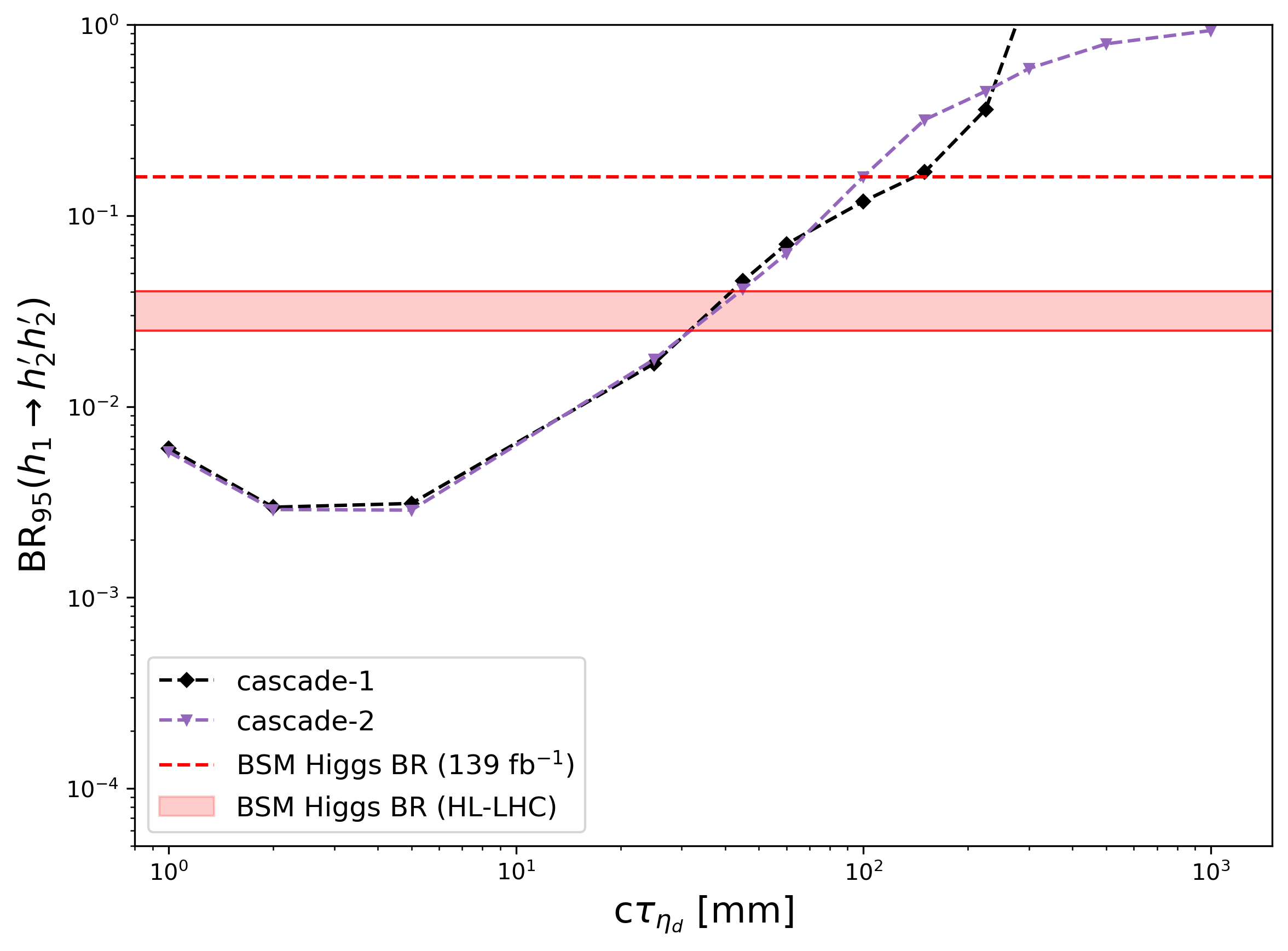} 
        \label{fig:limit_cas_cat1_R}
    \end{subfigure}
    \hfill
    \begin{subfigure}[b]{0.48\textwidth}
        \includegraphics[width=\textwidth]{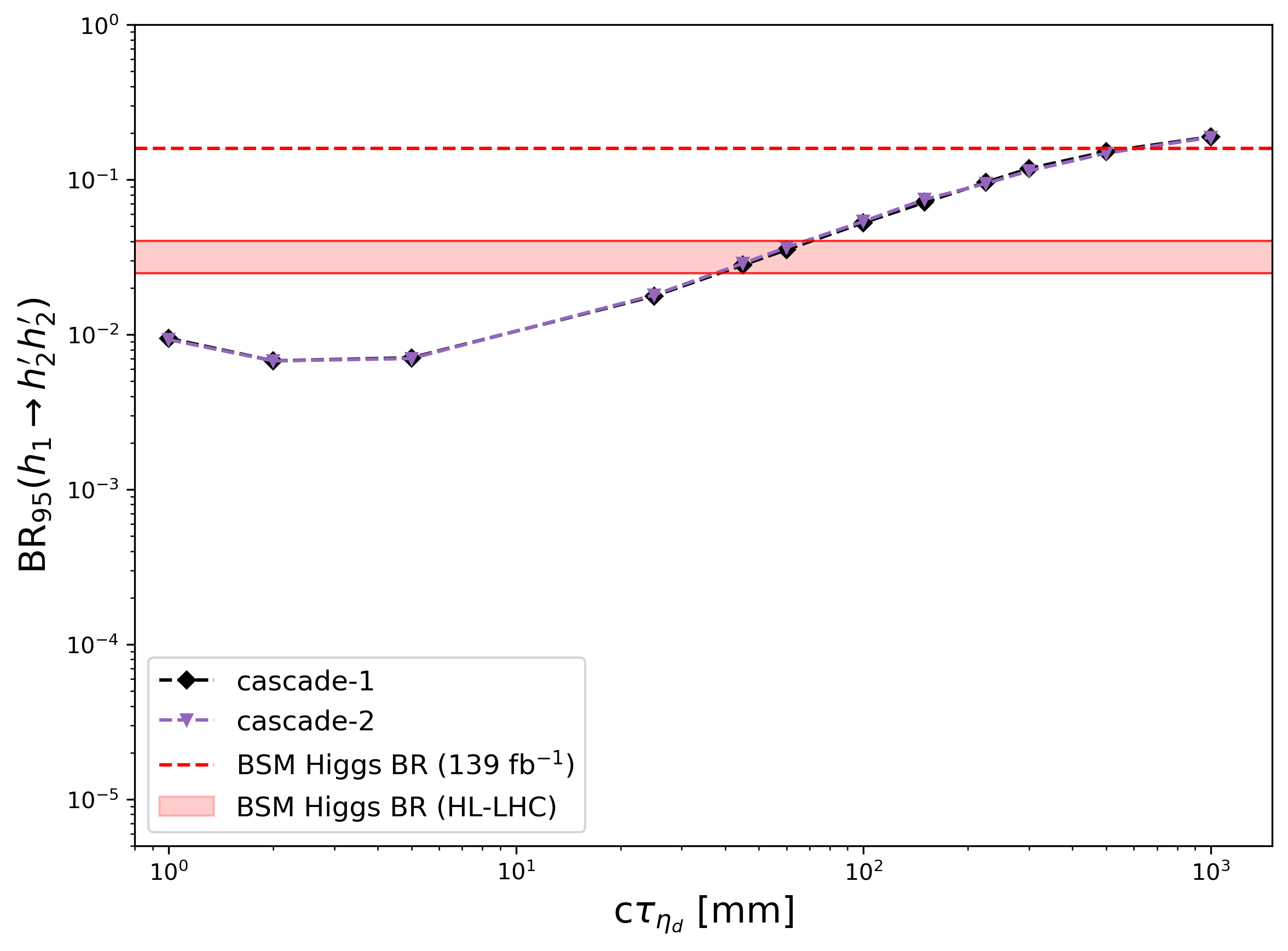} 
        \label{fig:limit_cas_cat2_R}
    \end{subfigure}
    
    \caption{Projected $95\%$ C.L. upper limits on exotic Higgs decay branching ratios using the \textit{R} strategy.
    The layout follows Fig.~\ref{fig:limits_14tev_grid}: dark shower (top) and cascade decay (bottom) scenarios under Category I (left) and Category II (right) triggers.}
    \label{fig:limits_14tev_grid_R}
\end{figure*}

Figure~\ref{fig:14tev_amax_R} displays the maximum signal acceptance ($A_{i,\text{max}}$) for the DS1 using the \textit{R} strategy as an example. Compared to the \textit{It5} results (Fig.~\ref{fig:14tev_amax}), the \textit{R} strategy exhibits a notably steeper decline in efficiency starting from $c\tau_{\eta_d} \approx 25$~mm. 
Moreover, the projected $95\%$ C.L. exclusion limits derived using the \textit{R} strategy are shown in Fig.~\ref{fig:limits_14tev_grid_R}. Consistent with the acceptance trends, the \textit{R} strategy yields generally weaker constraints than \textit{It5}. This difference is particularly pronounced for the cascade decay scenarios and in the large $\eta_d$ lifetime regime, where the limit curves rise much more rapidly as $c\tau_{\eta_d}$ increases. 

\section{$\Delta\phi(\vec{p}_T^{\,\text{miss}},\,\text{cluster})$ and $\Delta R(\text{jet, cluster})$ distributions for each Higgs production mode}
\label{appendix:C}

In this appendix, we present a detailed decomposition of the independent contributions from the six primary Higgs production channels to the total signal yield, corresponding to an integrated luminosity of $137~\text{fb}^{-1}$ at $\sqrt{s} = 13$~TeV. The six primary Higgs production channels include ggH, VBF, VH-had, VH-lep, $t\bar{t}$H-had \& $t\bar{t}$H-semi and $t\bar{t}$H-lep \& $t\bar{t}$H-semi processes. 

\begin{figure}[htbp]
    \centering
    \includegraphics[width=0.9\textwidth]{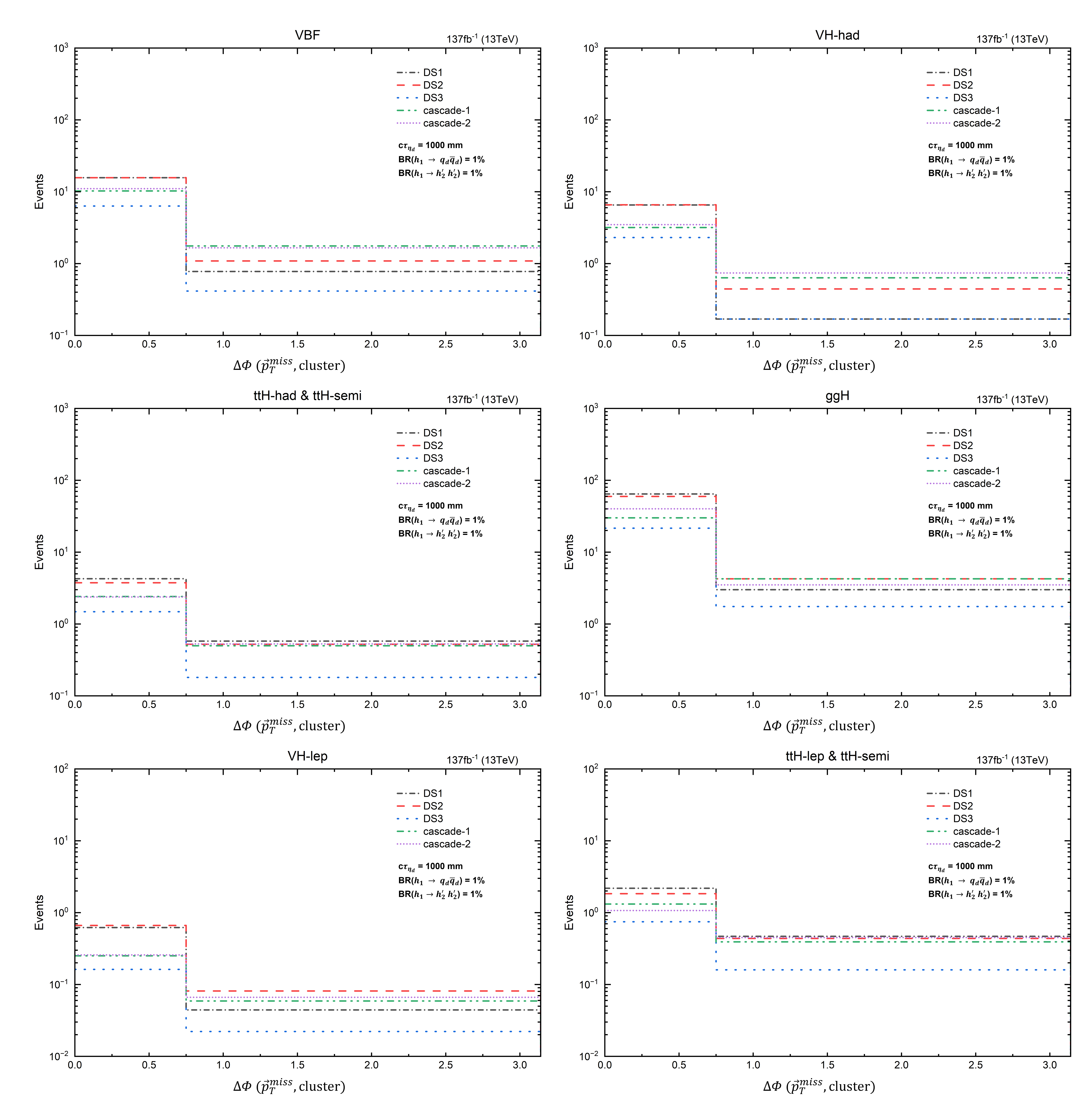}
    \caption{Decomposition of the $\Delta\phi(\vec{p}_T^{\,\text{miss}},\,\text{cluster})$ signal yield into the six constituent Higgs production channels at $c\tau_{\eta_d} = 1000~\mathrm{mm}$. For each channel, five benchmark points are compared: DS1, DS2, and DS3 (shown in black, red, and blue) and cascade-1 and cascade-2 (shown in green and purple). }
    \label{fig:deltaphi-individual}
\end{figure}

Figure~\ref{fig:deltaphi-individual} presents the decomposition of the $\Delta\phi(\vec{p}_{\text{T}}^{\text{miss}}, \text{cluster})$ distribution and all results are presented assuming a $\eta_d$ proper decay length of $c\tau_{\eta_d} = 1000$~mm. The corresponding expected number of signal events is given by: 
\begin{equation}
    N_i(m_{\eta_d}, c\tau_{\eta_d}) = \sigma_{\text{proc}}^{\text{SM}} \times \mathcal{B} \times A_i(c\tau_{\eta_d}, m_{\eta_d}) \times \mathcal{L}
    \label{eq:signal_yield}
\end{equation}
where $\mathcal{B}$ denotes the branching ratio of the exotic Higgs decay, corresponding to BR$(h_1 \to q_D \bar{q}_D)$ for the dark shower scenarios and BR$(h_1 \to h_2' h_2')$ for the cascade decay scenarios. In this study, a reference branching ratio of $\mathcal{B} = 1\%$ is assumed for the expected signal yields.

According to Eq.~\eqref{eq:signal_yield}, the expected signal yield scales linearly with the Higgs production cross-section. Consequently, the ggH channel, characterized by the largest SM cross-section, provides the dominant contribution to the total signal yield. For all Higgs production modes under consideration, the $\Delta\phi(\vec{p}_T^{\text{miss}}, \text{cluster})$ distributions across the five BPs in Fig.~\ref{fig:deltaphi-individual} exhibit kinematic features consistent with the inclusive results presented in Fig.~\ref{fig:delta phi all-0.75}.

\begin{figure}[htbp]
    \centering
    \includegraphics[width=0.8\textwidth]{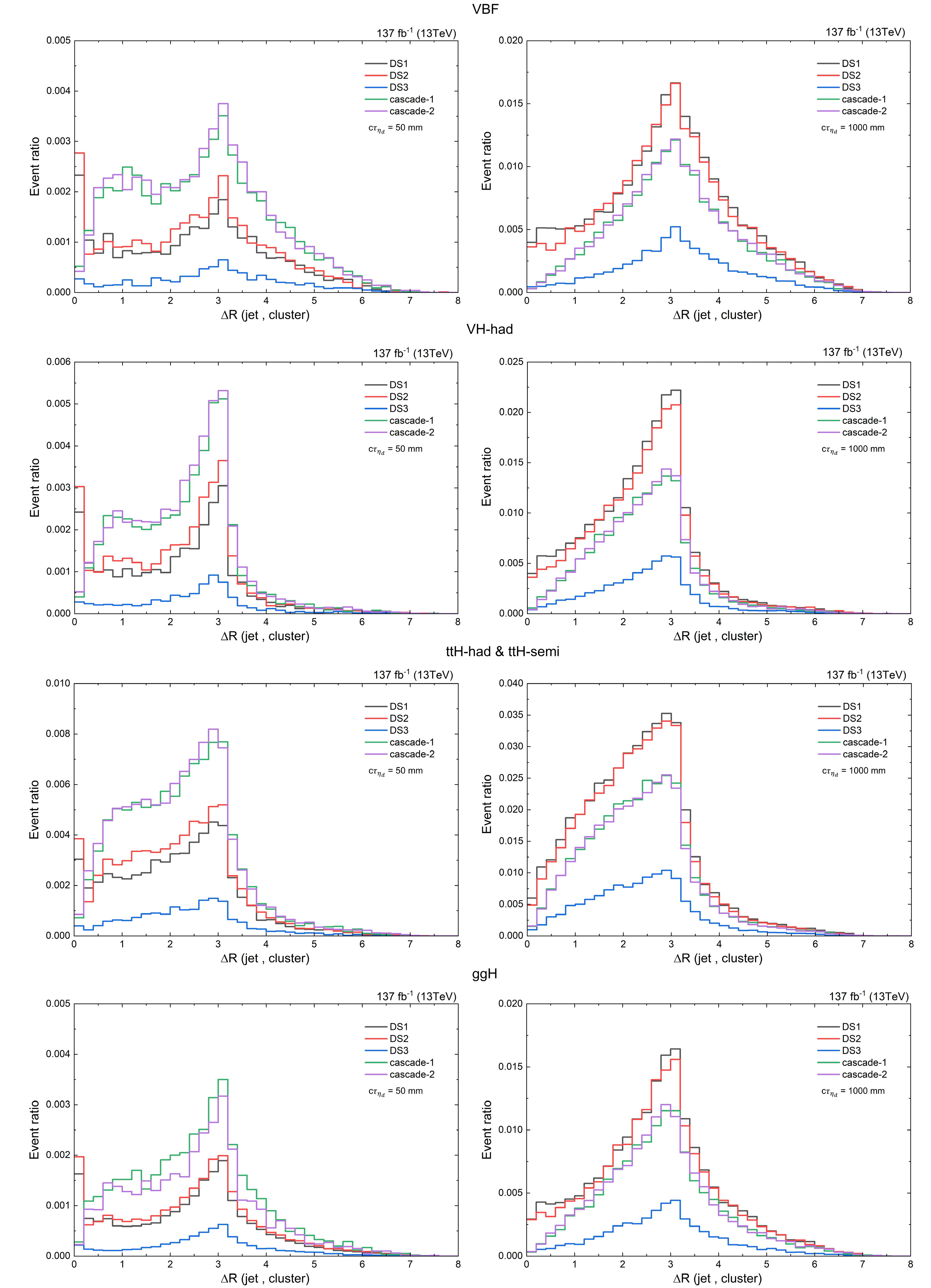}
    \caption{Distributions of the angular separation $\Delta R(\text{jet, cluster})$ for the hadronic final state production channels, including VBF, VH-had, and $t\bar{t}$H-had \& $t\bar{t}$H-semi and ggH. The panels illustrate the results evaluated at $c\tau_{\eta_d} = 50$~mm (left) and 1000~mm (right) across the five benchmark points: DS1 (black), DS2 (red), and DS3 (blue), alongside cascade-1 (green) and cascade-2 (purple). }
    \label{fig:deltaR-individual-had}
\end{figure}

\begin{figure}[htbp]
    \centering
    \includegraphics[width=1\textwidth]{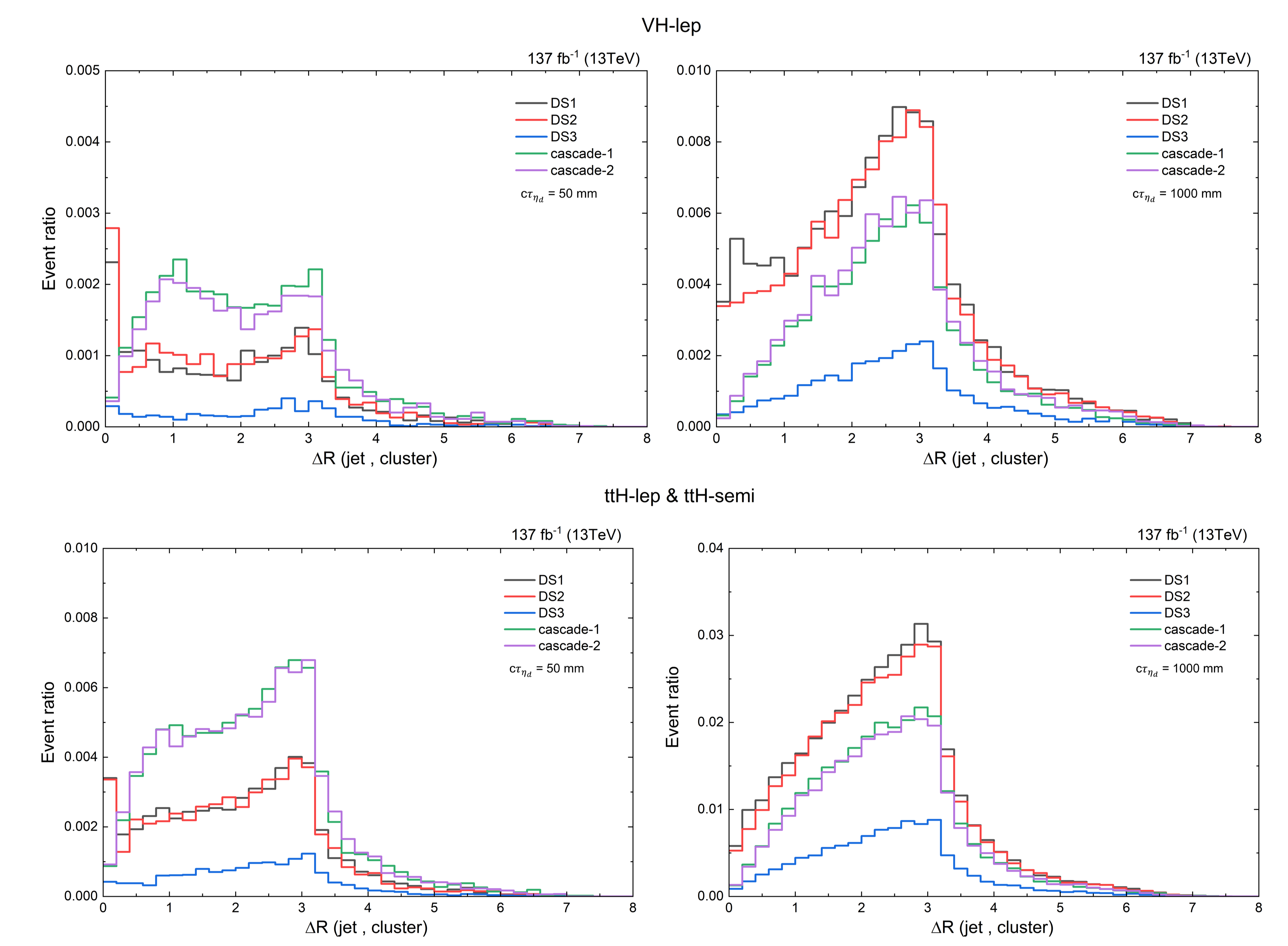}
    \caption{Similar to Fig.~\ref{fig:deltaR-individual-had}, but for leptonic final-state production channels, specifically VH-lep, $t\bar{t}$H-lep \& $t\bar{t}$H-semi. }
    \label{fig:deltaR-individual-lep}
\end{figure}

The following two figures provides a detailed decomposition of the $\Delta R(\text{jet, cluster})$ distributions, augmenting the aggregate results presented in Figure~\ref{fig:deltaR} by showing the specific contributions from individual signal components. Figure~\ref{fig:deltaR-individual-had} displays the $\Delta R(\text{jet, cluster})$ distributions for hadronic final-state production channels, while Fig.~\ref{fig:deltaR-individual-lep} illustrates the corresponding distributions for leptonic final-state channels. Both of them compare results across the five BPs for $c\tau_{\eta_d} = 50$~mm (left) and $1000$~mm (right), respectively.

The distributions for all individual Higgs production modes remain qualitatively consistent with the inclusive results shown in Fig.~\ref{fig:deltaR}. At $c\tau_{\eta_d} = 50$~mm, a pronounced peak at small $\Delta R(\text{jet, cluster})$ emerges in the DS1 and DS2 scenarios, suggesting a high degree of collimation where signal clusters frequently overlap or merge with reconstructed jets. In the DS3 scenario, despite a suppressed overall yield due to the increased fraction of invisible stable $\omega$ particles, a similar peak at small $\Delta R$ persists.

Conversely, the small-$\Delta R$ peak vanishes across all BPs at $c\tau_{\eta_d} = 1000$~mm. This attenuation is attributed to the increased decay length of $\eta_d$, which results in decay vertices significantly displaced from the tracker system, thereby decoupling the signal clusters from the primary jet activity. This kinematic evolution explains the reduced sensitivity in the Higgs exotic decay branching ratio exclusion limits at shorter $\eta_d$ lifetimes, where jet and cluster overlap significantly complicates signal-background discrimination.

\end{document}